\documentclass[floats,floatfix,amssymb,prd,twocolumn,superscriptaddress,nofootinbib,preprintnumbers]{revtex4-1}

\usepackage{subcaption}
\usepackage{ragged2e}
\DeclareCaptionJustification{justified}{\justifying}
\makeatletter
\newcommand{\subsetsim}{\mathrel{\mathpalette\subset@sim\relax}}
\newcommand{\subset@sim}[2]{%
  \vtop{\offinterlineskip\m@th
    \ialign{\hfil##\cr
      $#1\subset$\cr\noalign{\kern0.5pt}\scalebox{0.9}{$#1\sim$}\cr
    }%
  }%
}
\makeatother

\usepackage{amssymb,amsmath,verbatim,mathtools,needspace,enumitem,etoolbox,graphicx,physics,microtype,afterpage,bm}
\usepackage[dvipsnames, usenames]{xcolor}
\definecolor{linkcolor}{rgb}{0.0,0.3,0.5}
\usepackage{booktabs}
\usepackage[unicode, colorlinks=true, linkcolor=linkcolor, citecolor=linkcolor, filecolor=linkcolor,urlcolor=linkcolor, pdfusetitle]{hyperref}
\usepackage[all]{hypcap}
\usepackage[T1]{fontenc}
\usepackage[utf8]{inputenc}
\usepackage{tabularx}
\usepackage{float}
\usepackage{cleveref}

\usepackage{multirow}
\usepackage{pifont}
\usepackage{lmodern}
\usepackage{ulem}

\allowdisplaybreaks
\usepackage{tikz}
\usepackage{framed}
\usepackage{hyperref}
\hypersetup{colorlinks, citecolor=bluscuro, linkcolor=black, urlcolor=bluscuro}
\definecolor{bluscuro}{rgb}{0.15, 0.2, .85}

\newcommand{\be}{\begin{equation}}
\newcommand{\ee}{\end{equation}}

\newcommand{\btheta}{{\theta}}

\begin{document}

\title{
Forecasting the sensitivity of Pulsar Timing Arrays 
to gravitational wave backgrounds
}

\author{Stanislav Babak}
\email{stas@apc.in2p3.fr}
\affiliation{Astroparticule et Cosmologie, CNRS, Universit\'e Paris Cit\'e, F-75013 Paris, France}

\author{Mikel Falxa}
\email{falxa@apc.in2p3.fr}
\affiliation{LPC2E, CNRS, 3 Av. de la Recherche Scientifique, 45071 Orléans, France}
\affiliation{Astroparticule et Cosmologie, CNRS, Universit\'e Paris Cit\'e, F-75013 Paris, France}

\author{Gabriele Franciolini}
\email{gabriele.franciolini@cern.ch}
\affiliation{CERN, Theoretical Physics Department,
Esplanade des Particules 1, Geneva 1211, Switzerland} 

\author{Mauro Pieroni}
\email{mauro.pieroni@cern.ch}
\affiliation{CERN, Theoretical Physics Department,
Esplanade des Particules 1, Geneva 1211, Switzerland}

\begin{abstract}
Pulsar Timing Array (PTA) observations hinted towards the existence of a stochastic gravitational wave background (SGWB) in the nHz frequency band. Still, the nature of the SGWB signal cannot be confidently inferred from current data, and the leading explanation invokes mergers of supermassive black holes. If confirmed, such discovery would not only represent a turning point in our understanding of astrophysics, but it may severely limit the capability of searching for additional cosmological sources in the nHz frequency range. In this work, we build a simple framework to forecast the sensitivity of future PTA configurations and assess the parameter estimation of SGWB, which could consist of several contributions.
We release the python code \href{https://github.com/Mauropieroni/fastPTA/}{\texttt{fastPTA}} implementing this framework and ready to use. 
\end{abstract}

\maketitle

\preprint{CERN-TH-2024-039} 

{
  \hypersetup{linkcolor=black}
  \tableofcontents
}
\hypersetup{linkcolor=bluscuro}

\section{Introduction}
\label{sec:introduction}

Pulsar timing array~(PTAs) experiments are unique probes to study the gravitational-wave~(GW) spectrum in the nano-Hertz~(nHz) frequency range. Recently, the results reported by the NANOGrav~\cite{NANOGrav:2023gor,NANOGrav:2023hde}, EPTA (in combination with InPTA)\,\cite{EPTA:2023fyk,EPTA:2023sfo,EPTA:2023xxk}, PPTA\,\cite{Reardon:2023gzh,Zic:2023gta,Reardon:2023zen} and CPTA\,\cite{Xu:2023wog} collaborations hinted towards the existence of a common spectrum of a stochastic nature, with around 
$(2\divisionsymbol 4) \sigma$
significance for a Hellings-Down~(HD) angular correlation, consistent with the quadrupolar nature of GWs in general relativity~\cite{Hellings:1983fr}.

All collaborations reported the signal could be explained by a stochastic GW background (SGWB) with a preference for 
blue-tilted spectrum of energy density $\Omega_{\rm GW}(f)\propto f^{n_T}$, 
with $n_T\simeq 2$, and large uncertainties varying among different collaborations.
The astrophysical explanation of this signal would rely on the incoherent superposition of GW signals from a population of inspiralling supermassive black hole (SMBH) binaries on circular orbits, whose frequency spectrum is characterised by a scaling law $\Omega_{\rm GW} \propto f^{2/3}$~\cite{Phinney:2001di}.
Nevertheless, orbital eccentricity and stellar environment could affect the binary evolution which is coupled to the statistical fluctuation in the distribution of SMBHBs may play an important role and can lead to modified spectra (see, e.g.,~\cite{Sesana:2008mz,Kocsis:2010xa,Kelley:2016gse,Perrodin:2017bxr,Ellis:2023owy,NANOGrav:2023hfp,NANOGrav:2023hvm,Ghoshal:2023fhh, Bonetti:2017lnj}).
On the other hand, with current data, it is still impossible to rule out a cosmological explanation for the observed signal. The most studied scenarios capable of producing a SGWB in the nHz frequency range would be: first-order phase transitions~\cite{NANOGrav:2021flc,Xue:2021gyq,Nakai:2020oit,DiBari:2021dri,Sakharov:2021dim,Li:2021qer,Ashoorioon:2022raz,Benetti:2021uea,Barir:2022kzo,Hindmarsh:2022awe,Gouttenoire:2023naa,Baldes:2023fsp,Li:2023bxy}; cosmic strings and domain walls~\cite{Ellis:2020ena,Datta:2020bht,Samanta:2020cdk,Buchmuller:2020lbh,Blasi:2020mfx,Ramazanov:2021eya,Babichev:2021uvl,Gorghetto:2021fsn,Buchmuller:2021mbb,Blanco-Pillado:2021ygr,Ferreira:2022zzo,An:2023idh,Qiu:2023wbs,Zeng:2023jut,King:2023cgv,Babichev:2023pbf, Kitajima:2023cek,Barman:2023fad}; scalar-induced GWs generated from primordial fluctuations~\cite{Vaskonen:2020lbd,DeLuca:2020agl,Bhaumik:2020dor,Inomata:2020xad,Kohri:2020qqd,Domenech:2020ers,Namba:2020kij,Sugiyama:2020roc,Zhou:2020kkf,Lin:2021vwc,Rezazadeh:2021clf,Kawasaki:2021ycf,Ahmed:2021ucx,Yi:2022ymw,Yi:2022anu,Dandoy:2023jot,Zhao:2023xnh,Ferrante:2023bgz,Cai:2023uhc,Franciolini:2023pbf,Balaji:2023ehk, Liu:2023ymk}; 
tensor perturbations generated during inflation (see, e.g.,~\cite{Vagnozzi:2023lwo}) and many others (see also~\cite{Franciolini:2023wjm,Madge:2023dxc,Figueroa:2023zhu, Garcia-Bellido:2023ser, Murai:2023gkv, Konoplya:2023fmh, EPTA:2023xiy}).

Future observations will narrow down the range of models compatible with the data. For example, the forecasted sensitivity of future PTA experiments may improve by
various orders of magnitude compared to the ones achieved by current experiments. This will allow us to better understand the nature of the dominant source of GWB in the nHz while constraining signals from new physics appearing with amplitude below current sensitivity. 
Crucially, the presence of the currently observed foreground will greatly limit the constraining power of future observations on subdominant SGWB. 
In other words, regardless of the signal origin, currently hinted SGWB will constitute an unavoidable source of foreground noise when searching for subdominant contributions.

In this work, we aim to build a simplified framework capable of reliably estimating future PTA sensitivity without the need for expensive data simulation and analyses. 
We will remain agnostic on the nature of the signal, which we assume to be an isotropic, stationary SGWB characterized by the maximum likelihood power-law model resulting from the analysis of the recent PTA datasets~\cite{EPTA:2023sfo}, 
and we will model noises based on the currently observed pulsars.
We then forecast future PTA uncertainties and describe how they will shrink both with a longer observation time and a larger number of pulsars.
Finally, we will forecast the sensitivity to subdominant contribution to the SGWB, considering simple models as a proof of concept. Detailed analysis considering motivated cosmological SGWB, coming from new physics would be presented elsewhere.

The paper is organized as follows. 
In Sec.~\ref{sec:sensitivity}, we discuss how we build the PTA noise sensitivity, following Ref.~\cite{Hazboun:2019vhv} and with some additional simplifying assumptions. In Sec.~\ref{sec:ptasimulations} we describe the PTA data simulations we adopt to validate our proposed methodology against state-of-the-art pipelines for the PTA data analysis. 
In sec.~\ref{sec:results} we report the results while presenting conclusions and outlook in Sec.~\ref{sec:conclusions}.

At the repository linked in Ref.~\cite{code_repo}, we release the code \href{https://github.com/Mauropieroni/fastPTA/}{\texttt{fastPTA}} readily usable to estimate the sensitivities and the measurement uncertainties with future PTA configuration, with longer observation times and a simulated larger set of pulsars, which also features 
the possibility of including additional subdominant GWB contributions.

\section{Building the Pulsar timing array sensitivity}\label{sec:sensitivity}

In this section, we describe how, following Ref.~\cite{Anholm:2008wy,Chamberlin:2014ria,Rosado:2015epa,Hazboun:2019vhv}, we build the PTA sensitivity curves based on a few simplifying assumptions. In the following sections, we show that this framework reproduces measurement uncertainties estimated using current data (EPTA DR2, in particular,~\cite{EPTA:2023fyk}) and forecasted from mock data sets generated with state-of-the-art techniques.

As a starting point, we assume the combined PTA data consists of residuals of time of arrivals (TOAs) $d_{I}(t)$, where the index $I$ runs up the total number of pulsars observed $1, \dots, N_p$. Moreover, we will assume the GW signal and noise to be stationary\footnote{
In reality, we only need stationarity in the frequency band relevant to the targeted stochastic GW signal. See conclusions for a more comprehensive discussion of this point}
and define the Fourier domain data as
\begin{equation}
\tilde d_i^k  = \int_{-T_{\rm obs}/2}^{T_{\rm obs}/2} d_I(t) e^{-i 2 \pi f_k t } {\rm d }t .
\end{equation}
Given the finite duration of the data stream, we adopt a finite set of frequencies for the Fourier basis, $f_k = k/T_{\rm obs}$, starting from the inverse of the observation time $T_{\rm obs}$ and going up to the Nyquist frequency. 
We assume that the data contains the stochastic GW signal $\tilde s_I^k$ and noise $\tilde n_I^k$, as $\tilde d_I^k = \tilde s_I^k + \tilde n_I^k$, which we will further assume to be Gaussian with zero mean $\langle \tilde s_I^k \rangle = \langle \tilde n_I^k \rangle =0$.

Given the aforementioned assumptions,
one can write down the full covariance matrix, 
including contributions from both 
intrinsic pulsar noise, measurement process, and GWB as
\begin{equation}\label{e:C_IJ}
C_{IJ} = C_{n,IJ} + C_{h,IJ}\,.
\end{equation}
In the following, we will assume the noise
contribution to the covariant matrix to be dominated by the diagonal terms in the pulsar indices, i.e., to be uncorrelated among different pulsars. 
This allows us to approximate $C_{n,IJ} = \delta_{IJ} P_{n,I}$.
Additionally, the GW contribution presents the unique Hellings and Downs (HD) correlation pattern to be discussed in Sec.~\ref{sec:GW}.

\paragraph*{Transmission function.} 
The signal is extracted from timing residuals built by subtracting the expected time of arrival computed using the timing model.  Fitting for the parameters entering the timing model (i.e., describing each pulsar intrinsic rotation frequency, its derivative, proper motion, etc), results in a polynomial suppression of sensitivity at low frequencies.\footnote{Sensitivity to the GWs with frequencies below $1/T_\text{\tiny obs}$ can be retained in delays compared to higher order spin-down terms of the pulsar timing model (see e.g.,~\cite{DeRocco:2022irl,DeRocco:2023qae}). }  Crucially, the drastic decrease only happens around the frequency $1/T_{\rm obs}$.
This suppression of signal can effectively be described by a transmission function scaling as $1/f^6$ (for a quadratic spin-down model) below the frequency of $1/T_{\rm obs}$~\cite{Hazboun:2019vhv}. 
We model this transfer function as 
    \begin{equation}
        {\cal T}(f) \simeq \left[  1+ 
        1/\left (f T_\text{\tiny obs} \right ) \right ]^{-6}.
    \label{eq:t_f}
    \end{equation}
An additional loss of the sensitivity is induced around $f=1 /{\rm yr}$ from fitting the sky position and the proper motion and around $f = 2/ {\rm yr}$ due to parallax. This generates large spikes in the sensitivity curve.
Furthermore, if the pulsar is in a binary system with a period that falls within the frequency range, an additional dip in the transmission function will appear due to the fitting of the orbital parameters. 
As these losses typically appear at relatively high frequencies, we will neglect them in our estimates.

\subsection{Noise model}

Following the discussion in Ref.~\cite{Hazboun:2019vhv}, we isolate a few key features of the noise budget. For each pulsar, the noise contribution comes schematically from 
\begin{equation}
    \text{Noise budget}= \text{WN + RN + DM + SV},
\end{equation}
defined as follows:

\paragraph{White noise (WN).}
The time of arrival of the signal shows an overall white noise due to the finite signal-to-noise ratio (SNR) of the match filtering process adopted to extract them. It appears as a flat contribution to the power spectra of timing residuals. We characterize the white noise contribution for each pulsar as $P_I^{\rm WN} $, where $I$ is the index associated with each pulsar. 
    This is typically controlled by the parameters EQUAD, ECORR, and EFAC in PTA analyses~\cite{EPTA:2023sfo,NANOGrav:2023hde}.    
    This can be described by a time delay power spectrum 
        \begin{equation}
P_I^{\rm WN} =  2 \sigma^2 \Delta t,
    \end{equation}
where $\Delta t$ is the inverse of the observing cadence and $\sigma$ is the rms timing uncertainty.

\paragraph{Red noise (RN).}
The stochasticity in the pulsar's rotation causes an ``achromatic" red noise, which we model with a power-law power spectral density (PSD). This noise is independent of the frequency of the radio observation (hence ``achromatic'') and often dominates at low frequencies:
\begin{equation}\label{eq:RN}
P_I^{\rm RN} =   A^{\rm RN}_I  \left (\frac{f}{f_r} \right )^{\gamma^{\rm RN}_I};
    \end{equation}
where $f_r$ is a reference frequency often assumed to be 1/yr, and the amplitude $A^{\rm RN}_I$ and the spectral index $\gamma^{\rm  RN}_I$ is intrinsic to each pulsar $I$.

\paragraph{Chromatic noise components (DM, SV).}
Other important noise sources are temporal variations in dispersion measure (DM) and scattering variations (SV). Those components are chromatic and add time delays to the ToAs as $\propto \nu^{-2}$ (DM) and  $\propto \nu^{-4}$ (SV), where $\nu$ is the observed radio frequency, and both are caused by the time-varying electron
column density in the interstellar medium along the line of sight (see~\cite{EPTA:2021fqa} and references therein). The PSD for both noises is usually described by power-law similar to Eq.~\eqref{eq:RN} but with added chromaticity.

All in all, we can approximate the noise model as a white noise contribution with the addition of a low-frequency red power law (see e.g., Ref.~\cite{EPTA:2015ike}) of the form 
\begin{align}\label{eq:noisetemp}
P_{n,I}
= P_I^{\rm WN} 
+ A^{\rm RN}_I  \left (\frac{f}{f_r} \right )^{\gamma_I^{\rm RN}}
+ P_I^{\rm DM,SV} ,
\end{align}
where we arbitrarily fix the reference frequency at $f_r = f_{\rm yr}$ and $\gamma_I^{\rm RN}<0$.
$P_I^{\rm DM, \, SV}$ in the previous equation indicate the possible presence of DM and SV terms, modeled as power-laws analogously to eq.~\eqref{eq:RN}, only present for some of the pulsars.  In this work, we have neglected the chromaticity and treated  DM and SV on an equal footing as achromatic noise. The main reason is that there is still a significant coupling between chromatic and achromatic noise components in the currently observed data. We will re-discuss this point in the concluding section.

As we will see, the simplified modeling summarised above provides a sufficiently accurate description of current sensitivity, allowing the reproduction of the uncertainties obtained in current analyses. 
After validating this construction, we will use this framework to forecast future detector performance. 
In particular, we will investigate the limit in which the currently observed SGWB dominates over the experimental noise, in what is referred to as the signal-dominated regime. This further motivates the simplifying assumptions we made to describe the noise model.

\subsection{Stochastic GW background model}\label{sec:GW}

The spin-2 perturbations of the metric can be expressed in terms of plane GWs with frequency $f$, polarizations $\{+,\times\}$, 
and propagation directions $\hat k$ as
\begin{multline}
h_{ab}(t, \vec x) =
\int {\rm d}^2\Omega_{\hat k}\>
\int_{-\infty}^{\infty}{\rm d}f\>
\left[\tilde h_+(f,\hat k) e^+_{ab}(\hat{k})
\right.
\\
\left.
+\tilde h_\times(f,\hat k) e^\times_{ab}(\hat{k})\right]
e^{i2 \pi f(t - \hat k \cdot \vec x/c)}\,,
\label{e:hab(t,x)_stoch}
\end{multline}
where we introduced the polarization tensors $e^{+,\times}_{ab}(\hat k)$, with the two polarizations denoted $P=\{+,\times\}$ in the following.
We assume that the SGWB is stationary, unpolarized, and isotropic, which means 
\be
\langle \tilde h_P(f,\hat k) \tilde h_{P'}^*(f',\hat k') \rangle
= \frac{1}{16\pi}S_h(f)\delta(f-f')\delta_{PP'}\delta^2(\hat k,\hat k')\,,
\ee
where $S_h(f)$ is the (one-sided) strain power spectral density of the GWB. 
We define the SGWB energy density (per logarithmic frequency interval) as (see e.g.,~\cite{Caprini:2018mtu})
\begin{equation}\label{eq:omegaGWdefS}
    \Omega_{\rm GW} h^2 = \frac{h^2}{\rho_c}\frac{d \rho_{\rm GW}}{d \log f}
    \equiv 
    \frac{2 \pi^2 f^3}{3 H_0^2/h^2} S_h,
\end{equation}
where $\rho_c /h^2 =  3 (H_0/h)^2 /( 8 \pi G)$ is the Universe critical energy density and $H_0 /h = 1/(9.78 {\rm Gyr})$ is the Hubble parameter today. 
In the last equality, we have associated $ \Omega_{\rm GW}h^2$ 
to the strain power spectral density $S_h$.

The timing residual response of the signal coming from a pulsar $I$ to the SGWB can be expressed in Fourier space as~\cite{Hazboun:2019vhv}
\begin{align}\label{eq:response_f}
\tilde h_I(f) =
\int {\rm d}^2\Omega_{\hat k}\>
\sum_P
R_I^P(f,\hat k)\tilde h_P(f;\hat k), 
\end{align}
which is integrated over all propagation directions $\hat k$.
The response function for a pulsar located at a distance $D$ along the direction $\hat p_I$ is 
\begin{align}
R_I^{P}(f,\hat k)\equiv
\frac{\epsilon^{P}_{ab}(\hat k) }{i 4\pi f}
\frac{\hat{p_I}^a \hat{p_I}^b}{1+\hat{p_I}\cdot \hat k}
\left(1-e^{-i 2\pi f D(1+\hat k\cdot\hat p_I)/c}\right)\,.
\label{eq:response}
\end{align}
When dealing with PTA observations, it is easy to see that the characteristic frequency $f_* = (2 \pi D/c)^{-1}$ associated to a distance $D$ (which is of order kpc) turns out to be $f_* \simeq 2 \times 10^{-12}$Hz.
On the other hand, the minimum frequency accessible in our analysis is limited by the observation time $f \sim 1/T_{\rm obs} \sim $nHz. 
Therefore, one finds that for PTA experiments $f D \gg1$, and the frequency-dependent term in the response function~\eqref{eq:response} is well approximated by $R^P(f) \propto 1/f$, while the rapidly oscillating piece is negligible when averaged over the directions $\hat k$ in~\eqref{eq:response_f}.

It follows that the SGWB signal covariance matrix is
\be
C_{h,IJ}
\equiv\langle \tilde h_I \tilde h_J^{*T} \rangle 
= R_{IJ} S_h(f) ,
\label{e:CIJ}
\ee
where we introduced~\cite{Hazboun:2019vhv}
\begin{equation}
R_{IJ} = 
\chi_{IJ} \cdot  {\cal R}(f)  
\left [
{\cal T}_I(f)
{\cal T}_J(f)
{T_{IJ} }/{T_\text{obs} }
\right]^{1/2} ,
\end{equation}
and we conveniently defined the 
frequency dependent, sky-averaged, quadratic response function ${\cal R}(f) \equiv 1/12 \pi^2 f^2$. 
The first geometrical factor is the well-known HD correlation pattern as a function only of the angular separation between pulsars as required by symmetries (see e.g.~\cite{Kehagias:2024plp}), discussed below. In contrast, the time-dependent factor is introduced to account for each pulsar transmission function as well as the individual observation time of each pulsar $T_I$. 
This is because we can only include off-diagonal components correlating signals between different pulsars for an effective overlapping time defined as $T_{IJ}={\rm min}[T_{I},T_{J}] $.

\subsubsection{The Hellings-Downs correlation}\label{sec:HD correlation}
We introduced the HD function~\cite{Hellings:1983fr}
for a pair of pulsars separated by an angle 
$\zeta_{IJ} \equiv \arccos(\hat p_I \cdot \hat p_J)$, which takes the analytic expression 
\begin{equation}
\chi_{IJ} 
=
\frac{1}{2} +
\frac{3}{2} \xi_{IJ}
\left[\ln\xi_{IJ} - \frac{1}{6}\right]
+\frac{1}{2}\,\delta_{IJ}\,,
\label{e:HD}    
\end{equation}
where 
$\xi_{IJ} \equiv 
\left({1-\cos(\zeta_{IJ})}\right)/{2}$.

To estimate the sensitivity to the HD angular correlation function, it is convenient to decompose it on a suitable basis. Following Ref.~\cite{EPTA:2023fyk}, we adopt the following templates:
\begin{itemize}[leftmargin=3mm]
\item {\it Binned HD function:}
    we model the correlation function as a step-wise constant over $n$ bins in the angular variable $\xi_{IJ}$.
    This takes the form
    \begin{equation}\label{eq:binHDt}
\chi_{IJ} 
        = \sum_{i = 0}^{n-1}
        b_i \Theta(\zeta_{IJ} - \zeta^i)
        \Theta(\zeta^{i+1} -\zeta_{IJ} ),
    \end{equation}
where the bins in the angular variables are equally spaced $\zeta^i = i \pi /n$.\footnote{Notice that for a small number of pulsars, a sensible choice would be to divide the angular range unevenly and choose bins containing an equal number of pulsar pairs~\cite{EPTA:2023fyk}.} We also introduced the Heaviside theta function $\Theta$. We impose Gaussian priors on the coefficients $b_i$ with width $\sigma_{b_i} = 1$.
    
\item {\it HD expansion in Legendre polynomials:}
    following~\cite{Gair:2014rwa,EPTA:2023fyk}, we expand the HD function as
\begin{equation}\label{eq:legexpHD}
\chi_{IJ} 
=
\sum_{\ell = 0}^{n} a_\ell P_\ell(\cos \zeta_{IJ}).
\end{equation}
Using the standard normalization of the Legendre polynomials, the coefficients are found by~\cite{Gair:2014rwa,Romano:2023zhb}
\begin{align}
    a_\ell 
    &= \frac{2 \ell +1}{2} \int_{-1}^{1} 
    \chi_\text{\tiny HD} (x) P_\ell(x) \textrm{d} x 
    \nonumber \\ 
    &= 
    \frac{3}{2}
    \frac{ (2\ell +1) }{(\ell+2)(\ell+1)\ell(\ell-1)}.    
\end{align}
for $\ell \geq 2$, and $a_0 =a_1 = 0$.
We will estimate the sensitivity to the HD curve by constraining the expansion parameters $a_\ell$.
We impose priors on $a_\ell$ to be Gaussian with width $\sigma_{a_\ell} = 1$.
\end{itemize}

\subsubsection{Effective sensitivity}\label{sec:effectiveS}

We compute the SNR by
 \begin{equation}
 {\rm SNR}^2
    \equiv 
 \sum_{f_k}
\left [
C^{-1}_{IJ} C^{-1}_{KL} 
R_{JK} R_{LI}
\right ]
S_h ^2,
\end{equation}
where, here and throughout the work, we implicitly assume Einstein's summation convention on the pulsar indices in capital letters. 
It is instructive to define an effective sensitivity, as~\cite{Hazboun:2019vhv}
\begin{equation}
 S_{\rm eff}(f) = \left (C^{-1}_{IJ} C^{-1}_{KL} 
R_{JK} R_{LI}
\right ) ^{-1/2},
\end{equation}
in such a way that the SNR computation takes the familiar form ${\rm SNR}^2 =  T_\text{obs} \sum_k (S_h/S_\text{eff})^2$.

In the weak signal limit (which is in practice the limit in which  ${\cal R}S_h \ll  P_n$), 
one can expand the correlation matrix above to find 
\be
S_{\rm eff}(f)
=
\left(\sum_{I\neq J}
\frac{T_{IJ}}{T_\text{obs}}
\frac{
\mathcal{T}_I (f)\mathcal{T}_J (f)
\chi^2_{IJ}}{P_{n,I}(f)P_{n,J}(f)/\mathcal{R}^2(f)}
\right)^{-1/2}\,,
\label{e:Seff_gwb}
\ee
which includes contributions from the Hellings and
Downs factors $\chi_{IJ}$ and the individual pulsar
strain-noise power spectral densities.
This corresponds to the result of Ref.~\cite{Hazboun:2019vhv} and shown in Refs.~\cite{NANOGrav:2023ctt,InternationalPulsarTimingArray:2023mzf} to report current PTA sensitivities.
As we will see, when considering future PTA datasets (including longer observation times and a larger number of pulsars), it is important to adopt the full expression for the effective sensitivity to go beyond the weak signal approximation. 
We report examples of future effective sensitivity, in the presence of the currently observed GW background in Fig.~\ref{fig:Omega_sens} in the next section.

\subsection{Likelihood function and parameter estimation }
\label{sec:Fisher}

The log-likelihood for Gaussian and zero mean data $\tilde{d}^k$, with $k$ running over frequencies $f_k$, described only by their variance, dubbed $C_{IJ}(f_k,\btheta)$, can be written as
\cite{Contaldi:2020rht,Bond:1998zw}
\begin{equation}\label{eq:likelihood}
- \ln \mathcal{L}
(\tilde{d} \vert \btheta )
\propto
\sum_{k,IJ} 
 \ln \left [ C_{IJ}(f_k,\btheta)\right ] 
+ \tilde{d}_I^k C_{IJ}^{-1} (f_k,\btheta) \tilde{d}_J^{k*} , 
\end{equation}
which is also known as Whittle likelihood.
This likelihood function for $\btheta$ corresponds to the probability of the data $\tilde d$ given $\btheta$, the signal model parameters.

\paragraph{Fisher information matrix estimates.}

Here we briefly summarise the main ingredients of the Fisher Information Matrix (FIM) formalism, which is often used in the limit of large SNR to assess parameter uncertainties.  

The FIM $F_{\alpha \beta}$ is defined as
\begin{equation}
\label{eq:FIM_definition}
F_{\alpha \beta} \equiv - \left. \frac{\partial^2 \log \mathcal{L} }{ \partial \btheta^\alpha \partial \btheta^\beta } \right|_{\btheta =  \btheta_0} = \sum_k \textrm{Tr} \left[ C^{-1}  \frac{\partial C}{\partial \btheta^\alpha} C^{-1} \frac{\partial C}{\partial \btheta^\beta} \right]_{\btheta =  \btheta_0} \; ,
\end{equation}
where, $\btheta_0$ represents the maximum likelihood estimator of model parameters, determined by imposing
\begin{equation}
   \left. \frac{\partial \log \mathcal{L} }{ \partial \btheta^\alpha  } \right|_{\btheta =  \btheta_0 } \propto \sum_k \frac{ \partial  C}{\partial \btheta^\alpha} \left[ C^{-1} -  C^{-1} \;\tilde{d}^k \tilde{d}^{k*} \; C^{-1}  \right]   = 0  \; ,
\end{equation}
with $C(f_k, \btheta_0) = \tilde{d}^k \tilde{d}^{k*}$. In practice, the discrete sum over finite frequencies can be replaced with a continuous integral over the frequency range as
\begin{equation}
    \label{eq:FIM_final}
    F_{\alpha \beta} \equiv
    \sum_{f_k}
    \frac{\partial \log C}{\partial \btheta^\alpha} \frac{\partial \log C}{\partial \btheta^\beta} \,.
\end{equation}
Keeping all indices fully expressed, one obtains
\begin{equation}
    F_{\alpha \beta} 
    \equiv 
 \sum_{f_k}
 C^{-1}_{IJ} C^{-1}_{KL} 
\frac{\partial (R_{JK} S_h)}{\partial \btheta^\alpha}
\frac{\partial (R_{LI} S_h)}{\partial \btheta^\beta} .
\end{equation}
Finally, the covariance matrix $C_{\alpha \beta}$, is obtained by inverting the FIM, from which one can estimate uncertainties as 
$\sigma_\alpha \equiv \sqrt{F_{\alpha \alpha}^{-1}}$.

\paragraph{MCMC analyses.}

Given that the FIM is an unreliable estimator in the low SNR limit, where the validity of the Gaussian approximation is expected to degrade, we will test the FIM results with full-fledged Bayesian parameter estimation. 
For this purpose, we simulate frequency domain data $\tilde{d}^k \equiv \tilde{d}(f_k)$,  
where $k$ indexes the frequencies within the detector sensitivity range. 
We generate Gaussian realizations for the signal and all noise components, with zero mean and variances defined by their respective power spectral densities.
The parameter estimation is then performed by running 
a Monte Carlo Markov Chain (MCMC) using \texttt{emcee}~\cite{Foreman-Mackey:2012any}  with uninformative priors on signal parameters.

\section{Simulating future pulsar timing array networks}\label{sec:ptasimulations}

To test our method, we compare to currently available datasets, while we also propose to simulate future generation datasets. In the upcoming years, the Square Kilometer Array (SKA) will be a central protagonist among PTA collaborations~\cite{Janssen:2014dka}. At the moment, PTA datasets are limited by radiometer noise. The SKA will provide high-precision pulsar timing measurement with uncertainties below $\sim$100ns~\cite{Janssen:2014dka}, making it roughly 10 times better than current generation telescopes \cite{Lazio:2013mea}. 
With a reduction of the white noise at relatively high frequencies, it also comes with a better determination of the RN, relevant at low frequencies.
For our purposes, we will simulate two mock datasets:
\begin{itemize}
    \item {\rm An EPTA-like with 50 pulsars and 20 yr observations.}
    We will denote this as EPTA20 in the rest of the paper.
    \item {\rm A SKA-like with 50 pulsars and 10 yr observations.}
    We will denote this configuration as SKA10.
\end{itemize}

\subsection{Fake PTA data and time-domain Bayesian analysis}\label{sec:fakePTA}

\subsubsection{Time domain representation}

As mentioned in Sec.~\ref{sec:sensitivity}, we are dealing with the timing residuals obtained by fitting and subtracting the timing model from observed pulses' ToAs. The residuals from all pulsars are concatenated in a single array forming PTA data set  $\vec{d} = [d_1(t), d_2(t), ..., d_{N_{\rm pulsars}}(t)]$. We expect that the residual errors in fitting the timing model parameters are small and could be approximated by a linear model 
\begin{equation}
    d_I ' \rightarrow d_I - \sum_k \alpha_k \vec{M}_{Ik}, 
\end{equation}
where $M_{Ik}$ is a design matrix (representing the derivative of the timing model w.r.t timing model parameters), and $\alpha_k$ are residual timing model errors, which we consider as random variables. We have already used the design matrix in the frequency domain while building the transmission function. 

For the description of the noise components, we use Gaussian process representation (see \cite{vanHaasteren:2014qva} for a detailed description) with a discrete Fourier basis $f_i = i/T_I$ with $i=[1, ..., N_f]$ as the eigenfunctions: 
\begin{equation}
    n_I(t) = \sum_i  X_{i, I} \cos(2\pi f_i t) + Y_{i, I} \sin(2\pi f_i t),
\label{eq:noise_gp}
\end{equation}
where $X_{i,I}, Y_{i,I} \sim \mathcal{N}(0, S(f_i)\Delta f_i)$, $S(f)$ the one-sided PSD of the noise and $\Delta f_i = f_{i+1} - f_i$.  The weights $X_{i,I}, Y_{i,I}$ for the 
pulsar intrinsic noise components are uncorrelated among the pulsars. 
 For GWB signals, they are spatially correlated between pulsars $I$ and $J$ and are distributed as a $N_{\rm pulsars}$ dimensional multivariate normal distribution $\vec{X}_i, \vec{Y}_i \sim \mathcal{N}(0, \chi_{IJ}S(f_i)\Delta f_i)$ where $\chi_{IJ}$ are the introduced above HD correlation coefficients. In general, the noise covariance matrix is given as 
 \begin{eqnarray*}
    C_I^n & = & \langle n_I (t) n_I (t) \rangle \\
& = & \begin{bmatrix}
\cos (2\pi f_1 \vec{t}_I) \\
\sin (2\pi f_1 \vec{t}_I) \\
\vdots \\
\cos (2\pi f_{N_f} \vec{t}_I) \\
\sin (2\pi f_{N_f} \vec{t}_I)
\end{bmatrix}
\cdot
\begin{bmatrix}
\langle X_{1,I}^2 \rangle \\
\langle Y_{1,I}^2 \rangle \\
\vdots \\
\langle X_{N_f,I}^2 \rangle \\
\langle Y_{N_f,I}^2 \rangle
\end{bmatrix}
\cdot
\begin{bmatrix}
\cos (2\pi f_1 \vec{t}_I) \\
\sin (2\pi f_1 \vec{t}_I) \\
\vdots \\
\cos (2\pi f_{N_f} \vec{t}_I) \\
\sin (2\pi f_{N_f} \vec{t}_I)
\end{bmatrix},
\end{eqnarray*}
where $\langle X_{1,I}^2 \rangle$ being the variance of the $X_{i,I}, Y_{i,I}$ as defined by their Gaussian distribution.

\subsubsection{Generating fake data}
\label{subsubsec:fake_data}
We simulate the data in the time domain\footnote{The time-domain simulations were made using \url{https://github.com/mfalxa/fakepta}.}. We start with choosing the pulsar array and epochs of observation, which we populate with the white noise with dispersion $\sigma$.
We simulated the red (spin) noise of each pulsar using Gaussian Process according to \eqref{eq:noise_gp} where the weights $\vec{X}_{i,I}$ and $\vec{Y}_{i,I}$  are drawn from their respective normal probability distributions with variance defined by PSD, $S(f)$.
The simulated timing model includes only five components with the following design matrix:
\begin{itemize}
    \item \textit{Offset} : $\vec{M}_0 \propto \vec{1}$,
    \item \textit{Spin rate} : $\vec{M}_1 \propto \vec{t}_I$,
    \item \textit{Quadratic spin-down rate} : $\vec{M}_2 \propto \vec{t}_I ^2$,
    \item \textit{RA} : $\vec{M}_3 \propto \sin (2\pi f_{yr} \vec{t}_I)$,
    \item \textit{DEC} : $\vec{M}_4 \propto \cos (2\pi f_{yr} \vec{t}_I)$.
\end{itemize}
The overall offset, spin rate, and quadratic spindown rate correspond to the inaccuracy in the pulsar rotation. The offset in the pulsar sky position is modeled by RA and DEC terms, giving annual modulation due to the Earth's orbital motion.  During the Bayesian PTA analysis, we marginalize over the coefficients $\alpha_k$ of the linear timing model, assuming non-informative improper prior.  This marginalization is responsible for the transmission function described in~\eqref{eq:t_f}~\cite{Hazboun:2019vhv}.

\subsection{Mock datasets}
\label{subsec:mock_datasets}

We simulated two PTA mock datasets:
\begin{itemize}[leftmargin=3mm]
    \item \textbf{EPTA20 :} built from the 25 pulsars of the EPTA DR2new 10.4 yr dataset with the same noise properties (as in~\cite{EPTA:2023akd}) but with a doubled observation time, to which we append 25 additional pulsars with the same noise properties and observation time as EPTA DR2new but randomized positions drawn uniformly in the sky. The noises are re-injected according to their maximum likelihood values in~\cite{EPTA:2023akd}.
    \item \textbf{SKA10 :} SKA-like PTA with 50 pulsars and 10 yr observations. Every pulsar is simulated with a 2-week cadence of observation, $\sigma=$100ns timing uncertainty, and random position uniformly drawn in the sky. Time-correlated noises with power-law spectra are included (red noise and dispersion measure noise) with log-amplitude $\log_{10}A$ and spectral index $\gamma$ drawn uniformly between $\log_{10}A=[-17, -13]$ and $\gamma=[1, 5]$.
\end{itemize}
These properties are summarised in Tab.~\ref{tab:summarypropfake}.
\begin{table}[h]
\begin{tabular}{c|c|c}
    & EPTA20 & SKA10 \\
    \hline
    \hline
    $N_{\rm pulsars}$ & 25 + 25 & 50\\
    $\nu$ [GHz] & \cite{EPTA:2023akd} & 1.4 - 3 \\
    $T_{\rm  obs}$ [yr] & 20.8 + 10.4 & 10\\
    $\sigma$ [s] & $\sim 10^{-6}$ & $10^{-7}$\\
    $\Delta t$ [days] & $\sim 3$ & 14\\
    TM & TM$_5$ + \cite{EPTA:2023akd} & TM$_5$\\
    $\log_{10} A_{\rm RN}$ &  \cite{EPTA:2023akd} & [-17, -13]\\
    $\gamma_{\rm RN}$ &  \cite{EPTA:2023akd} & [1, 5]\\
    $\log_{10} A_{\rm DM}$ &  \cite{EPTA:2023akd} & [-17, -13]\\
    $\gamma_{\rm DM}$ &  \cite{EPTA:2023akd} & [1, 5]\\
    $\log_{10} A_{\rm SV}$ &  \cite{EPTA:2023akd} & -\\
    $\gamma_{\rm SV}$ &  \cite{EPTA:2023akd} & - \\
\end{tabular}
\caption{Summary of mock datasets properties. $N_{\rm pulsars}$ is the number of pulsars, $\nu$ are the observing frequency sub-bands, $T_{\rm obs}$ is the time of observation, $\sigma$ is the level of white noise, $\Delta t$ is the cadence, TM is the timing model where TM$_5$ is the five-component model described in~\ref{subsubsec:fake_data}, $\log_{10} A$ and $\gamma$ give the range of the uniform distribution used to draw the red noise parameters. The entries marked by Ref.~\cite{EPTA:2023akd} correspond to the choice of parameters consistent with ERPTA DR2new analysis. 
}\label{tab:summarypropfake}
\end{table}
\begin{figure*}[!t]\centering
\includegraphics[width=0.32\textwidth]{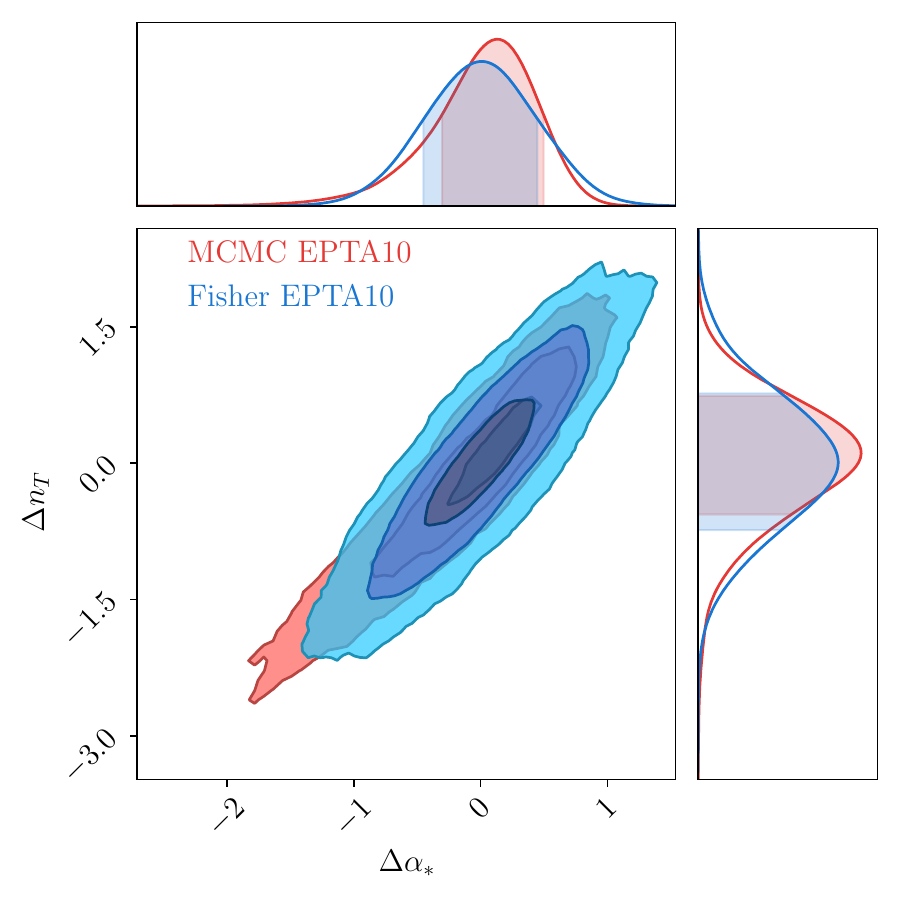}
\includegraphics[width=0.32\textwidth]{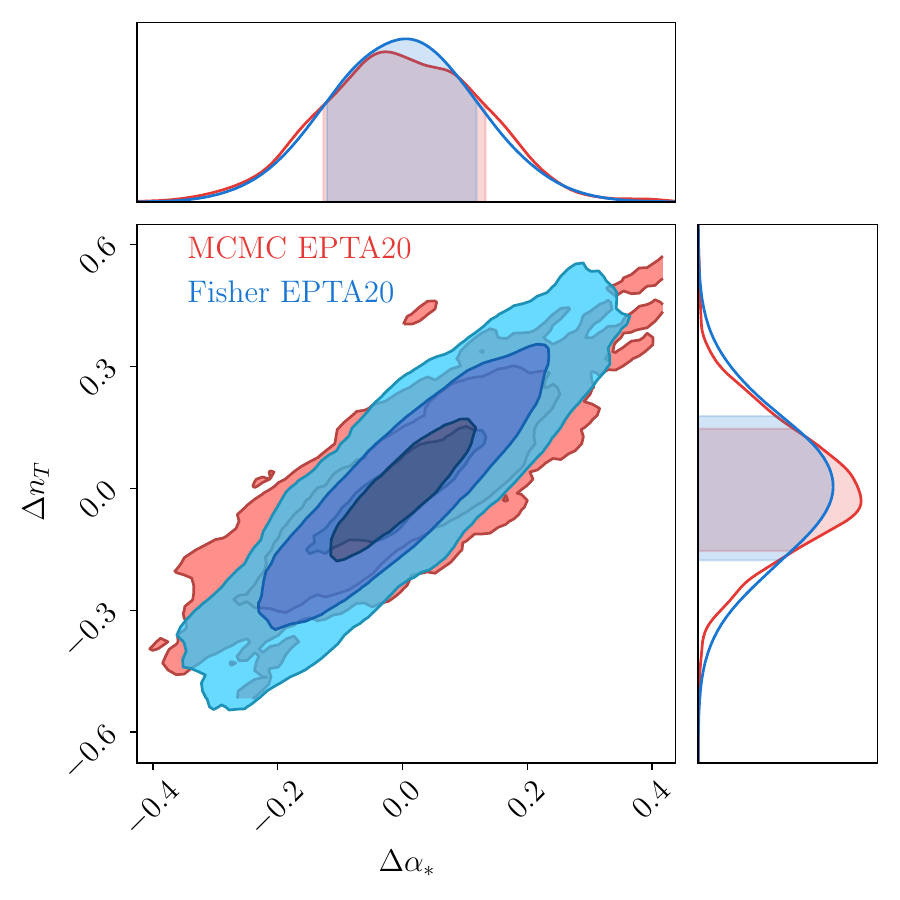}
\includegraphics[width=0.32\textwidth]{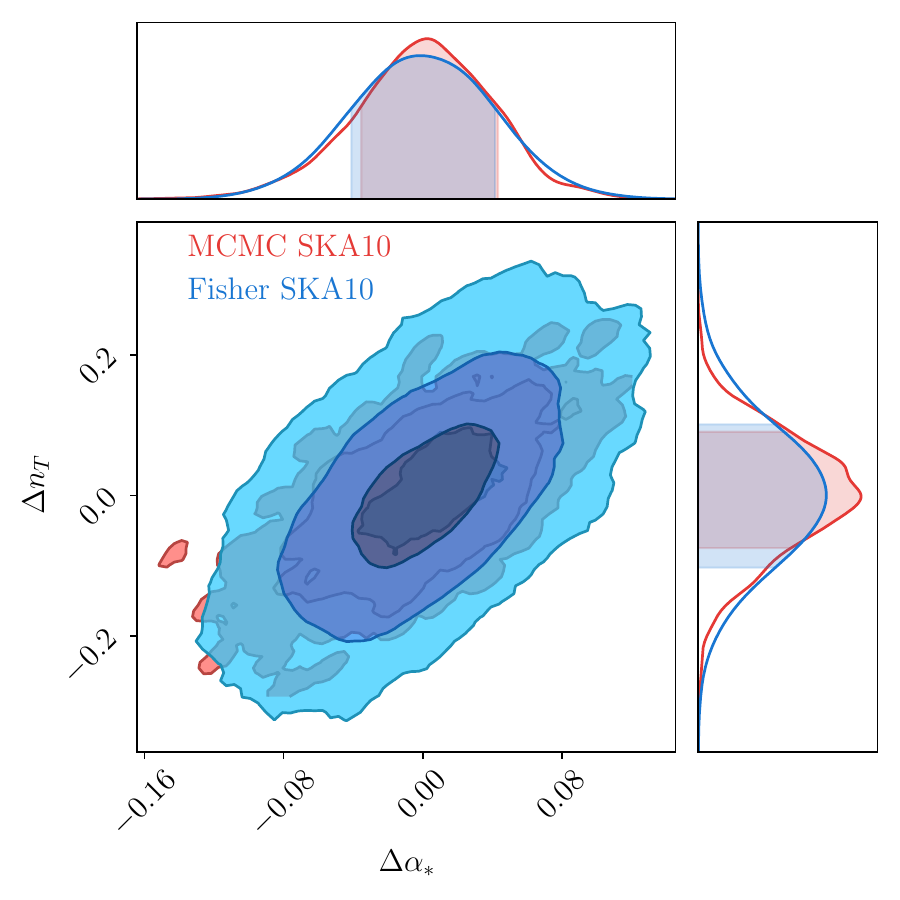}
\caption{
Comparison between the measurement uncertainties estimated using FIM (blue) and time-domain Bayesian parameter estimation (red, see Sec.~\ref{sec:fakePTA} for more details). 
For visualization purposes, the FIM result is shown by building samples drawn from a multivariate Gaussian distribution with covariance $FIM^{-1}$.
For both parameters, we remove the mean observed values to compare with the FIM which does not provide information on it. 
In the two-dimensional panels, we indicate 68\%, 95\%, and 99\% C.I. 
{\it Left panel:} current posterior obtained with EPTADR2new datasets (EPTA10).
{\it Center panel:} forecasted EPTA 20 yr dataset.
{\it Right panel:} forecasted SKA 10 yr dataset.
}
\label{fig:compare results}
\end{figure*}
In both datasets, we inject a stochastic GW signal with HD correlations and a power-law spectrum with $\log_{10} A=-14$ and $\gamma=3$ to reproduce the signal observed in~\cite{EPTA:2023fyk, NANOGrav:2023gor}, potentially originating from the population of SMBHB in the Universe. This signal will act as foreground, and we will also inject additional stochastic GW signals into the datasets to test our ability to detect them in the presence of a foreground (see Sec.~\ref{subsec:sensitivty_sub_SGWB}).

\subsection{Time-domain Bayesian analysis}\label{sec:timedomainBayesian}

The timing model marginalized Gaussian likelihood used in time-domain PTA analysis~\cite{vanHaasteren:2014qva} can be expressed as (analogously to the previously defined likelihood~\eqref{eq:likelihood})
\begin{equation}\label{eq:likelihood_time}
- \ln \mathcal{L}
(d \vert \btheta )
\propto \textrm{det}\{C_{IJ} \} + 
\sum_{IJ} d_I(t) C_{IJ}^{-1} (\btheta) d_J^\intercal (t) , 
\end{equation}
where $C_{IJ} = C_I \delta_{IJ} + C_{h,IJ}$ has a block structure of size $N_{\rm pulsars} \times N_{\rm pulsars}$ in which the GW signal $C_{h,IJ} = \chi_{IJ} C_h$ ($I \neq J$) appears off diagonal. The individual pulsar' noise components form the diagonal $I=J$ where $C_I = C_I^{TM} + C_I^{RN} + C_I^{DM} + C_I^{SV}$ (corresponding to the timing model, the red noise, the dispersion measure noise, and the chromatic noise respectively).

We perform Bayesian analysis using the software ENTERPRISE~\cite{enterprise,enterprise_ext} and sampler PTMCMC~\cite{justin_ellis_2017_1037579}. ENTERPRISE provides computation of the likelihood and prior based on the chosen model. All noise components are modelled as Gaussian Processes with incomplete Fourier basis as eigenfunctions \cite{vanHaasteren:2014qva}. We are using PTMCMC without a parallel tempering option, which is sufficient for our purpose.  The resulting posterior distributions for both EPTA20 and SKA10 are shown in Fig.~\ref{fig:compare results} (central and right panels, red color).

\section{Results}\label{sec:results}

Throughout this section, we will assume that the SGWB observed by the various collaborations is described by a power-law (PL)
\begin{equation}\label{eq:bestfitSGWB}
h^2 \Omega_{\rm GW} ^{\rm PL}(f)
= 
A_{\rm GWB} \left (\frac{f}{f_{\rm yr}} \right)^{n_{T}}
\end{equation}
with $A_{\rm GWB} \simeq 6.3 \times 10^{-8}$ and $n_{T} \simeq  2$.
For definiteness, these parameters are fixed to the maximum likelihood values obtained by the EPTA~\cite{EPTA:2023fyk}, which is in good agreement with the other observations~\cite{InternationalPulsarTimingArray:2023mzf}.
To make contact with other quantities used in the literature, this corresponds to a strain amplitude ($h_c \equiv \sqrt{f S_h}$) at $f = f_\text{yr}$ of $A_{h_c}\simeq 10^{-14}$ and pulsar timing residuals spectral index $\gamma = 5-n_T = 3$.
When performing parameter estimation, it is convenient to define the log of the amplitude $\alpha_* \equiv \log_{10}(A_{\rm GWB})$.

Assuming this SGWB, we obtain SNR = 4.5 with the current EPTA (DRNew) pulsar configuration, which is compatible with the observations~\cite{EPTA:2023fyk}. On the other hand, we obtain ${\rm SNR} = 184 $ for EPTA20 and ${\rm SNR} = 292 $ for SKA10.

To validate our framework based on the approximated likelihood introduced in~\cref{sec:Fisher}, we compare our results to the ones obtained with state-of-the-art techniques described in~\ref{sec:timedomainBayesian}. 
In Fig.~\ref{fig:compare results}, we compare the uncertainties obtained with our simplified FIM approach to actual PTA parameter estimation. 
We remove the mean from the posterior distribution of both amplitude and tilt showing the deviations from the mean values.
In the left panel, we compare results for the current EPTA DR2new configuration, assuming $T_\text{obs} = 10.33$yrs of data with the 25 pulsars adopted by EPTA in their analysis~\cite{EPTA:2023sfo}.
In the center and right panels, we compare our forecasted sensitivity to signal parameters with the simulations for EPTA20 and SKA10,
with $T_\text{obs} = 20$yrs and $T_\text{obs} = 10$yrs, respectively and $N_{\rm pulsars} = 50$, built as described in Sec.~\ref{sec:ptasimulations}. 
In all cases, we find very good agreement between both results, showing the validity of our assumptions in Sec.~\ref{sec:sensitivity} for our purposes.

\subsection{Future effective sensitivity}

\begin{figure}[t]\centering
\includegraphics[width=\columnwidth]{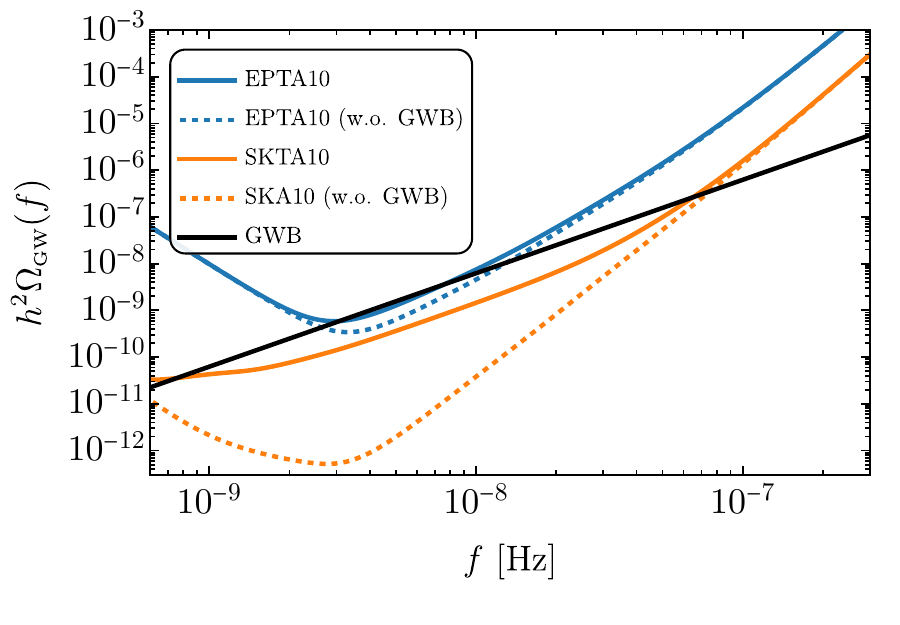}
\vspace{-2em}
\caption{
Current sensitivity of the EPTADR2new dataset, denoted as EPTA10,  (blue lines), and forecasted future sensitivity achieved by the SKA 10 yr (yellow lines). Both scenarios are reported assuming the presence (absence) of an SGWB in a solid (dashed) line. 
The black solid line corresponds to the maximum likelihood power-law signal~\eqref{eq:bestfitSGWB}.
}
  \label{fig:Omega_sens}
\end{figure}

Using the metrics defined in sec.~\ref{sec:effectiveS}, we can forecast the effective sensitivity achieved by current and future experiments. 
In particular, we plot $S_{\rm eff}(f)$ Fig.~\ref{fig:Omega_sens} for different detector designs, i.e., EPTA10 and SKA10, translated in the GW energy density using Eq.~\eqref{eq:omegaGWdefS}. 
Notice we plot the effective sensitivity as a function of frequencies also below $f\sim 1/(10 {\rm yrs})$ for presentation purposes. 
In the same plot, we show the maximum likelihood power-law SGWB for comparison. These results assume the HD function to be fixed and described by Eq.~\eqref{e:HD} (i.e., we do not infer it from the data).

In both cases, the dashed lines report the putative sensitivity obtained without injecting an SGWB. 
The effective sensitivity reaches a minimum of around $f\sim 3$nHz, with a steep high-frequency slope mostly induced by the large WN components. 
Indeed, in the high-frequency part of the sensitivity curve, we observe $\Omega_{\rm GW}(f) \sim f^5$, as expected from a WN-dominated regime, including frequency-dependent factors coming from the definition of energy density and the response function. 
Instead, at low frequencies, the sensitivity is degraded with a tilt steeper than $\propto 1/f$, expected from the behavior of transmission functions alone, showing the relevance of the red-noise components included in the pulsar noise budget. 
SKA10 sensitivity greatly surpasses EPTA10 due to the much smaller expected WN (with an improvement that is more than an order of magnitude at high frequencies).

The solid lines report the effective sensitivity obtained including the SGWB background as in Eq.~\eqref{eq:bestfitSGWB}. 
While including the SGWB does not significantly affect the sensitivity of EPTA10, showing that one still falls in the weak signal regime, the SKA10 effective sensitivity strongly degrades, driven by the large SGWB foreground compared to the pulsar noises.

In practice, the take-home message is that searches for subdominant contributions to the SGWB below the solid lines would be inconclusive, at odds with the conclusion drawn considering the effective sensitivity with neglected foregrounds. 
This is one of the main results of this paper. 
We will show examples of multi-SGWB analyses in the following sections, corroborating the qualitative conclusions shown here.

\begin{figure*}[t!]\centering
\includegraphics[width=0.9\textwidth]{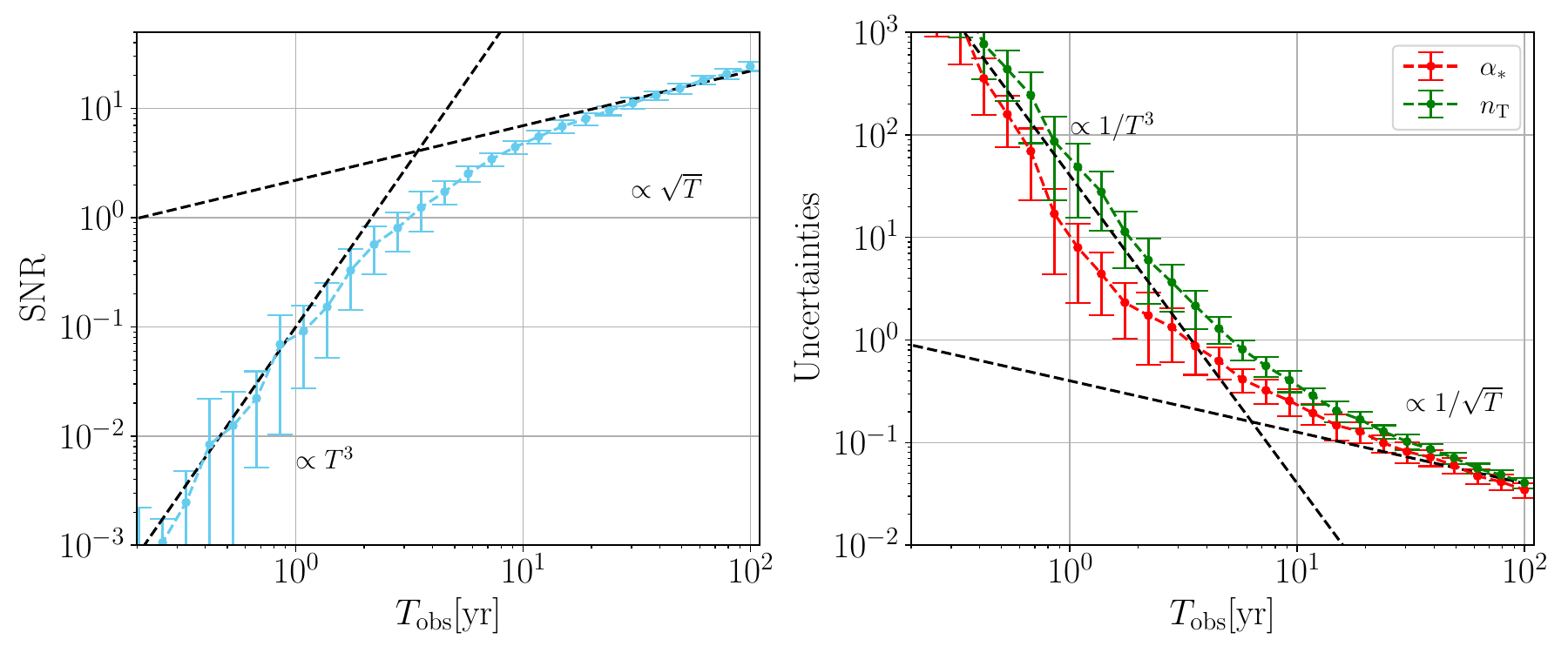}
\caption{
{\it Left panel:}
scaling of the SNR as a function of the observation time $T_\text{\tiny obs}$ assuming a PTA network composed of 30 pulsars. 
The vertical error bars indicate the standard deviation of the result obtained by varying over a random realization of the PTA array (both for pulsar positions and noises).
The black dashed lines indicate the two limiting scaling regimes expected for the SNR. The low-$T_{\rm obs} $ regime is dictated by our choice of the spectrum with $n_T = 2$, while the high-$T_{\rm obs}$ regime is universal.
{\it Right panel:} scaling of the uncertainties on the amplitude and tilt of the SGWB as a function of observation time. 
}
  \label{fig:TScaling}
\end{figure*}

\begin{figure*}[t]\centering
\includegraphics[width=0.9\textwidth]{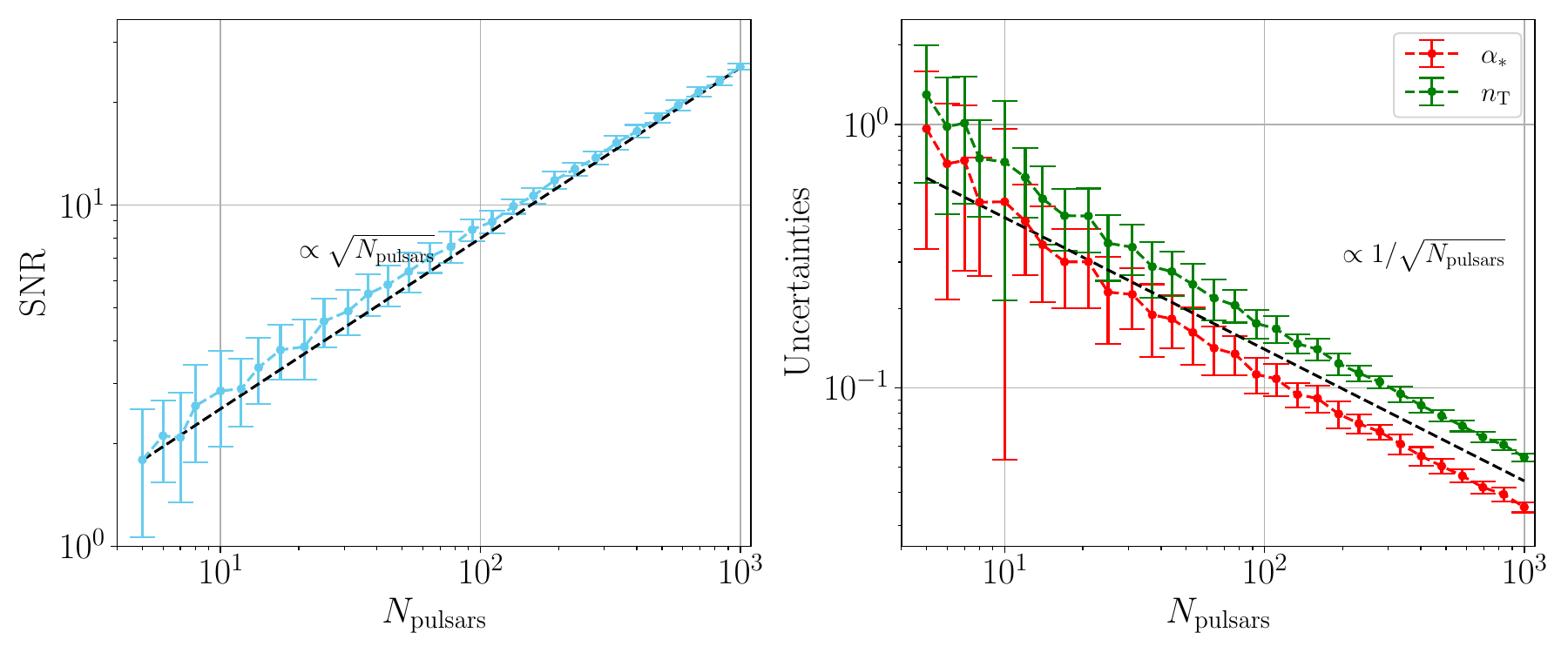}
\caption{
Scaling of the SNR (left panel) and uncertainties (right panel) as a function of the number of pulsars, assuming fixed $T_\text{\tiny obs} =  20$yrs. 
The error band indicates the standard deviation of both results induced by producing multiple realizations of the pulsar sample. 
}
  \label{fig:NScaling}
\end{figure*}

\subsection{Scaling with time and number of pulsars}

With the proposed framework, we can forecast the evolution with time and the number of pulsars of both the SNR and measurement precision.

In Fig.~\ref{fig:TScaling}, we report the results of different random realizations of $30$ pulsars with variable observation time $T_\text{obs}$.
We indicate with the uncertainty bands the corresponding spread obtained from randomly selecting pulsars with different sky localization (assuming isotropic distribution) and noises.\footnote{For both Figs.~\ref{fig:TScaling} and~\ref{fig:NScaling}, we generated $30$ different realizations for each value of $T_\text{obs}$ and $N_\text{pulsars}$, to estimate the scatter induced by randomly choosing noise parameter for each pulsar.}
The parameters determining the noise of each pulsar are sampled from distributions built out of the currently observed ones~\cite{EPTA:2023sfo}. 

In these plots, we aim to test the scaling of the relevant quantities. For this reason, we show FIM results also for low-SNR values where FIM predictions are unreliable. It should be kept in mind that results obtained in this regime only serve the purpose of testing the scalings and should not be used to deduce the detectability of the signal. Also, in this case, these results assume the HD function to be fixed and described by Eq.~\eqref{e:HD} (i.e., we do not infer it from the data). We do not expect this assumption to modify the scaling obtained.

\begin{figure*}[t]\centering
\includegraphics[width=0.45\textwidth]{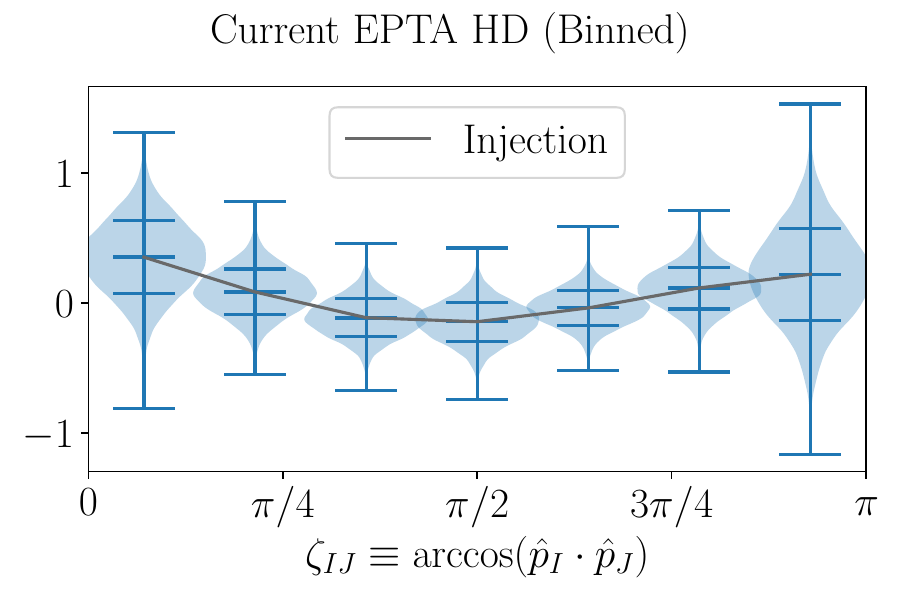}
\includegraphics[width=0.45\textwidth]{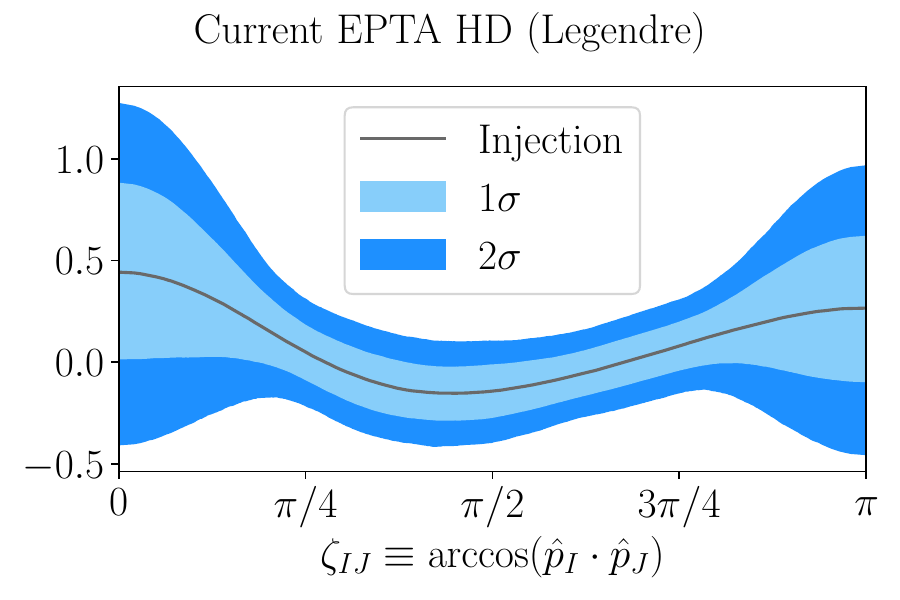}
\caption{
Uncertainties on the HD curve were obtained using the Fisher matrix analysis and assuming a power-law signal.
{\it Left panel:} constraints on the coefficients of the binned HD function. We divide the angular direction into 7 equally spaced bins. Uncertainties are comparable with the one observed found in EPTADR2 analyses~\cite{EPTA:2023fyk}.
By construction, within the FIM formalism, the central value corresponds to the injection. 
{\it Right panel:}
Posterior predictive distribution of the HD function obtained from the uncertainties on the coefficients of the Legendre polynomial expansion in Eq.~\eqref{eq:legexpHD}.
}
  \label{fig:HDEPTA}
\end{figure*}

\begin{figure*}[t]\centering
\includegraphics[width=0.45\textwidth]{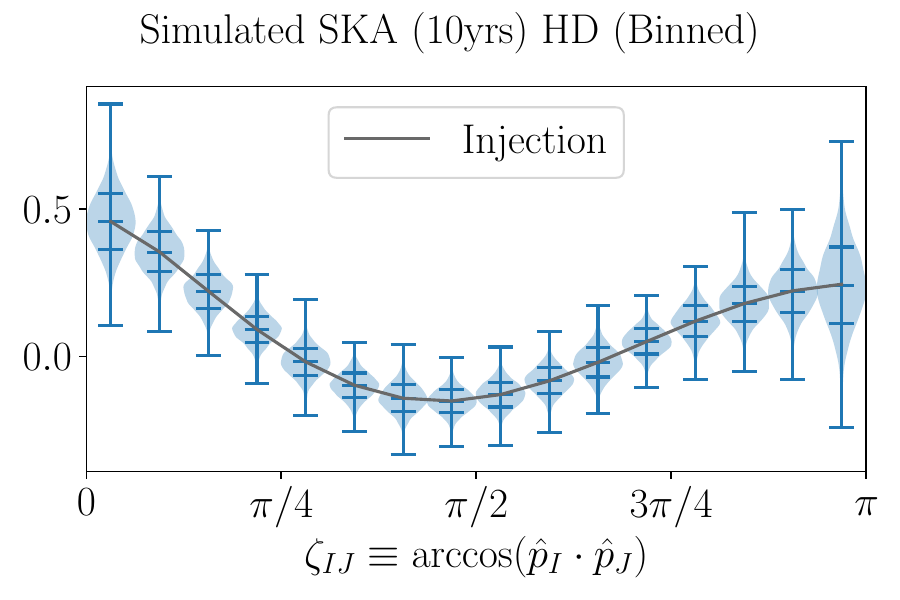}
\includegraphics[width=0.45\textwidth]{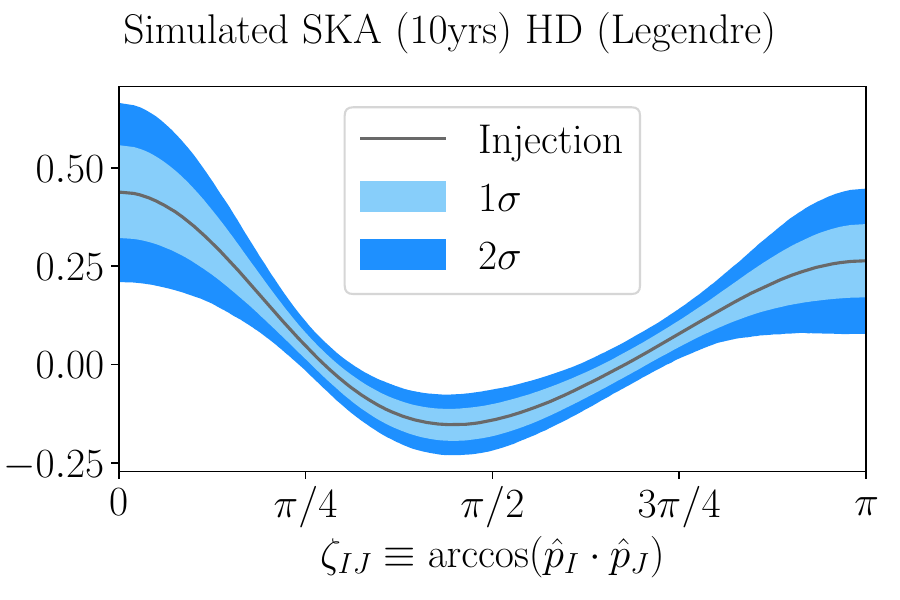}
\caption{Same as Fig.~\ref{fig:HDEPTA}, but assuming the SKA10 future  configuration. }
  \label{fig:HDSKA}
\end{figure*}

Focusing on the evolution of SNR, we observe that at small observation times, it grows rapidly as SNR$\sim T_\text{\tiny obs}^3$. 
This can be explained as follows (see analytical considerations reported in Ref.~\cite{Siemens:2013zla}).
Assuming WN dominates at the initially relatively high frequencies $f_\text{min} \sim 1/T_\text{\tiny obs}$, and that $\gamma >0$, 
the integral for the SNR is dominated by the contribution close to $f_\text{min}$. 
Longer observation times allow to probe lower frequencies, enhancing the ratio $S_h/S_\text{eff} \sim f_\text{min}^{-\gamma} \sim T_\text{obs}^{\gamma}$.
Including the additional pre-factor and frequency summation, this gives SNR$\propto T_\text{obs}^{\gamma}$.
When the effective sensitivity falls below the GW amplitude at the minimum frequency, one transitions to the intermediate regime, and the above scaling is broken. 
Eventually, the SNR growth with time converges to $\propto \sqrt{T_\text{obs}}$ as predicted in the strong signal regime, which we observe on the right side of the plot.

An analogous behavior is observed in the evolution of uncertainties on signal parameters $\alpha_*$ and $n_T$. By construction, they are expected to scale inversely with the SNR, in the high SNR regime.
We observe that both feature an initially steep decline 
that converges towards a $1/\sqrt{T_{\rm obs}}$ in the far future. 
We stress that current PTA experiments are currently probing the intermediate regime between the two asymptotic scalings, seen in the plot around $T_{\rm obs} \sim 10$yrs (see also~\cite{NANOGrav:2020spf}). 
Additionally, we see that the uncertainty on the tilt reaches the scaling regime more slowly because more frequency modes need to be observed to better pin down the spectral tilt compared to the overall amplitude. 

In Fig.~\ref{fig:NScaling} we show the corresponding scaling as a function of the number of pulsars. 
In the left panel, we see that the SNR grows as $\sqrt{N}$ as a function of number of pulsars. This scaling originates from including auto-correlation terms in our analysis (i.e., diagonal terms in the pulsar indices of the covariant matrix). 
Correspondingly, the uncertainties on SGWB spectral amplitude and tilt decrease as $1/\sqrt{N_{\rm pulsars}}$.
Had we only included the off-diagonal components, featuring the HD correlation, we would have obtained a scaling ${\rm SNR}\sim \left ( \sum_{I\neq J} \chi_{IJ}^2 \right )^{1/2}$
which behaves as ${\rm SNR} \sim \left [ N_{\rm pulsars}(N_{\rm pulsars}-1)\right ]^{1/2}\sim N_{\rm pulsars}$.

\begin{figure*}[t]\centering
\includegraphics[width=0.49\textwidth]{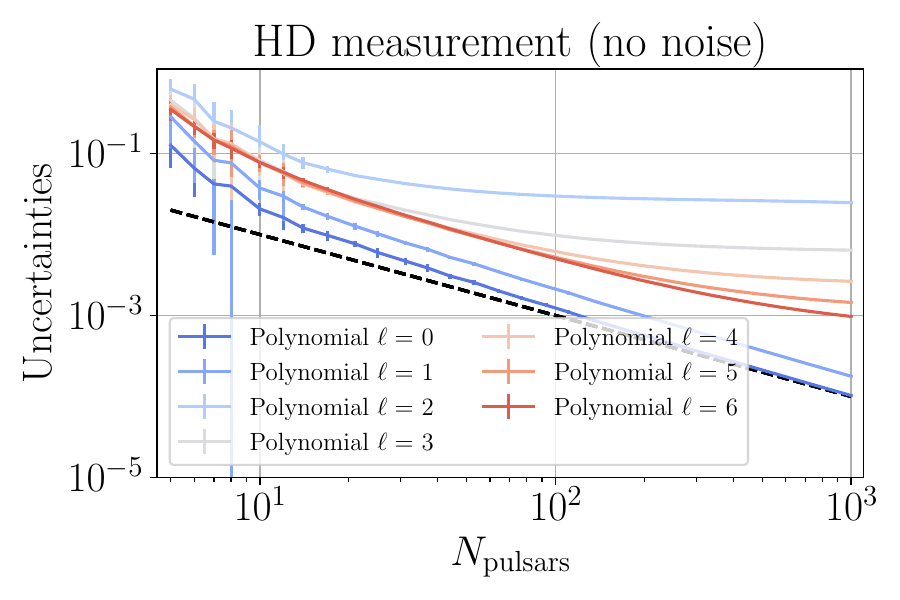}
\includegraphics[width=0.49\textwidth]{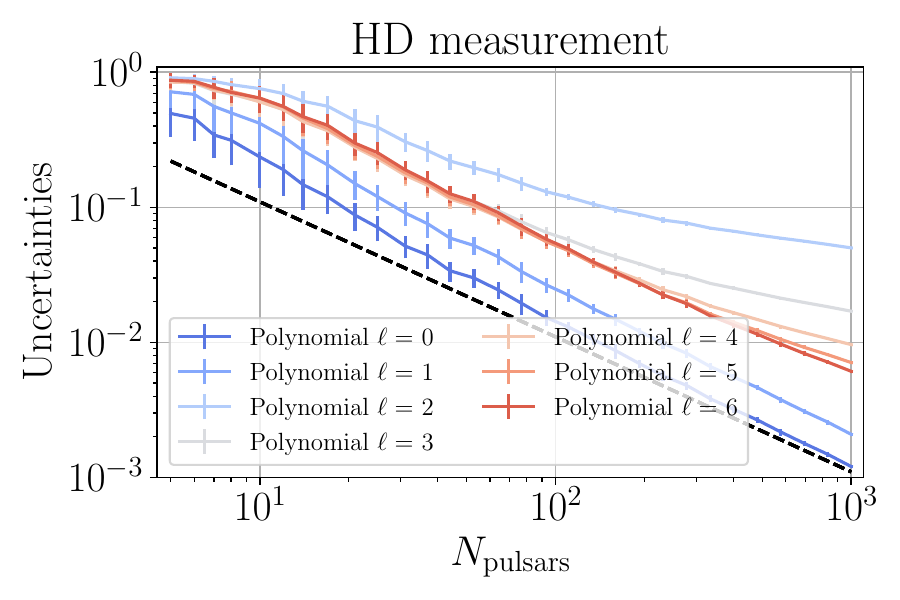}
\caption{
Uncertainties on the fifth-order expansion coefficients of the HD curve in Legendre's polynomials $a_\ell$ as a function of the number of pulsars $N_{\rm pulsars}$. We adopt $N_{\rm modes} = 30$.
{\it Left panel:} zero noise limit. The $0^{\rm th}$ and $1^{\rm st}$ orders are zero, and the corresponding uncertainty scales as $\propto 1/N_{\rm pulsars}$ (indicated by the dashed black line).  
The higher-order pieces reach the plateau predicted by cosmic variance in Eq.~\eqref{eq:cosmicvar}. See the discussion in the main text.  
{\it Right panel:} 
Same as the left panel, but includes noise. The plateau at small values of $N_{\rm pulsars}$ is reached when information is small enough that our prior on $\sigma_{a_\ell} =1$ becomes dominant.}
  \label{fig:HDevoN}
\end{figure*}

\subsection{Sensitivity to the HD correlation pattern}

In this section, we show how the formalism developed in the current work can also forecast the sensitivity to the HD correlation. This is a crucial observable to identify GW signals and could be used to test modified theories of gravity (e.g.,~\cite{Liang:2023ary}). For this purpose, we allow $\chi_{IJ}$ to vary, and, adopting the HD curve parameterizations described in Sec.~\ref{sec:HD correlation}, we assess the ability of different PTA configurations to constrain the expansion coefficients.

First of all, we consider the case of the EPTA10 configuration. 
In Fig.~\ref{fig:HDEPTA} (left panel), we report the uncertainties on the coefficient $b_i$ of the binned parameterization~\eqref{eq:binHDt} found using FIM.
For this analysis, we considered 7 bins as well as the signal amplitude and slope as free parameters. 
As one can notice, the uncertainties are large both at narrow and wide angular separations, where fewer pulsar pairs are present, while reaching down to values around $\Delta b_i\sim 0.3$ at $68\%$ C.L. in the central bins. These results are in good agreement with the posterior distributions reported by the EPTA collaboration~\cite{EPTA:2023fyk}, although they choose unequally spaced bins to evenly distribute the number of pulsar pairs as a function of angular separation.

In the right panel of Fig.~\ref{fig:HDEPTA}, we show the posterior predictive distribution of $\chi_{IJ}$
obtained using the uncertainties estimated with FIM for the current EPTA10 configuration. In this case, the FIM analysis adopts the 
coefficients of the Legendre polynomial expansion~\eqref{eq:legexpHD} up to the sixth order.
As expected, relatively tight bounds are obtained in the central portion of the angular separation variable (which contains a larger number of pulsar pairs), with constraints that degrade at the edges of the plot.

Having checked the validity of our framework to estimate the sensitivity to HD correlations, we forecast the corresponding uncertainties obtained with the future SKA10 dataset. In Fig.~\ref{fig:HDSKA}, we show the corresponding posterior on binned expansion parameters (left panel) as well as posterior predictive distribution reconstructed using Legendre expansion (right panel).
As expected, the improved SNR ratio in the SKA10 forecast leads to significantly narrower uncertainties on the HD correlation function. In the central bins, one finds uncertainties around $\Delta b_i\sim 0.04$ at $68\%$ C.L.

\paragraph*{Cosmic variance.}

For a Gaussian background, the multipole moments are drawn from a Gaussian distribution with variance $a_\ell$ defined in Eq.~\eqref{eq:legexpHD}.
Even neglecting noise, and assuming perfect measurement precision, the estimate of the variance of the distribution is limited by 
the number of modes observed for each $\ell$.
As described in Refs.~\cite{Roebber:2016jzl,Allen:2022dzg}, assuming single mode observation and no noise, the uncertainty on $a_\ell$ should plateau towards the value
\begin{equation}
\Delta a_{\ell} = \frac{a_\ell}{\sqrt{2\ell + 1} } \; .
    \label{eq:cosmicvar}
\end{equation} 
In reality, multiple frequency modes are observed by PTA, and one expects the limit on the uncertainties due to cosmic variance to scale as $\Delta a_\ell \sim \left ( N_{\rm modes} \right )^{-1/2}$.

In Fig.~\ref{fig:HDevoN}, we show how the uncertainty on $a_\ell$ scales as a function of the number of pulsars. To better show the qualitative behavior, we first compute the results assuming no noise and $30$ observed frequencies (left plot) and $T_{\rm obs} = 10$yr, only including terms up to $\ell =6$ in the FIM.
As one can see, the uncertainty for $\ell = 0,1$ scales as $1/N_{\rm pulsars}$, without showing any significant departure at high $N_{\rm pulsars}$. This is because, $a_{0,1} = 0$ due to the quadruple nature of GR, and their inference does not suffer from self-noise. 
On the other hand, the uncertainty on high multiples converges towards a plateau, which is dictated by cosmic variance~\eqref{eq:cosmicvar}. 
The plateau is reached at an increasing number of pulsars for higher $\ell \geq 2 $
due to the increasingly larger $N_{\rm pulsars} $ needed to cover the sky positions with sufficient finer angular resolution to saturate the information available. 

In the right panel of Fig.~\ref{fig:HDevoN}, we show analogous results assuming the noise to be present. In this case, we simulate 30 pulsars observed for $T_\text{obs} = 10$yrs.
We see that uncertainties on $a_\ell$ feature a plateau at low values of $N_{\rm pulsars}$, where their posterior distribution is dominated by the prior (assumed to be Gaussian with $\sigma_{a_\ell} =1$). When $N_{\rm pulsars}\gtrsim 10$ (incidentally close to the currently analyzed sample of pulsars by EPTA), uncertainties start to decrease, especially at lower multipoles. Qualitatively, we observe the same behavior obtained in the left panel for the idealized case of no noise. 
The $\ell =0,1$ uncertainties keep the $1/N_{\rm pulsars}$ scaling, while starting from the lowest orders in $\ell$ (especially $\ell = 2,3$) we notice a flattening of the curve already from $N_{\rm pulsars} \sim {\cal O} ({\rm few}\cdot  10^2) $.
We do not push the computations to even larger $N_{\rm pulsars}$ for two main reasons: 1) this configuration becomes increasingly optimistic for any future scenario, and 2) to retain a reasonable computational cost.

\subsection{Sensitivity to subdominant SGWBs}
\label{subsec:sensitivty_sub_SGWB}

In this section, we forecast future sensitivity to subdominant contribution to the nHz SGWB. 
For presentation purposes, we consider the scenario in which the additional subdominant GWB contribution is described by a lognormal (LN) shape 
    \begin{equation}\label{eq:LNSGWB}
       h^2 \Omega_{\rm CGW} (f) =
        10^{\alpha_{\rm LN}}
       e^{\left [
        - \ln^2(f/(10^{\gamma_{\rm LN}} {\rm Hz}))/
        (2 \cdot 10^{2 \beta_{\rm LN}})
        \right ]}.
\end{equation}
This should serve as a toy model we adopt to test our framework.
Typical new physics scenarios of interest for PTA observations predict the existence of a blue tilted spectrum which reaches a peak close to the nHz frequencies, also to comply with $\Delta N_{\rm eff}$ upper bounds on cosmological (i.e., from the early universe) GWs, forcing the total primordial GW abundance $\Omega_{\rm GW}h^2 < 1.6 \cdot 10^{-6} \left ( \Delta N_{\rm eff} /0.28 \right )$ at 95\% C.L.~\cite{Planck:2018vyg}.
Nonetheless, flatter spectra in the nHz range are predicted by cosmic string scenarios, as well as second order induced GWs from enhanced and flat curvature power spectra. 
In-depth analyses which assume specific spectral shapes motivated by new physics cosmological scenarios (see e.g.,~\cite{NANOGrav:2023hvm,EPTA:2023xxk, Madge:2023dxc,Figueroa:2023zhu,Ellis:2023oxs}) will be presented elsewhere~\cite{inprepCGWB}.

\begin{figure}[t]\centering
\includegraphics[width=0.5\textwidth]{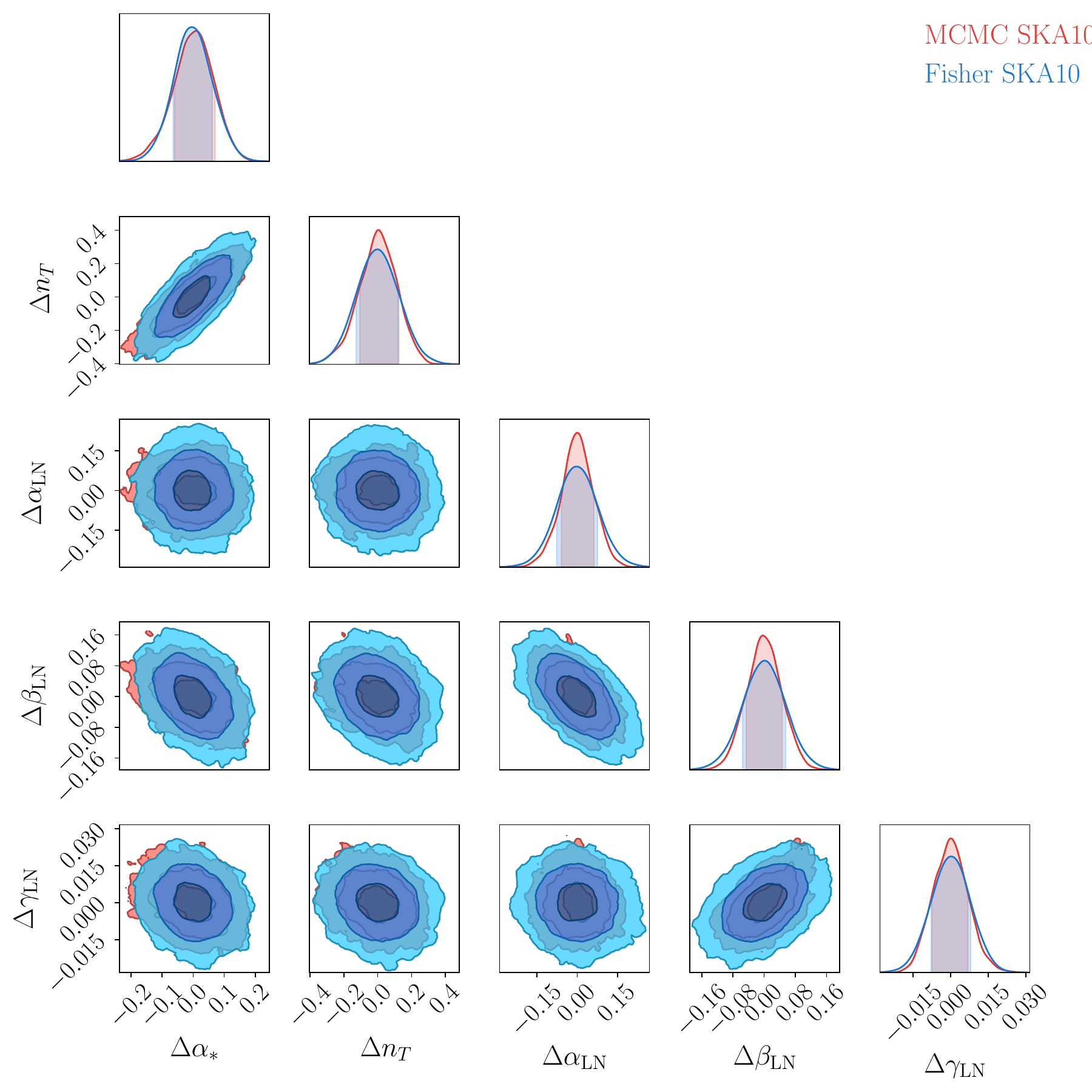}
\caption{ Same as Fig.~\ref{fig:compare results}, but injecting a SGWB from multiple sources: {\it i) } the dominant GW signal parametrized as a PL with $\alpha_* = -7.2$ and $n_T = 2$ and {\it ii) } subdominant LN bump controlled by
$\alpha_{\rm LN} = -7$, $\beta_{\rm LN} = -1$ and $\gamma_{\rm LN} = -7.5$. 
}
  \label{fig:multi_signal}
\end{figure}

Again, to validate our framework in the case of multiple SGWB contributions, we generate a synthetic dataset as described in Sec.~\ref{subsec:mock_datasets} with the SKA10 configuration (50 pulsars and $T_{\rm obs}=10$yrs) of observation 
of an SGWB which takes contributions from two sources
\begin{equation}
h^2 \Omega_{\rm GW}= 
h^2 \Omega_{\rm GW}^\text{PL}+
h^2 \Omega_{\rm GW}^\text{LN}.
\end{equation}
Collectively, the injected signal features five parameters, which are fixed to 
$\{\alpha_{*}, n_T,
\alpha_{\rm LN}, \beta_{\rm LN}, \gamma_{\rm LN}\}  = \{-7.2, 2, -7, -1, -7.5 \} $.
In Fig.~\ref{fig:multi_signal}, we compare the FIM estimates to the state-of-the-art multi-components injection and inference. As in Fig.~\ref{fig:compare results}, one observes excellent agreement between both results, validating our framework even in the presence of multiple contributions to the SGWB. 

\begin{figure*}[t]\centering
\includegraphics[width=0.9\textwidth]{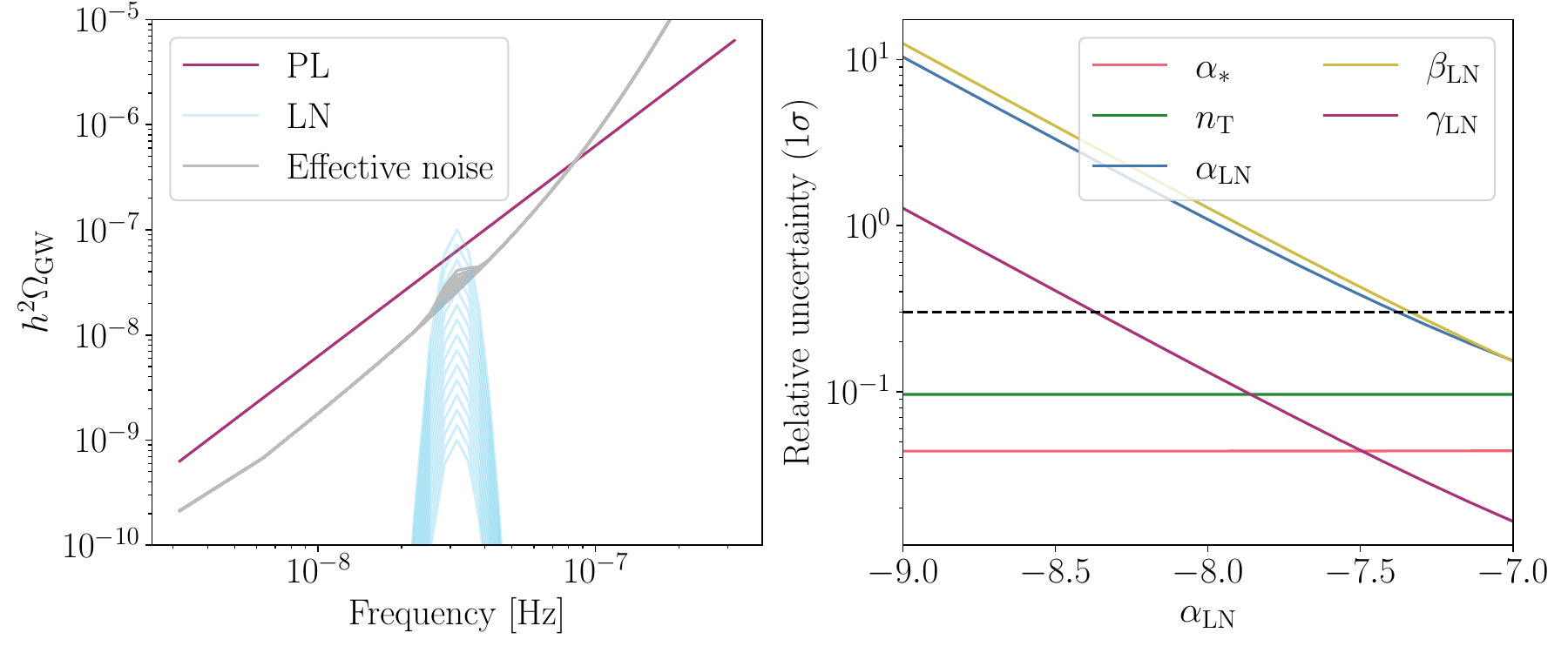}
\caption{
{\it Left panel:}
Injected SGWB including PL (purple, Eq.~\eqref{eq:bestfitSGWB}) and LN (cyan, Eq.~\eqref{eq:LNSGWB}).
The grey solid lines show the effective sensitivity obtained with the future SKA10 configuration (with $N_{\rm pulsars} = 100$ and $T_{\rm obs} = 10$yrs). Different values of the amplitude are shown, varying $\alpha_{\rm LN} \in [-9,-7]$ with regular spacing.
{\it Right panel:}
Relative uncertainties on the model parameters, injecting the same values shown in the left panel. While uncertainties on PL parameters remain practically constant, we see improved accuracy on the LN parameters with larger $\alpha_{\rm LN}$. 
}
\label{fig:multi_signal_scan}
\end{figure*}

We further explore the sensitivity to an LN subdominant contribution to the SGWB in Fig.~\ref{fig:multi_signal_scan}.
We simulate different scenarios in which the amplitude of the LN signal is varied, keeping fixed remaining parameters $\{\alpha_*, n_T,
 \beta_{\rm LN}, \gamma_{\rm LN}\}  = \{-7.2, 2 ,-1, -7.5 \} $.
 Uncertainties on PL parameters remain practically constant. This is because most of the injections feature a negligible contribution from the LN bump. Only large values of $\alpha_{\rm LN}$ the uncertainty grow slightly, due to the partial contamination around LN peak frequencies. 
 On the other hand, we observe increasing precision on the LN parameters with larger $\alpha_{\rm LN}$. Interestingly, the relative uncertainty on the signal amplitude falls below $\sim 30\%$ around $\alpha_{\rm LN} \simeq -7.4$, indicating detection capability, when the LN amplitude crosses the effective sensitivity.

\section{Conclusions}\label{sec:conclusions}

Pulsar timing array experiments provide an unprecedented opportunity to search for the existence of GWs in the nHZ frequency range. This will allow to constrain both astrophysical and cosmological sources of such signatures, with increasing precision. 
Current data report growing evidence for the existence of a signal with an amplitude $\Omega_{\rm GW} h^2\sim 10^{-9}$ at the best constrained frequencies $f\sim 1/(10 {\rm yrs})$.
In the future, longer observation time and the growing number of pulsars available will drastically expand PTA sensitivity.

In this work, we have demonstrated how to achieve fast estimation of detectability and parameter estimation for a stochastic GW signal which one could expect for a particular Pulsar Timing Array. Despite the simplistic assumptions adopted, we find that, compared to a full Bayesian analysis using the software {\rm{ENTERPRISE}} on the EPTA DR2new and simulated data, this approach gives reliable results. Moreover, we have demonstrated that this method recovers an expected scaling of the PTA sensitivity with observation time and the number of pulsars. We can also study the spatial correlation in the pulsar's pairs either using a binned consideration or Legendre decomposition. Finally, the formalism allows us to investigate multi-component stochastic GW signals of different strengths, going beyond the weak signal approximation so far adopted in the literature, providing fast and reliable forecasts on the constraining power of PTA configurations on both astrophysical and cosmological SGWBs.  We release the code \href{https://github.com/Mauropieroni/fastPTA/}{\texttt{fastPTA}} which implements this framework.

If we neglect a pulsar term, extending this work to include the resolvable GW signals from individual sources (e.g., continuous GW (CGW) from inspiralling SMBHBs) would be quite simple. In this case, assuming that the CGW candidates are identified, we can include CGW in the transmission function, which will appear as a notch at the CGW frequency. 
To demonstrate this, we can write the CGW in the form $s = \sum_{k} a_k, h_k(t, f_{\rm{CGW}}, \rm{sky})$ (see~\cite{Babak:2011mr,Ellis:2012zv}) which is similar to the timing (linearised) model and will give a contribution to the $G$ matrix~\cite{Hazboun:2019vhv} and, therefore, to the transmission ${\cal T}$. Extension of this approach to CGWs with pulsar terms is somewhat more complicated and will be discussed elsewhere. 

As promised, we want to say a few words about mimicking the chromatic noises in our approach. We can restore the dependence on the radio frequency, as it will become important for the simultaneous ultra-broad band observations. One possibility on how it can be done is presented in~\cite{Hazboun:2019vhv} using the DMX model. In the case of Gaussian Process representation, we need to translate it to DMX-compatible representation. The DM variations through the DMX model appear in the transmission function.

We conclude by mentioning possible further future directions:
\begin{itemize}[leftmargin=3mm]
\item Astrophysical models predict the SGWB comes from the superposition of unresolved signals. In this case, spectral fluctuations due to finite source numbers within frequency bins may appear in the observable band at a relatively "high" frequency. It would be interesting to estimate the future sensitivity as well as frequency resolutions to such spikes, which are not expected in cosmological scenarios. 
\item Throughout our work, we have assumed the dominant GWB to be stationary and isotropic. In particular, the effective sensitivity reported in our work assumes this background cannot be reduced. If the signal is of astrophysical origin, with sufficient sensitivity one should be able to resolve individual SMBH mergers and remove some CGW. This could improve the effective sensitivity to subdominant GWBs.
\item With our code, one could also test the sensitivity to different correlation patterns beyond HD (e.g. motivated by modified theories of gravity). 
\item Finally, it would be interesting to extend this framework to include SGWB anisotropies (see e.g.,~\cite{Ali-Haimoud:2020ozu,Ali-Haimoud:2020iyz,Cruz:2024svc} for work in this direction), going beyond the weak signal approximation. 
\end{itemize}

\begin{acknowledgments}
We thank Bruce Allen, Paul Frederik Depta, and Valerie Domcke for valuable discussions on the HD cosmic variance. MP thanks Rutger van Haasteren for the useful discussion on PTA data analysis.

SB acknowledges support from  funding
from the French National Research Agency (grant ANR21-CE31-0026, project MBH waves), the European
Union’s Horizon 2020 research and innovation program
under the Marie Sk\l{}odowska-Curie grant agreement No.
101007855. MP acknowledges the hospitality of Imperial College London, which provided office space during some
parts of this project.
\end{acknowledgments}

\bibliography{main}

\begin{thebibliography}{119}%
\makeatletter
\providecommand \@ifxundefined [1]{%
 \@ifx{#1\undefined}
}%
\providecommand \@ifnum [1]{%
 \ifnum #1\expandafter \@firstoftwo
 \else \expandafter \@secondoftwo
 \fi
}%
\providecommand \@ifx [1]{%
 \ifx #1\expandafter \@firstoftwo
 \else \expandafter \@secondoftwo
 \fi
}%
\providecommand \natexlab [1]{#1}%
\providecommand \enquote  [1]{``#1''}%
\providecommand \bibnamefont  [1]{#1}%
\providecommand \bibfnamefont [1]{#1}%
\providecommand \citenamefont [1]{#1}%
\providecommand \href@noop [0]{\@secondoftwo}%
\providecommand \href [0]{\begingroup \@sanitize@url \@href}%
\providecommand \@href[1]{\@@startlink{#1}\@@href}%
\providecommand \@@href[1]{\endgroup#1\@@endlink}%
\providecommand \@sanitize@url [0]{\catcode `\\12\catcode `\$12\catcode
  `\&12\catcode `\#12\catcode `\^12\catcode `\_12\catcode `\%12\relax}%
\providecommand \@@startlink[1]{}%
\providecommand \@@endlink[0]{}%
\providecommand \url  [0]{\begingroup\@sanitize@url \@url }%
\providecommand \@url [1]{\endgroup\@href {#1}{\urlprefix }}%
\providecommand \urlprefix  [0]{URL }%
\providecommand \Eprint [0]{\href }%
\providecommand \doibase [0]{http://dx.doi.org/}%
\providecommand \selectlanguage [0]{\@gobble}%
\providecommand \bibinfo  [0]{\@secondoftwo}%
\providecommand \bibfield  [0]{\@secondoftwo}%
\providecommand \translation [1]{[#1]}%
\providecommand \BibitemOpen [0]{}%
\providecommand \bibitemStop [0]{}%
\providecommand \bibitemNoStop [0]{.\EOS\space}%
\providecommand \EOS [0]{\spacefactor3000\relax}%
\providecommand \BibitemShut  [1]{\csname bibitem#1\endcsname}%
\let\auto@bib@innerbib\@empty
\bibitem [{\citenamefont {Agazie}\ \emph
  {et~al.}(2023{\natexlab{a}})\citenamefont {Agazie} \emph
  {et~al.}}]{NANOGrav:2023gor}%
  \BibitemOpen
  \bibfield  {author} {\bibinfo {author} {\bibfnamefont {G.}~\bibnamefont
  {Agazie}} \emph {et~al.} (\bibinfo {collaboration} {NANOGrav}),\ }\href
  {\doibase 10.3847/2041-8213/acdac6} {\bibfield  {journal} {\bibinfo
  {journal} {Astrophys. J. Lett.}\ }\textbf {\bibinfo {volume} {951}},\
  \bibinfo {pages} {L8} (\bibinfo {year} {2023}{\natexlab{a}})},\ \Eprint
  {http://arxiv.org/abs/2306.16213} {arXiv:2306.16213 [astro-ph.HE]}
  \BibitemShut {NoStop}%
\bibitem [{\citenamefont {Agazie}\ \emph
  {et~al.}(2023{\natexlab{b}})\citenamefont {Agazie} \emph
  {et~al.}}]{NANOGrav:2023hde}%
  \BibitemOpen
  \bibfield  {author} {\bibinfo {author} {\bibfnamefont {G.}~\bibnamefont
  {Agazie}} \emph {et~al.} (\bibinfo {collaboration} {NANOGrav}),\ }\href
  {\doibase 10.3847/2041-8213/acda9a} {\bibfield  {journal} {\bibinfo
  {journal} {Astrophys. J. Lett.}\ }\textbf {\bibinfo {volume} {951}},\
  \bibinfo {pages} {L9} (\bibinfo {year} {2023}{\natexlab{b}})},\ \Eprint
  {http://arxiv.org/abs/2306.16217} {arXiv:2306.16217 [astro-ph.HE]}
  \BibitemShut {NoStop}%
\bibitem [{\citenamefont {Antoniadis}\ \emph
  {et~al.}(2023{\natexlab{a}})\citenamefont {Antoniadis} \emph
  {et~al.}}]{EPTA:2023fyk}%
  \BibitemOpen
  \bibfield  {author} {\bibinfo {author} {\bibfnamefont {J.}~\bibnamefont
  {Antoniadis}} \emph {et~al.} (\bibinfo {collaboration} {EPTA, InPTA:}),\
  }\href {\doibase 10.1051/0004-6361/202346844} {\bibfield  {journal} {\bibinfo
   {journal} {Astron. Astrophys.}\ }\textbf {\bibinfo {volume} {678}},\
  \bibinfo {pages} {A50} (\bibinfo {year} {2023}{\natexlab{a}})},\ \Eprint
  {http://arxiv.org/abs/2306.16214} {arXiv:2306.16214 [astro-ph.HE]}
  \BibitemShut {NoStop}%
\bibitem [{\citenamefont {Antoniadis}\ \emph
  {et~al.}(2023{\natexlab{b}})\citenamefont {Antoniadis} \emph
  {et~al.}}]{EPTA:2023sfo}%
  \BibitemOpen
  \bibfield  {author} {\bibinfo {author} {\bibfnamefont {J.}~\bibnamefont
  {Antoniadis}} \emph {et~al.} (\bibinfo {collaboration} {EPTA}),\ }\href
  {\doibase 10.1051/0004-6361/202346841} {\bibfield  {journal} {\bibinfo
  {journal} {Astron. Astrophys.}\ }\textbf {\bibinfo {volume} {678}},\ \bibinfo
  {pages} {A48} (\bibinfo {year} {2023}{\natexlab{b}})},\ \Eprint
  {http://arxiv.org/abs/2306.16224} {arXiv:2306.16224 [astro-ph.HE]}
  \BibitemShut {NoStop}%
\bibitem [{\citenamefont {Antoniadis}\ \emph
  {et~al.}(2023{\natexlab{c}})\citenamefont {Antoniadis} \emph
  {et~al.}}]{EPTA:2023xxk}%
  \BibitemOpen
  \bibfield  {author} {\bibinfo {author} {\bibfnamefont {J.}~\bibnamefont
  {Antoniadis}} \emph {et~al.} (\bibinfo {collaboration} {EPTA}),\ }\href@noop
  {} {\  (\bibinfo {year} {2023}{\natexlab{c}})},\ \Eprint
  {http://arxiv.org/abs/2306.16227} {arXiv:2306.16227 [astro-ph.CO]}
  \BibitemShut {NoStop}%
\bibitem [{\citenamefont {Reardon}\ \emph
  {et~al.}(2023{\natexlab{a}})\citenamefont {Reardon} \emph
  {et~al.}}]{Reardon:2023gzh}%
  \BibitemOpen
  \bibfield  {author} {\bibinfo {author} {\bibfnamefont {D.~J.}\ \bibnamefont
  {Reardon}} \emph {et~al.},\ }\href {\doibase 10.3847/2041-8213/acdd02}
  {\bibfield  {journal} {\bibinfo  {journal} {Astrophys. J. Lett.}\ }\textbf
  {\bibinfo {volume} {951}},\ \bibinfo {pages} {L6} (\bibinfo {year}
  {2023}{\natexlab{a}})},\ \Eprint {http://arxiv.org/abs/2306.16215}
  {arXiv:2306.16215 [astro-ph.HE]} \BibitemShut {NoStop}%
\bibitem [{\citenamefont {Zic}\ \emph {et~al.}(2023)\citenamefont {Zic} \emph
  {et~al.}}]{Zic:2023gta}%
  \BibitemOpen
  \bibfield  {author} {\bibinfo {author} {\bibfnamefont {A.}~\bibnamefont
  {Zic}} \emph {et~al.},\ }\href {\doibase 10.1017/pasa.2023.36} {\bibfield
  {journal} {\bibinfo  {journal} {Publ. Astron. Soc. Austral.}\ }\textbf
  {\bibinfo {volume} {40}},\ \bibinfo {pages} {e049} (\bibinfo {year}
  {2023})},\ \Eprint {http://arxiv.org/abs/2306.16230} {arXiv:2306.16230
  [astro-ph.HE]} \BibitemShut {NoStop}%
\bibitem [{\citenamefont {Reardon}\ \emph
  {et~al.}(2023{\natexlab{b}})\citenamefont {Reardon} \emph
  {et~al.}}]{Reardon:2023zen}%
  \BibitemOpen
  \bibfield  {author} {\bibinfo {author} {\bibfnamefont {D.~J.}\ \bibnamefont
  {Reardon}} \emph {et~al.},\ }\href {\doibase 10.3847/2041-8213/acdd03}
  {\bibfield  {journal} {\bibinfo  {journal} {Astrophys. J. Lett.}\ }\textbf
  {\bibinfo {volume} {951}},\ \bibinfo {pages} {L7} (\bibinfo {year}
  {2023}{\natexlab{b}})},\ \Eprint {http://arxiv.org/abs/2306.16229}
  {arXiv:2306.16229 [astro-ph.HE]} \BibitemShut {NoStop}%
\bibitem [{\citenamefont {Xu}\ \emph {et~al.}(2023)\citenamefont {Xu} \emph
  {et~al.}}]{Xu:2023wog}%
  \BibitemOpen
  \bibfield  {author} {\bibinfo {author} {\bibfnamefont {H.}~\bibnamefont {Xu}}
  \emph {et~al.},\ }\href {\doibase 10.1088/1674-4527/acdfa5} {\bibfield
  {journal} {\bibinfo  {journal} {Res. Astron. Astrophys.}\ }\textbf {\bibinfo
  {volume} {23}},\ \bibinfo {pages} {075024} (\bibinfo {year} {2023})},\
  \Eprint {http://arxiv.org/abs/2306.16216} {arXiv:2306.16216 [astro-ph.HE]}
  \BibitemShut {NoStop}%
\bibitem [{\citenamefont {Hellings}\ and\ \citenamefont
  {Downs}(1983)}]{Hellings:1983fr}%
  \BibitemOpen
  \bibfield  {author} {\bibinfo {author} {\bibfnamefont {R.~w.}\ \bibnamefont
  {Hellings}}\ and\ \bibinfo {author} {\bibfnamefont {G.~s.}\ \bibnamefont
  {Downs}},\ }\href {\doibase 10.1086/183954} {\bibfield  {journal} {\bibinfo
  {journal} {Astrophys. J. Lett.}\ }\textbf {\bibinfo {volume} {265}},\
  \bibinfo {pages} {L39} (\bibinfo {year} {1983})}\BibitemShut {NoStop}%
\bibitem [{\citenamefont {Phinney}(2001)}]{Phinney:2001di}%
  \BibitemOpen
  \bibfield  {author} {\bibinfo {author} {\bibfnamefont {E.~S.}\ \bibnamefont
  {Phinney}},\ }\href@noop {} {\  (\bibinfo {year} {2001})},\ \Eprint
  {http://arxiv.org/abs/astro-ph/0108028} {arXiv:astro-ph/0108028} \BibitemShut
  {NoStop}%
\bibitem [{\citenamefont {Sesana}\ \emph {et~al.}(2008)\citenamefont {Sesana},
  \citenamefont {Vecchio},\ and\ \citenamefont {Colacino}}]{Sesana:2008mz}%
  \BibitemOpen
  \bibfield  {author} {\bibinfo {author} {\bibfnamefont {A.}~\bibnamefont
  {Sesana}}, \bibinfo {author} {\bibfnamefont {A.}~\bibnamefont {Vecchio}}, \
  and\ \bibinfo {author} {\bibfnamefont {C.~N.}\ \bibnamefont {Colacino}},\
  }\href {\doibase 10.1111/j.1365-2966.2008.13682.x} {\bibfield  {journal}
  {\bibinfo  {journal} {Mon. Not. Roy. Astron. Soc.}\ }\textbf {\bibinfo
  {volume} {390}},\ \bibinfo {pages} {192} (\bibinfo {year} {2008})},\ \Eprint
  {http://arxiv.org/abs/0804.4476} {arXiv:0804.4476 [astro-ph]} \BibitemShut
  {NoStop}%
\bibitem [{\citenamefont {Kocsis}\ and\ \citenamefont
  {Sesana}(2011)}]{Kocsis:2010xa}%
  \BibitemOpen
  \bibfield  {author} {\bibinfo {author} {\bibfnamefont {B.}~\bibnamefont
  {Kocsis}}\ and\ \bibinfo {author} {\bibfnamefont {A.}~\bibnamefont
  {Sesana}},\ }\href {\doibase 10.1111/j.1365-2966.2010.17782.x} {\bibfield
  {journal} {\bibinfo  {journal} {Mon. Not. Roy. Astron. Soc.}\ }\textbf
  {\bibinfo {volume} {411}},\ \bibinfo {pages} {1467} (\bibinfo {year}
  {2011})},\ \Eprint {http://arxiv.org/abs/1002.0584} {arXiv:1002.0584
  [astro-ph.CO]} \BibitemShut {NoStop}%
\bibitem [{\citenamefont {Kelley}\ \emph {et~al.}(2017)\citenamefont {Kelley},
  \citenamefont {Blecha},\ and\ \citenamefont {Hernquist}}]{Kelley:2016gse}%
  \BibitemOpen
  \bibfield  {author} {\bibinfo {author} {\bibfnamefont {L.~Z.}\ \bibnamefont
  {Kelley}}, \bibinfo {author} {\bibfnamefont {L.}~\bibnamefont {Blecha}}, \
  and\ \bibinfo {author} {\bibfnamefont {L.}~\bibnamefont {Hernquist}},\ }\href
  {\doibase 10.1093/mnras/stw2452} {\bibfield  {journal} {\bibinfo  {journal}
  {Mon. Not. Roy. Astron. Soc.}\ }\textbf {\bibinfo {volume} {464}},\ \bibinfo
  {pages} {3131} (\bibinfo {year} {2017})},\ \Eprint
  {http://arxiv.org/abs/1606.01900} {arXiv:1606.01900 [astro-ph.HE]}
  \BibitemShut {NoStop}%
\bibitem [{\citenamefont {Perrodin}\ and\ \citenamefont
  {Sesana}(2018)}]{Perrodin:2017bxr}%
  \BibitemOpen
  \bibfield  {author} {\bibinfo {author} {\bibfnamefont {D.}~\bibnamefont
  {Perrodin}}\ and\ \bibinfo {author} {\bibfnamefont {A.}~\bibnamefont
  {Sesana}},\ }\href {\doibase 10.1007/978-3-319-97616-7_3} {\bibfield
  {journal} {\bibinfo  {journal} {Astrophys. Space Sci. Libr.}\ }\textbf
  {\bibinfo {volume} {457}},\ \bibinfo {pages} {95} (\bibinfo {year} {2018})},\
  \Eprint {http://arxiv.org/abs/1709.02816} {arXiv:1709.02816 [astro-ph.HE]}
  \BibitemShut {NoStop}%
\bibitem [{\citenamefont {Ellis}\ \emph {et~al.}(2023)\citenamefont {Ellis},
  \citenamefont {Fairbairn}, \citenamefont {H\"utsi}, \citenamefont {Raidal},
  \citenamefont {Urrutia}, \citenamefont {Vaskonen},\ and\ \citenamefont
  {Veerm\"ae}}]{Ellis:2023owy}%
  \BibitemOpen
  \bibfield  {author} {\bibinfo {author} {\bibfnamefont {J.}~\bibnamefont
  {Ellis}}, \bibinfo {author} {\bibfnamefont {M.}~\bibnamefont {Fairbairn}},
  \bibinfo {author} {\bibfnamefont {G.}~\bibnamefont {H\"utsi}}, \bibinfo
  {author} {\bibfnamefont {M.}~\bibnamefont {Raidal}}, \bibinfo {author}
  {\bibfnamefont {J.}~\bibnamefont {Urrutia}}, \bibinfo {author} {\bibfnamefont
  {V.}~\bibnamefont {Vaskonen}}, \ and\ \bibinfo {author} {\bibfnamefont
  {H.}~\bibnamefont {Veerm\"ae}},\ }\href {\doibase
  10.1051/0004-6361/202346268} {\bibfield  {journal} {\bibinfo  {journal}
  {Astron. Astrophys.}\ }\textbf {\bibinfo {volume} {676}},\ \bibinfo {pages}
  {A38} (\bibinfo {year} {2023})},\ \Eprint {http://arxiv.org/abs/2301.13854}
  {arXiv:2301.13854 [astro-ph.CO]} \BibitemShut {NoStop}%
\bibitem [{\citenamefont {Agazie}\ \emph
  {et~al.}(2023{\natexlab{c}})\citenamefont {Agazie} \emph
  {et~al.}}]{NANOGrav:2023hfp}%
  \BibitemOpen
  \bibfield  {author} {\bibinfo {author} {\bibfnamefont {G.}~\bibnamefont
  {Agazie}} \emph {et~al.} (\bibinfo {collaboration} {NANOGrav}),\ }\href
  {\doibase 10.3847/2041-8213/ace18b} {\bibfield  {journal} {\bibinfo
  {journal} {Astrophys. J. Lett.}\ }\textbf {\bibinfo {volume} {952}},\
  \bibinfo {pages} {L37} (\bibinfo {year} {2023}{\natexlab{c}})},\ \Eprint
  {http://arxiv.org/abs/2306.16220} {arXiv:2306.16220 [astro-ph.HE]}
  \BibitemShut {NoStop}%
\bibitem [{\citenamefont {Afzal}\ \emph {et~al.}(2023)\citenamefont {Afzal}
  \emph {et~al.}}]{NANOGrav:2023hvm}%
  \BibitemOpen
  \bibfield  {author} {\bibinfo {author} {\bibfnamefont {A.}~\bibnamefont
  {Afzal}} \emph {et~al.} (\bibinfo {collaboration} {NANOGrav}),\ }\href
  {\doibase 10.3847/2041-8213/acdc91} {\bibfield  {journal} {\bibinfo
  {journal} {Astrophys. J. Lett.}\ }\textbf {\bibinfo {volume} {951}},\
  \bibinfo {pages} {L11} (\bibinfo {year} {2023})},\ \Eprint
  {http://arxiv.org/abs/2306.16219} {arXiv:2306.16219 [astro-ph.HE]}
  \BibitemShut {NoStop}%
\bibitem [{\citenamefont {Ghoshal}\ and\ \citenamefont
  {Strumia}(2024)}]{Ghoshal:2023fhh}%
  \BibitemOpen
  \bibfield  {author} {\bibinfo {author} {\bibfnamefont {A.}~\bibnamefont
  {Ghoshal}}\ and\ \bibinfo {author} {\bibfnamefont {A.}~\bibnamefont
  {Strumia}},\ }\href {\doibase 10.1088/1475-7516/2024/02/054} {\bibfield
  {journal} {\bibinfo  {journal} {JCAP}\ }\textbf {\bibinfo {volume} {02}},\
  \bibinfo {pages} {054} (\bibinfo {year} {2024})},\ \Eprint
  {http://arxiv.org/abs/2306.17158} {arXiv:2306.17158 [astro-ph.CO]}
  \BibitemShut {NoStop}%
\bibitem [{\citenamefont {Bonetti}\ \emph {et~al.}(2018)\citenamefont
  {Bonetti}, \citenamefont {Sesana}, \citenamefont {Barausse},\ and\
  \citenamefont {Haardt}}]{Bonetti:2017lnj}%
  \BibitemOpen
  \bibfield  {author} {\bibinfo {author} {\bibfnamefont {M.}~\bibnamefont
  {Bonetti}}, \bibinfo {author} {\bibfnamefont {A.}~\bibnamefont {Sesana}},
  \bibinfo {author} {\bibfnamefont {E.}~\bibnamefont {Barausse}}, \ and\
  \bibinfo {author} {\bibfnamefont {F.}~\bibnamefont {Haardt}},\ }\href
  {\doibase 10.1093/mnras/sty874} {\bibfield  {journal} {\bibinfo  {journal}
  {Mon. Not. Roy. Astron. Soc.}\ }\textbf {\bibinfo {volume} {477}},\ \bibinfo
  {pages} {2599} (\bibinfo {year} {2018})},\ \Eprint
  {http://arxiv.org/abs/1709.06095} {arXiv:1709.06095 [astro-ph.GA]}
  \BibitemShut {NoStop}%
\bibitem [{\citenamefont {Arzoumanian}\ \emph {et~al.}(2021)\citenamefont
  {Arzoumanian} \emph {et~al.}}]{NANOGrav:2021flc}%
  \BibitemOpen
  \bibfield  {author} {\bibinfo {author} {\bibfnamefont {Z.}~\bibnamefont
  {Arzoumanian}} \emph {et~al.} (\bibinfo {collaboration} {NANOGrav}),\ }\href
  {\doibase 10.1103/PhysRevLett.127.251302} {\bibfield  {journal} {\bibinfo
  {journal} {Phys. Rev. Lett.}\ }\textbf {\bibinfo {volume} {127}},\ \bibinfo
  {pages} {251302} (\bibinfo {year} {2021})},\ \Eprint
  {http://arxiv.org/abs/2104.13930} {arXiv:2104.13930 [astro-ph.CO]}
  \BibitemShut {NoStop}%
\bibitem [{\citenamefont {Xue}\ \emph {et~al.}(2021)\citenamefont {Xue} \emph
  {et~al.}}]{Xue:2021gyq}%
  \BibitemOpen
  \bibfield  {author} {\bibinfo {author} {\bibfnamefont {X.}~\bibnamefont
  {Xue}} \emph {et~al.},\ }\href {\doibase 10.1103/PhysRevLett.127.251303}
  {\bibfield  {journal} {\bibinfo  {journal} {Phys. Rev. Lett.}\ }\textbf
  {\bibinfo {volume} {127}},\ \bibinfo {pages} {251303} (\bibinfo {year}
  {2021})},\ \Eprint {http://arxiv.org/abs/2110.03096} {arXiv:2110.03096
  [astro-ph.CO]} \BibitemShut {NoStop}%
\bibitem [{\citenamefont {Nakai}\ \emph {et~al.}(2021)\citenamefont {Nakai},
  \citenamefont {Suzuki}, \citenamefont {Takahashi},\ and\ \citenamefont
  {Yamada}}]{Nakai:2020oit}%
  \BibitemOpen
  \bibfield  {author} {\bibinfo {author} {\bibfnamefont {Y.}~\bibnamefont
  {Nakai}}, \bibinfo {author} {\bibfnamefont {M.}~\bibnamefont {Suzuki}},
  \bibinfo {author} {\bibfnamefont {F.}~\bibnamefont {Takahashi}}, \ and\
  \bibinfo {author} {\bibfnamefont {M.}~\bibnamefont {Yamada}},\ }\href
  {\doibase 10.1016/j.physletb.2021.136238} {\bibfield  {journal} {\bibinfo
  {journal} {Phys. Lett. B}\ }\textbf {\bibinfo {volume} {816}},\ \bibinfo
  {pages} {136238} (\bibinfo {year} {2021})},\ \Eprint
  {http://arxiv.org/abs/2009.09754} {arXiv:2009.09754 [astro-ph.CO]}
  \BibitemShut {NoStop}%
\bibitem [{\citenamefont {Di~Bari}\ \emph {et~al.}(2021)\citenamefont
  {Di~Bari}, \citenamefont {Marfatia},\ and\ \citenamefont
  {Zhou}}]{DiBari:2021dri}%
  \BibitemOpen
  \bibfield  {author} {\bibinfo {author} {\bibfnamefont {P.}~\bibnamefont
  {Di~Bari}}, \bibinfo {author} {\bibfnamefont {D.}~\bibnamefont {Marfatia}}, \
  and\ \bibinfo {author} {\bibfnamefont {Y.-L.}\ \bibnamefont {Zhou}},\ }\href
  {\doibase 10.1007/JHEP10(2021)193} {\bibfield  {journal} {\bibinfo  {journal}
  {JHEP}\ }\textbf {\bibinfo {volume} {10}},\ \bibinfo {pages} {193} (\bibinfo
  {year} {2021})},\ \Eprint {http://arxiv.org/abs/2106.00025} {arXiv:2106.00025
  [hep-ph]} \BibitemShut {NoStop}%
\bibitem [{\citenamefont {Sakharov}\ \emph {et~al.}(2021)\citenamefont
  {Sakharov}, \citenamefont {Eroshenko},\ and\ \citenamefont
  {Rubin}}]{Sakharov:2021dim}%
  \BibitemOpen
  \bibfield  {author} {\bibinfo {author} {\bibfnamefont {A.~S.}\ \bibnamefont
  {Sakharov}}, \bibinfo {author} {\bibfnamefont {Y.~N.}\ \bibnamefont
  {Eroshenko}}, \ and\ \bibinfo {author} {\bibfnamefont {S.~G.}\ \bibnamefont
  {Rubin}},\ }\href {\doibase 10.1103/PhysRevD.104.043005} {\bibfield
  {journal} {\bibinfo  {journal} {Phys. Rev. D}\ }\textbf {\bibinfo {volume}
  {104}},\ \bibinfo {pages} {043005} (\bibinfo {year} {2021})},\ \Eprint
  {http://arxiv.org/abs/2104.08750} {arXiv:2104.08750 [hep-ph]} \BibitemShut
  {NoStop}%
\bibitem [{\citenamefont {Li}\ \emph {et~al.}(2021)\citenamefont {Li},
  \citenamefont {Shao}, \citenamefont {Wu},\ and\ \citenamefont
  {Yu}}]{Li:2021qer}%
  \BibitemOpen
  \bibfield  {author} {\bibinfo {author} {\bibfnamefont {S.-L.}\ \bibnamefont
  {Li}}, \bibinfo {author} {\bibfnamefont {L.}~\bibnamefont {Shao}}, \bibinfo
  {author} {\bibfnamefont {P.}~\bibnamefont {Wu}}, \ and\ \bibinfo {author}
  {\bibfnamefont {H.}~\bibnamefont {Yu}},\ }\href {\doibase
  10.1103/PhysRevD.104.043510} {\bibfield  {journal} {\bibinfo  {journal}
  {Phys. Rev. D}\ }\textbf {\bibinfo {volume} {104}},\ \bibinfo {pages}
  {043510} (\bibinfo {year} {2021})},\ \Eprint
  {http://arxiv.org/abs/2101.08012} {arXiv:2101.08012 [astro-ph.CO]}
  \BibitemShut {NoStop}%
\bibitem [{\citenamefont {Ashoorioon}\ \emph {et~al.}(2022)\citenamefont
  {Ashoorioon}, \citenamefont {Rezazadeh},\ and\ \citenamefont
  {Rostami}}]{Ashoorioon:2022raz}%
  \BibitemOpen
  \bibfield  {author} {\bibinfo {author} {\bibfnamefont {A.}~\bibnamefont
  {Ashoorioon}}, \bibinfo {author} {\bibfnamefont {K.}~\bibnamefont
  {Rezazadeh}}, \ and\ \bibinfo {author} {\bibfnamefont {A.}~\bibnamefont
  {Rostami}},\ }\href {\doibase 10.1016/j.physletb.2022.137542} {\bibfield
  {journal} {\bibinfo  {journal} {Phys. Lett. B}\ }\textbf {\bibinfo {volume}
  {835}},\ \bibinfo {pages} {137542} (\bibinfo {year} {2022})},\ \Eprint
  {http://arxiv.org/abs/2202.01131} {arXiv:2202.01131 [astro-ph.CO]}
  \BibitemShut {NoStop}%
\bibitem [{\citenamefont {Benetti}\ \emph {et~al.}(2022)\citenamefont
  {Benetti}, \citenamefont {Graef},\ and\ \citenamefont
  {Vagnozzi}}]{Benetti:2021uea}%
  \BibitemOpen
  \bibfield  {author} {\bibinfo {author} {\bibfnamefont {M.}~\bibnamefont
  {Benetti}}, \bibinfo {author} {\bibfnamefont {L.~L.}\ \bibnamefont {Graef}},
  \ and\ \bibinfo {author} {\bibfnamefont {S.}~\bibnamefont {Vagnozzi}},\
  }\href {\doibase 10.1103/PhysRevD.105.043520} {\bibfield  {journal} {\bibinfo
   {journal} {Phys. Rev. D}\ }\textbf {\bibinfo {volume} {105}},\ \bibinfo
  {pages} {043520} (\bibinfo {year} {2022})},\ \Eprint
  {http://arxiv.org/abs/2111.04758} {arXiv:2111.04758 [astro-ph.CO]}
  \BibitemShut {NoStop}%
\bibitem [{\citenamefont {Barir}\ \emph {et~al.}(2023)\citenamefont {Barir},
  \citenamefont {Geller}, \citenamefont {Sun},\ and\ \citenamefont
  {Volansky}}]{Barir:2022kzo}%
  \BibitemOpen
  \bibfield  {author} {\bibinfo {author} {\bibfnamefont {J.}~\bibnamefont
  {Barir}}, \bibinfo {author} {\bibfnamefont {M.}~\bibnamefont {Geller}},
  \bibinfo {author} {\bibfnamefont {C.}~\bibnamefont {Sun}}, \ and\ \bibinfo
  {author} {\bibfnamefont {T.}~\bibnamefont {Volansky}},\ }\href {\doibase
  10.1103/PhysRevD.108.115016} {\bibfield  {journal} {\bibinfo  {journal}
  {Phys. Rev. D}\ }\textbf {\bibinfo {volume} {108}},\ \bibinfo {pages}
  {115016} (\bibinfo {year} {2023})},\ \Eprint
  {http://arxiv.org/abs/2203.00693} {arXiv:2203.00693 [hep-ph]} \BibitemShut
  {NoStop}%
\bibitem [{\citenamefont {Hindmarsh}\ and\ \citenamefont
  {Kume}(2023)}]{Hindmarsh:2022awe}%
  \BibitemOpen
  \bibfield  {author} {\bibinfo {author} {\bibfnamefont {M.}~\bibnamefont
  {Hindmarsh}}\ and\ \bibinfo {author} {\bibfnamefont {J.}~\bibnamefont
  {Kume}},\ }\href {\doibase 10.1088/1475-7516/2023/04/045} {\bibfield
  {journal} {\bibinfo  {journal} {JCAP}\ }\textbf {\bibinfo {volume} {04}},\
  \bibinfo {pages} {045} (\bibinfo {year} {2023})},\ \Eprint
  {http://arxiv.org/abs/2210.06178} {arXiv:2210.06178 [astro-ph.CO]}
  \BibitemShut {NoStop}%
\bibitem [{\citenamefont {Gouttenoire}\ and\ \citenamefont
  {Volansky}(2023)}]{Gouttenoire:2023naa}%
  \BibitemOpen
  \bibfield  {author} {\bibinfo {author} {\bibfnamefont {Y.}~\bibnamefont
  {Gouttenoire}}\ and\ \bibinfo {author} {\bibfnamefont {T.}~\bibnamefont
  {Volansky}},\ }\href@noop {} {\  (\bibinfo {year} {2023})},\ \Eprint
  {http://arxiv.org/abs/2305.04942} {arXiv:2305.04942 [hep-ph]} \BibitemShut
  {NoStop}%
\bibitem [{\citenamefont {Baldes}\ \emph {et~al.}(2023)\citenamefont {Baldes},
  \citenamefont {Dichtl}, \citenamefont {Gouttenoire},\ and\ \citenamefont
  {Sala}}]{Baldes:2023fsp}%
  \BibitemOpen
  \bibfield  {author} {\bibinfo {author} {\bibfnamefont {I.}~\bibnamefont
  {Baldes}}, \bibinfo {author} {\bibfnamefont {M.}~\bibnamefont {Dichtl}},
  \bibinfo {author} {\bibfnamefont {Y.}~\bibnamefont {Gouttenoire}}, \ and\
  \bibinfo {author} {\bibfnamefont {F.}~\bibnamefont {Sala}},\ }\href@noop {}
  {\  (\bibinfo {year} {2023})},\ \Eprint {http://arxiv.org/abs/2306.15555}
  {arXiv:2306.15555 [hep-ph]} \BibitemShut {NoStop}%
\bibitem [{\citenamefont {Li}\ and\ \citenamefont {Xie}(2023)}]{Li:2023bxy}%
  \BibitemOpen
  \bibfield  {author} {\bibinfo {author} {\bibfnamefont {S.-P.}\ \bibnamefont
  {Li}}\ and\ \bibinfo {author} {\bibfnamefont {K.-P.}\ \bibnamefont {Xie}},\
  }\href {\doibase 10.1103/PhysRevD.108.055018} {\bibfield  {journal} {\bibinfo
   {journal} {Phys. Rev. D}\ }\textbf {\bibinfo {volume} {108}},\ \bibinfo
  {pages} {055018} (\bibinfo {year} {2023})},\ \Eprint
  {http://arxiv.org/abs/2307.01086} {arXiv:2307.01086 [hep-ph]} \BibitemShut
  {NoStop}%
\bibitem [{\citenamefont {Ellis}\ and\ \citenamefont
  {Lewicki}(2021)}]{Ellis:2020ena}%
  \BibitemOpen
  \bibfield  {author} {\bibinfo {author} {\bibfnamefont {J.}~\bibnamefont
  {Ellis}}\ and\ \bibinfo {author} {\bibfnamefont {M.}~\bibnamefont
  {Lewicki}},\ }\href {\doibase 10.1103/PhysRevLett.126.041304} {\bibfield
  {journal} {\bibinfo  {journal} {Phys. Rev. Lett.}\ }\textbf {\bibinfo
  {volume} {126}},\ \bibinfo {pages} {041304} (\bibinfo {year} {2021})},\
  \Eprint {http://arxiv.org/abs/2009.06555} {arXiv:2009.06555 [astro-ph.CO]}
  \BibitemShut {NoStop}%
\bibitem [{\citenamefont {Datta}\ \emph {et~al.}(2021)\citenamefont {Datta},
  \citenamefont {Ghosal},\ and\ \citenamefont {Samanta}}]{Datta:2020bht}%
  \BibitemOpen
  \bibfield  {author} {\bibinfo {author} {\bibfnamefont {S.}~\bibnamefont
  {Datta}}, \bibinfo {author} {\bibfnamefont {A.}~\bibnamefont {Ghosal}}, \
  and\ \bibinfo {author} {\bibfnamefont {R.}~\bibnamefont {Samanta}},\ }\href
  {\doibase 10.1088/1475-7516/2021/08/021} {\bibfield  {journal} {\bibinfo
  {journal} {JCAP}\ }\textbf {\bibinfo {volume} {08}},\ \bibinfo {pages} {021}
  (\bibinfo {year} {2021})},\ \Eprint {http://arxiv.org/abs/2012.14981}
  {arXiv:2012.14981 [hep-ph]} \BibitemShut {NoStop}%
\bibitem [{\citenamefont {Samanta}\ and\ \citenamefont
  {Datta}(2021)}]{Samanta:2020cdk}%
  \BibitemOpen
  \bibfield  {author} {\bibinfo {author} {\bibfnamefont {R.}~\bibnamefont
  {Samanta}}\ and\ \bibinfo {author} {\bibfnamefont {S.}~\bibnamefont
  {Datta}},\ }\href {\doibase 10.1007/JHEP05(2021)211} {\bibfield  {journal}
  {\bibinfo  {journal} {JHEP}\ }\textbf {\bibinfo {volume} {05}},\ \bibinfo
  {pages} {211} (\bibinfo {year} {2021})},\ \Eprint
  {http://arxiv.org/abs/2009.13452} {arXiv:2009.13452 [hep-ph]} \BibitemShut
  {NoStop}%
\bibitem [{\citenamefont {Buchmuller}\ \emph {et~al.}(2020)\citenamefont
  {Buchmuller}, \citenamefont {Domcke},\ and\ \citenamefont
  {Schmitz}}]{Buchmuller:2020lbh}%
  \BibitemOpen
  \bibfield  {author} {\bibinfo {author} {\bibfnamefont {W.}~\bibnamefont
  {Buchmuller}}, \bibinfo {author} {\bibfnamefont {V.}~\bibnamefont {Domcke}},
  \ and\ \bibinfo {author} {\bibfnamefont {K.}~\bibnamefont {Schmitz}},\ }\href
  {\doibase 10.1016/j.physletb.2020.135914} {\bibfield  {journal} {\bibinfo
  {journal} {Phys. Lett. B}\ }\textbf {\bibinfo {volume} {811}},\ \bibinfo
  {pages} {135914} (\bibinfo {year} {2020})},\ \Eprint
  {http://arxiv.org/abs/2009.10649} {arXiv:2009.10649 [astro-ph.CO]}
  \BibitemShut {NoStop}%
\bibitem [{\citenamefont {Blasi}\ \emph {et~al.}(2021)\citenamefont {Blasi},
  \citenamefont {Brdar},\ and\ \citenamefont {Schmitz}}]{Blasi:2020mfx}%
  \BibitemOpen
  \bibfield  {author} {\bibinfo {author} {\bibfnamefont {S.}~\bibnamefont
  {Blasi}}, \bibinfo {author} {\bibfnamefont {V.}~\bibnamefont {Brdar}}, \ and\
  \bibinfo {author} {\bibfnamefont {K.}~\bibnamefont {Schmitz}},\ }\href
  {\doibase 10.1103/PhysRevLett.126.041305} {\bibfield  {journal} {\bibinfo
  {journal} {Phys. Rev. Lett.}\ }\textbf {\bibinfo {volume} {126}},\ \bibinfo
  {pages} {041305} (\bibinfo {year} {2021})},\ \Eprint
  {http://arxiv.org/abs/2009.06607} {arXiv:2009.06607 [astro-ph.CO]}
  \BibitemShut {NoStop}%
\bibitem [{\citenamefont {Ramazanov}\ \emph {et~al.}(2022)\citenamefont
  {Ramazanov}, \citenamefont {Babichev}, \citenamefont {Gorbunov},\ and\
  \citenamefont {Vikman}}]{Ramazanov:2021eya}%
  \BibitemOpen
  \bibfield  {author} {\bibinfo {author} {\bibfnamefont {S.}~\bibnamefont
  {Ramazanov}}, \bibinfo {author} {\bibfnamefont {E.}~\bibnamefont {Babichev}},
  \bibinfo {author} {\bibfnamefont {D.}~\bibnamefont {Gorbunov}}, \ and\
  \bibinfo {author} {\bibfnamefont {A.}~\bibnamefont {Vikman}},\ }\href
  {\doibase 10.1103/PhysRevD.105.063530} {\bibfield  {journal} {\bibinfo
  {journal} {Phys. Rev. D}\ }\textbf {\bibinfo {volume} {105}},\ \bibinfo
  {pages} {063530} (\bibinfo {year} {2022})},\ \Eprint
  {http://arxiv.org/abs/2104.13722} {arXiv:2104.13722 [hep-ph]} \BibitemShut
  {NoStop}%
\bibitem [{\citenamefont {Babichev}\ \emph {et~al.}(2022)\citenamefont
  {Babichev}, \citenamefont {Gorbunov}, \citenamefont {Ramazanov},\ and\
  \citenamefont {Vikman}}]{Babichev:2021uvl}%
  \BibitemOpen
  \bibfield  {author} {\bibinfo {author} {\bibfnamefont {E.}~\bibnamefont
  {Babichev}}, \bibinfo {author} {\bibfnamefont {D.}~\bibnamefont {Gorbunov}},
  \bibinfo {author} {\bibfnamefont {S.}~\bibnamefont {Ramazanov}}, \ and\
  \bibinfo {author} {\bibfnamefont {A.}~\bibnamefont {Vikman}},\ }\href
  {\doibase 10.1088/1475-7516/2022/04/028} {\bibfield  {journal} {\bibinfo
  {journal} {JCAP}\ }\textbf {\bibinfo {volume} {04}},\ \bibinfo {pages} {028}
  (\bibinfo {year} {2022})},\ \Eprint {http://arxiv.org/abs/2112.12608}
  {arXiv:2112.12608 [hep-ph]} \BibitemShut {NoStop}%
\bibitem [{\citenamefont {Gorghetto}\ \emph {et~al.}(2021)\citenamefont
  {Gorghetto}, \citenamefont {Hardy},\ and\ \citenamefont
  {Nicolaescu}}]{Gorghetto:2021fsn}%
  \BibitemOpen
  \bibfield  {author} {\bibinfo {author} {\bibfnamefont {M.}~\bibnamefont
  {Gorghetto}}, \bibinfo {author} {\bibfnamefont {E.}~\bibnamefont {Hardy}}, \
  and\ \bibinfo {author} {\bibfnamefont {H.}~\bibnamefont {Nicolaescu}},\
  }\href {\doibase 10.1088/1475-7516/2021/06/034} {\bibfield  {journal}
  {\bibinfo  {journal} {JCAP}\ }\textbf {\bibinfo {volume} {06}},\ \bibinfo
  {pages} {034} (\bibinfo {year} {2021})},\ \Eprint
  {http://arxiv.org/abs/2101.11007} {arXiv:2101.11007 [hep-ph]} \BibitemShut
  {NoStop}%
\bibitem [{\citenamefont {Buchmuller}\ \emph {et~al.}(2021)\citenamefont
  {Buchmuller}, \citenamefont {Domcke},\ and\ \citenamefont
  {Schmitz}}]{Buchmuller:2021mbb}%
  \BibitemOpen
  \bibfield  {author} {\bibinfo {author} {\bibfnamefont {W.}~\bibnamefont
  {Buchmuller}}, \bibinfo {author} {\bibfnamefont {V.}~\bibnamefont {Domcke}},
  \ and\ \bibinfo {author} {\bibfnamefont {K.}~\bibnamefont {Schmitz}},\ }\href
  {\doibase 10.1088/1475-7516/2021/12/006} {\bibfield  {journal} {\bibinfo
  {journal} {JCAP}\ }\textbf {\bibinfo {volume} {12}},\ \bibinfo {pages} {006}
  (\bibinfo {year} {2021})},\ \Eprint {http://arxiv.org/abs/2107.04578}
  {arXiv:2107.04578 [hep-ph]} \BibitemShut {NoStop}%
\bibitem [{\citenamefont {Blanco-Pillado}\ \emph {et~al.}(2021)\citenamefont
  {Blanco-Pillado}, \citenamefont {Olum},\ and\ \citenamefont
  {Wachter}}]{Blanco-Pillado:2021ygr}%
  \BibitemOpen
  \bibfield  {author} {\bibinfo {author} {\bibfnamefont {J.~J.}\ \bibnamefont
  {Blanco-Pillado}}, \bibinfo {author} {\bibfnamefont {K.~D.}\ \bibnamefont
  {Olum}}, \ and\ \bibinfo {author} {\bibfnamefont {J.~M.}\ \bibnamefont
  {Wachter}},\ }\href {\doibase 10.1103/PhysRevD.103.103512} {\bibfield
  {journal} {\bibinfo  {journal} {Phys. Rev. D}\ }\textbf {\bibinfo {volume}
  {103}},\ \bibinfo {pages} {103512} (\bibinfo {year} {2021})},\ \Eprint
  {http://arxiv.org/abs/2102.08194} {arXiv:2102.08194 [astro-ph.CO]}
  \BibitemShut {NoStop}%
\bibitem [{\citenamefont {Ferreira}\ \emph {et~al.}(2023)\citenamefont
  {Ferreira}, \citenamefont {Notari}, \citenamefont {Pujolas},\ and\
  \citenamefont {Rompineve}}]{Ferreira:2022zzo}%
  \BibitemOpen
  \bibfield  {author} {\bibinfo {author} {\bibfnamefont {R.~Z.}\ \bibnamefont
  {Ferreira}}, \bibinfo {author} {\bibfnamefont {A.}~\bibnamefont {Notari}},
  \bibinfo {author} {\bibfnamefont {O.}~\bibnamefont {Pujolas}}, \ and\
  \bibinfo {author} {\bibfnamefont {F.}~\bibnamefont {Rompineve}},\ }\href
  {\doibase 10.1088/1475-7516/2023/02/001} {\bibfield  {journal} {\bibinfo
  {journal} {JCAP}\ }\textbf {\bibinfo {volume} {02}},\ \bibinfo {pages} {001}
  (\bibinfo {year} {2023})},\ \Eprint {http://arxiv.org/abs/2204.04228}
  {arXiv:2204.04228 [astro-ph.CO]} \BibitemShut {NoStop}%
\bibitem [{\citenamefont {An}\ and\ \citenamefont {Yang}(2023)}]{An:2023idh}%
  \BibitemOpen
  \bibfield  {author} {\bibinfo {author} {\bibfnamefont {H.}~\bibnamefont
  {An}}\ and\ \bibinfo {author} {\bibfnamefont {C.}~\bibnamefont {Yang}},\
  }\href@noop {} {\  (\bibinfo {year} {2023})},\ \Eprint
  {http://arxiv.org/abs/2304.02361} {arXiv:2304.02361 [hep-ph]} \BibitemShut
  {NoStop}%
\bibitem [{\citenamefont {Qiu}\ and\ \citenamefont {Yu}(2023)}]{Qiu:2023wbs}%
  \BibitemOpen
  \bibfield  {author} {\bibinfo {author} {\bibfnamefont {Z.-Y.}\ \bibnamefont
  {Qiu}}\ and\ \bibinfo {author} {\bibfnamefont {Z.-H.}\ \bibnamefont {Yu}},\
  }\href {\doibase 10.1088/1674-1137/acd9bf} {\bibfield  {journal} {\bibinfo
  {journal} {Chin. Phys. C}\ }\textbf {\bibinfo {volume} {47}},\ \bibinfo
  {pages} {085104} (\bibinfo {year} {2023})},\ \Eprint
  {http://arxiv.org/abs/2304.02506} {arXiv:2304.02506 [hep-ph]} \BibitemShut
  {NoStop}%
\bibitem [{\citenamefont {Zeng}\ \emph {et~al.}(2023)\citenamefont {Zeng},
  \citenamefont {Liu},\ and\ \citenamefont {Guo}}]{Zeng:2023jut}%
  \BibitemOpen
  \bibfield  {author} {\bibinfo {author} {\bibfnamefont {Z.-M.}\ \bibnamefont
  {Zeng}}, \bibinfo {author} {\bibfnamefont {J.}~\bibnamefont {Liu}}, \ and\
  \bibinfo {author} {\bibfnamefont {Z.-K.}\ \bibnamefont {Guo}},\ }\href
  {\doibase 10.1103/PhysRevD.108.063005} {\bibfield  {journal} {\bibinfo
  {journal} {Phys. Rev. D}\ }\textbf {\bibinfo {volume} {108}},\ \bibinfo
  {pages} {063005} (\bibinfo {year} {2023})},\ \Eprint
  {http://arxiv.org/abs/2301.07230} {arXiv:2301.07230 [astro-ph.CO]}
  \BibitemShut {NoStop}%
\bibitem [{\citenamefont {King}\ \emph {et~al.}(2024)\citenamefont {King},
  \citenamefont {Marfatia},\ and\ \citenamefont {Rahat}}]{King:2023cgv}%
  \BibitemOpen
  \bibfield  {author} {\bibinfo {author} {\bibfnamefont {S.~F.}\ \bibnamefont
  {King}}, \bibinfo {author} {\bibfnamefont {D.}~\bibnamefont {Marfatia}}, \
  and\ \bibinfo {author} {\bibfnamefont {M.~H.}\ \bibnamefont {Rahat}},\ }\href
  {\doibase 10.1103/PhysRevD.109.035014} {\bibfield  {journal} {\bibinfo
  {journal} {Phys. Rev. D}\ }\textbf {\bibinfo {volume} {109}},\ \bibinfo
  {pages} {035014} (\bibinfo {year} {2024})},\ \Eprint
  {http://arxiv.org/abs/2306.05389} {arXiv:2306.05389 [hep-ph]} \BibitemShut
  {NoStop}%
\bibitem [{\citenamefont {Babichev}\ \emph {et~al.}(2023)\citenamefont
  {Babichev}, \citenamefont {Gorbunov}, \citenamefont {Ramazanov},
  \citenamefont {Samanta},\ and\ \citenamefont {Vikman}}]{Babichev:2023pbf}%
  \BibitemOpen
  \bibfield  {author} {\bibinfo {author} {\bibfnamefont {E.}~\bibnamefont
  {Babichev}}, \bibinfo {author} {\bibfnamefont {D.}~\bibnamefont {Gorbunov}},
  \bibinfo {author} {\bibfnamefont {S.}~\bibnamefont {Ramazanov}}, \bibinfo
  {author} {\bibfnamefont {R.}~\bibnamefont {Samanta}}, \ and\ \bibinfo
  {author} {\bibfnamefont {A.}~\bibnamefont {Vikman}},\ }\href {\doibase
  10.1103/PhysRevD.108.123529} {\bibfield  {journal} {\bibinfo  {journal}
  {Phys. Rev. D}\ }\textbf {\bibinfo {volume} {108}},\ \bibinfo {pages}
  {123529} (\bibinfo {year} {2023})},\ \Eprint
  {http://arxiv.org/abs/2307.04582} {arXiv:2307.04582 [hep-ph]} \BibitemShut
  {NoStop}%
\bibitem [{\citenamefont {Kitajima}\ \emph {et~al.}(2024)\citenamefont
  {Kitajima}, \citenamefont {Lee}, \citenamefont {Murai}, \citenamefont
  {Takahashi},\ and\ \citenamefont {Yin}}]{Kitajima:2023cek}%
  \BibitemOpen
  \bibfield  {author} {\bibinfo {author} {\bibfnamefont {N.}~\bibnamefont
  {Kitajima}}, \bibinfo {author} {\bibfnamefont {J.}~\bibnamefont {Lee}},
  \bibinfo {author} {\bibfnamefont {K.}~\bibnamefont {Murai}}, \bibinfo
  {author} {\bibfnamefont {F.}~\bibnamefont {Takahashi}}, \ and\ \bibinfo
  {author} {\bibfnamefont {W.}~\bibnamefont {Yin}},\ }\href {\doibase
  10.1016/j.physletb.2024.138586} {\bibfield  {journal} {\bibinfo  {journal}
  {Phys. Lett. B}\ }\textbf {\bibinfo {volume} {851}},\ \bibinfo {pages}
  {138586} (\bibinfo {year} {2024})},\ \Eprint
  {http://arxiv.org/abs/2306.17146} {arXiv:2306.17146 [hep-ph]} \BibitemShut
  {NoStop}%
\bibitem [{\citenamefont {Barman}\ \emph {et~al.}(2023)\citenamefont {Barman},
  \citenamefont {Borah}, \citenamefont {Jyoti~Das},\ and\ \citenamefont
  {Saha}}]{Barman:2023fad}%
  \BibitemOpen
  \bibfield  {author} {\bibinfo {author} {\bibfnamefont {B.}~\bibnamefont
  {Barman}}, \bibinfo {author} {\bibfnamefont {D.}~\bibnamefont {Borah}},
  \bibinfo {author} {\bibfnamefont {S.}~\bibnamefont {Jyoti~Das}}, \ and\
  \bibinfo {author} {\bibfnamefont {I.}~\bibnamefont {Saha}},\ }\href {\doibase
  10.1088/1475-7516/2023/10/053} {\bibfield  {journal} {\bibinfo  {journal}
  {JCAP}\ }\textbf {\bibinfo {volume} {10}},\ \bibinfo {pages} {053} (\bibinfo
  {year} {2023})},\ \Eprint {http://arxiv.org/abs/2307.00656} {arXiv:2307.00656
  [hep-ph]} \BibitemShut {NoStop}%
\bibitem [{\citenamefont {Vaskonen}\ and\ \citenamefont
  {Veerm\"ae}(2021)}]{Vaskonen:2020lbd}%
  \BibitemOpen
  \bibfield  {author} {\bibinfo {author} {\bibfnamefont {V.}~\bibnamefont
  {Vaskonen}}\ and\ \bibinfo {author} {\bibfnamefont {H.}~\bibnamefont
  {Veerm\"ae}},\ }\href {\doibase 10.1103/PhysRevLett.126.051303} {\bibfield
  {journal} {\bibinfo  {journal} {Phys. Rev. Lett.}\ }\textbf {\bibinfo
  {volume} {126}},\ \bibinfo {pages} {051303} (\bibinfo {year} {2021})},\
  \Eprint {http://arxiv.org/abs/2009.07832} {arXiv:2009.07832 [astro-ph.CO]}
  \BibitemShut {NoStop}%
\bibitem [{\citenamefont {De~Luca}\ \emph {et~al.}(2021)\citenamefont
  {De~Luca}, \citenamefont {Franciolini},\ and\ \citenamefont
  {Riotto}}]{DeLuca:2020agl}%
  \BibitemOpen
  \bibfield  {author} {\bibinfo {author} {\bibfnamefont {V.}~\bibnamefont
  {De~Luca}}, \bibinfo {author} {\bibfnamefont {G.}~\bibnamefont
  {Franciolini}}, \ and\ \bibinfo {author} {\bibfnamefont {A.}~\bibnamefont
  {Riotto}},\ }\href {\doibase 10.1103/PhysRevLett.126.041303} {\bibfield
  {journal} {\bibinfo  {journal} {Phys. Rev. Lett.}\ }\textbf {\bibinfo
  {volume} {126}},\ \bibinfo {pages} {041303} (\bibinfo {year} {2021})},\
  \Eprint {http://arxiv.org/abs/2009.08268} {arXiv:2009.08268 [astro-ph.CO]}
  \BibitemShut {NoStop}%
\bibitem [{\citenamefont {Bhaumik}\ and\ \citenamefont
  {Jain}(2021)}]{Bhaumik:2020dor}%
  \BibitemOpen
  \bibfield  {author} {\bibinfo {author} {\bibfnamefont {N.}~\bibnamefont
  {Bhaumik}}\ and\ \bibinfo {author} {\bibfnamefont {R.~K.}\ \bibnamefont
  {Jain}},\ }\href {\doibase 10.1103/PhysRevD.104.023531} {\bibfield  {journal}
  {\bibinfo  {journal} {Phys. Rev. D}\ }\textbf {\bibinfo {volume} {104}},\
  \bibinfo {pages} {023531} (\bibinfo {year} {2021})},\ \Eprint
  {http://arxiv.org/abs/2009.10424} {arXiv:2009.10424 [astro-ph.CO]}
  \BibitemShut {NoStop}%
\bibitem [{\citenamefont {Inomata}\ \emph {et~al.}(2021)\citenamefont
  {Inomata}, \citenamefont {Kawasaki}, \citenamefont {Mukaida},\ and\
  \citenamefont {Yanagida}}]{Inomata:2020xad}%
  \BibitemOpen
  \bibfield  {author} {\bibinfo {author} {\bibfnamefont {K.}~\bibnamefont
  {Inomata}}, \bibinfo {author} {\bibfnamefont {M.}~\bibnamefont {Kawasaki}},
  \bibinfo {author} {\bibfnamefont {K.}~\bibnamefont {Mukaida}}, \ and\
  \bibinfo {author} {\bibfnamefont {T.~T.}\ \bibnamefont {Yanagida}},\ }\href
  {\doibase 10.1103/PhysRevLett.126.131301} {\bibfield  {journal} {\bibinfo
  {journal} {Phys. Rev. Lett.}\ }\textbf {\bibinfo {volume} {126}},\ \bibinfo
  {pages} {131301} (\bibinfo {year} {2021})},\ \Eprint
  {http://arxiv.org/abs/2011.01270} {arXiv:2011.01270 [astro-ph.CO]}
  \BibitemShut {NoStop}%
\bibitem [{\citenamefont {Kohri}\ and\ \citenamefont
  {Terada}(2021)}]{Kohri:2020qqd}%
  \BibitemOpen
  \bibfield  {author} {\bibinfo {author} {\bibfnamefont {K.}~\bibnamefont
  {Kohri}}\ and\ \bibinfo {author} {\bibfnamefont {T.}~\bibnamefont {Terada}},\
  }\href {\doibase 10.1016/j.physletb.2020.136040} {\bibfield  {journal}
  {\bibinfo  {journal} {Phys. Lett. B}\ }\textbf {\bibinfo {volume} {813}},\
  \bibinfo {pages} {136040} (\bibinfo {year} {2021})},\ \Eprint
  {http://arxiv.org/abs/2009.11853} {arXiv:2009.11853 [astro-ph.CO]}
  \BibitemShut {NoStop}%
\bibitem [{\citenamefont {Dom\`enech}\ and\ \citenamefont
  {Pi}(2022)}]{Domenech:2020ers}%
  \BibitemOpen
  \bibfield  {author} {\bibinfo {author} {\bibfnamefont {G.}~\bibnamefont
  {Dom\`enech}}\ and\ \bibinfo {author} {\bibfnamefont {S.}~\bibnamefont
  {Pi}},\ }\href {\doibase 10.1007/s11433-021-1839-6} {\bibfield  {journal}
  {\bibinfo  {journal} {Sci. China Phys. Mech. Astron.}\ }\textbf {\bibinfo
  {volume} {65}},\ \bibinfo {pages} {230411} (\bibinfo {year} {2022})},\
  \Eprint {http://arxiv.org/abs/2010.03976} {arXiv:2010.03976 [astro-ph.CO]}
  \BibitemShut {NoStop}%
\bibitem [{\citenamefont {Namba}\ and\ \citenamefont
  {Suzuki}(2020)}]{Namba:2020kij}%
  \BibitemOpen
  \bibfield  {author} {\bibinfo {author} {\bibfnamefont {R.}~\bibnamefont
  {Namba}}\ and\ \bibinfo {author} {\bibfnamefont {M.}~\bibnamefont {Suzuki}},\
  }\href {\doibase 10.1103/PhysRevD.102.123527} {\bibfield  {journal} {\bibinfo
   {journal} {Phys. Rev. D}\ }\textbf {\bibinfo {volume} {102}},\ \bibinfo
  {pages} {123527} (\bibinfo {year} {2020})},\ \Eprint
  {http://arxiv.org/abs/2009.13909} {arXiv:2009.13909 [astro-ph.CO]}
  \BibitemShut {NoStop}%
\bibitem [{\citenamefont {Sugiyama}\ \emph {et~al.}(2021)\citenamefont
  {Sugiyama}, \citenamefont {Takhistov}, \citenamefont {Vitagliano},
  \citenamefont {Kusenko}, \citenamefont {Sasaki},\ and\ \citenamefont
  {Takada}}]{Sugiyama:2020roc}%
  \BibitemOpen
  \bibfield  {author} {\bibinfo {author} {\bibfnamefont {S.}~\bibnamefont
  {Sugiyama}}, \bibinfo {author} {\bibfnamefont {V.}~\bibnamefont {Takhistov}},
  \bibinfo {author} {\bibfnamefont {E.}~\bibnamefont {Vitagliano}}, \bibinfo
  {author} {\bibfnamefont {A.}~\bibnamefont {Kusenko}}, \bibinfo {author}
  {\bibfnamefont {M.}~\bibnamefont {Sasaki}}, \ and\ \bibinfo {author}
  {\bibfnamefont {M.}~\bibnamefont {Takada}},\ }\href {\doibase
  10.1016/j.physletb.2021.136097} {\bibfield  {journal} {\bibinfo  {journal}
  {Phys. Lett. B}\ }\textbf {\bibinfo {volume} {814}},\ \bibinfo {pages}
  {136097} (\bibinfo {year} {2021})},\ \Eprint
  {http://arxiv.org/abs/2010.02189} {arXiv:2010.02189 [astro-ph.CO]}
  \BibitemShut {NoStop}%
\bibitem [{\citenamefont {Zhou}\ \emph {et~al.}(2020)\citenamefont {Zhou},
  \citenamefont {Jiang}, \citenamefont {Cai}, \citenamefont {Sasaki},\ and\
  \citenamefont {Pi}}]{Zhou:2020kkf}%
  \BibitemOpen
  \bibfield  {author} {\bibinfo {author} {\bibfnamefont {Z.}~\bibnamefont
  {Zhou}}, \bibinfo {author} {\bibfnamefont {J.}~\bibnamefont {Jiang}},
  \bibinfo {author} {\bibfnamefont {Y.-F.}\ \bibnamefont {Cai}}, \bibinfo
  {author} {\bibfnamefont {M.}~\bibnamefont {Sasaki}}, \ and\ \bibinfo {author}
  {\bibfnamefont {S.}~\bibnamefont {Pi}},\ }\href {\doibase
  10.1103/PhysRevD.102.103527} {\bibfield  {journal} {\bibinfo  {journal}
  {Phys. Rev. D}\ }\textbf {\bibinfo {volume} {102}},\ \bibinfo {pages}
  {103527} (\bibinfo {year} {2020})},\ \Eprint
  {http://arxiv.org/abs/2010.03537} {arXiv:2010.03537 [astro-ph.CO]}
  \BibitemShut {NoStop}%
\bibitem [{\citenamefont {Lin}\ \emph {et~al.}(2023)\citenamefont {Lin},
  \citenamefont {Gao}, \citenamefont {Gong}, \citenamefont {Lu}, \citenamefont
  {Wang},\ and\ \citenamefont {Zhang}}]{Lin:2021vwc}%
  \BibitemOpen
  \bibfield  {author} {\bibinfo {author} {\bibfnamefont {J.}~\bibnamefont
  {Lin}}, \bibinfo {author} {\bibfnamefont {S.}~\bibnamefont {Gao}}, \bibinfo
  {author} {\bibfnamefont {Y.}~\bibnamefont {Gong}}, \bibinfo {author}
  {\bibfnamefont {Y.}~\bibnamefont {Lu}}, \bibinfo {author} {\bibfnamefont
  {Z.}~\bibnamefont {Wang}}, \ and\ \bibinfo {author} {\bibfnamefont
  {F.}~\bibnamefont {Zhang}},\ }\href {\doibase 10.1103/PhysRevD.107.043517}
  {\bibfield  {journal} {\bibinfo  {journal} {Phys. Rev. D}\ }\textbf {\bibinfo
  {volume} {107}},\ \bibinfo {pages} {043517} (\bibinfo {year} {2023})},\
  \Eprint {http://arxiv.org/abs/2111.01362} {arXiv:2111.01362 [gr-qc]}
  \BibitemShut {NoStop}%
\bibitem [{\citenamefont {Rezazadeh}\ \emph {et~al.}(2022)\citenamefont
  {Rezazadeh}, \citenamefont {Teimoori}, \citenamefont {Karimi},\ and\
  \citenamefont {Karami}}]{Rezazadeh:2021clf}%
  \BibitemOpen
  \bibfield  {author} {\bibinfo {author} {\bibfnamefont {K.}~\bibnamefont
  {Rezazadeh}}, \bibinfo {author} {\bibfnamefont {Z.}~\bibnamefont {Teimoori}},
  \bibinfo {author} {\bibfnamefont {S.}~\bibnamefont {Karimi}}, \ and\ \bibinfo
  {author} {\bibfnamefont {K.}~\bibnamefont {Karami}},\ }\href {\doibase
  10.1140/epjc/s10052-022-10735-w} {\bibfield  {journal} {\bibinfo  {journal}
  {Eur. Phys. J. C}\ }\textbf {\bibinfo {volume} {82}},\ \bibinfo {pages} {758}
  (\bibinfo {year} {2022})},\ \Eprint {http://arxiv.org/abs/2110.01482}
  {arXiv:2110.01482 [gr-qc]} \BibitemShut {NoStop}%
\bibitem [{\citenamefont {Kawasaki}\ and\ \citenamefont
  {Nakatsuka}(2021)}]{Kawasaki:2021ycf}%
  \BibitemOpen
  \bibfield  {author} {\bibinfo {author} {\bibfnamefont {M.}~\bibnamefont
  {Kawasaki}}\ and\ \bibinfo {author} {\bibfnamefont {H.}~\bibnamefont
  {Nakatsuka}},\ }\href {\doibase 10.1088/1475-7516/2021/05/023} {\bibfield
  {journal} {\bibinfo  {journal} {JCAP}\ }\textbf {\bibinfo {volume} {05}},\
  \bibinfo {pages} {023} (\bibinfo {year} {2021})},\ \Eprint
  {http://arxiv.org/abs/2101.11244} {arXiv:2101.11244 [astro-ph.CO]}
  \BibitemShut {NoStop}%
\bibitem [{\citenamefont {Ahmed}\ \emph {et~al.}(2022)\citenamefont {Ahmed},
  \citenamefont {Junaid},\ and\ \citenamefont {Zubair}}]{Ahmed:2021ucx}%
  \BibitemOpen
  \bibfield  {author} {\bibinfo {author} {\bibfnamefont {W.}~\bibnamefont
  {Ahmed}}, \bibinfo {author} {\bibfnamefont {M.}~\bibnamefont {Junaid}}, \
  and\ \bibinfo {author} {\bibfnamefont {U.}~\bibnamefont {Zubair}},\ }\href
  {\doibase 10.1016/j.nuclphysb.2022.115968} {\bibfield  {journal} {\bibinfo
  {journal} {Nucl. Phys. B}\ }\textbf {\bibinfo {volume} {984}},\ \bibinfo
  {pages} {115968} (\bibinfo {year} {2022})},\ \Eprint
  {http://arxiv.org/abs/2109.14838} {arXiv:2109.14838 [astro-ph.CO]}
  \BibitemShut {NoStop}%
\bibitem [{\citenamefont {Yi}\ and\ \citenamefont {Fei}(2023)}]{Yi:2022ymw}%
  \BibitemOpen
  \bibfield  {author} {\bibinfo {author} {\bibfnamefont {Z.}~\bibnamefont
  {Yi}}\ and\ \bibinfo {author} {\bibfnamefont {Q.}~\bibnamefont {Fei}},\
  }\href {\doibase 10.1140/epjc/s10052-023-11233-3} {\bibfield  {journal}
  {\bibinfo  {journal} {Eur. Phys. J. C}\ }\textbf {\bibinfo {volume} {83}},\
  \bibinfo {pages} {82} (\bibinfo {year} {2023})},\ \Eprint
  {http://arxiv.org/abs/2210.03641} {arXiv:2210.03641 [astro-ph.CO]}
  \BibitemShut {NoStop}%
\bibitem [{\citenamefont {Yi}(2023)}]{Yi:2022anu}%
  \BibitemOpen
  \bibfield  {author} {\bibinfo {author} {\bibfnamefont {Z.}~\bibnamefont
  {Yi}},\ }\href {\doibase 10.1088/1475-7516/2023/03/048} {\bibfield  {journal}
  {\bibinfo  {journal} {JCAP}\ }\textbf {\bibinfo {volume} {03}},\ \bibinfo
  {pages} {048} (\bibinfo {year} {2023})},\ \Eprint
  {http://arxiv.org/abs/2206.01039} {arXiv:2206.01039 [gr-qc]} \BibitemShut
  {NoStop}%
\bibitem [{\citenamefont {Dandoy}\ \emph {et~al.}(2023)\citenamefont {Dandoy},
  \citenamefont {Domcke},\ and\ \citenamefont {Rompineve}}]{Dandoy:2023jot}%
  \BibitemOpen
  \bibfield  {author} {\bibinfo {author} {\bibfnamefont {V.}~\bibnamefont
  {Dandoy}}, \bibinfo {author} {\bibfnamefont {V.}~\bibnamefont {Domcke}}, \
  and\ \bibinfo {author} {\bibfnamefont {F.}~\bibnamefont {Rompineve}},\ }\href
  {\doibase 10.21468/SciPostPhysCore.6.3.060} {\bibfield  {journal} {\bibinfo
  {journal} {SciPost Phys. Core}\ }\textbf {\bibinfo {volume} {6}},\ \bibinfo
  {pages} {060} (\bibinfo {year} {2023})},\ \Eprint
  {http://arxiv.org/abs/2302.07901} {arXiv:2302.07901 [astro-ph.CO]}
  \BibitemShut {NoStop}%
\bibitem [{\citenamefont {Zhao}\ \emph {et~al.}(2023)\citenamefont {Zhao},
  \citenamefont {Liu},\ and\ \citenamefont {Li}}]{Zhao:2023xnh}%
  \BibitemOpen
  \bibfield  {author} {\bibinfo {author} {\bibfnamefont {J.-X.}\ \bibnamefont
  {Zhao}}, \bibinfo {author} {\bibfnamefont {X.-H.}\ \bibnamefont {Liu}}, \
  and\ \bibinfo {author} {\bibfnamefont {N.}~\bibnamefont {Li}},\ }\href
  {\doibase 10.1103/PhysRevD.107.043515} {\bibfield  {journal} {\bibinfo
  {journal} {Phys. Rev. D}\ }\textbf {\bibinfo {volume} {107}},\ \bibinfo
  {pages} {043515} (\bibinfo {year} {2023})},\ \Eprint
  {http://arxiv.org/abs/2302.06886} {arXiv:2302.06886 [astro-ph.CO]}
  \BibitemShut {NoStop}%
\bibitem [{\citenamefont {Ferrante}\ \emph {et~al.}(2023)\citenamefont
  {Ferrante}, \citenamefont {Franciolini}, \citenamefont {Iovino},\ and\
  \citenamefont {Urbano}}]{Ferrante:2023bgz}%
  \BibitemOpen
  \bibfield  {author} {\bibinfo {author} {\bibfnamefont {G.}~\bibnamefont
  {Ferrante}}, \bibinfo {author} {\bibfnamefont {G.}~\bibnamefont
  {Franciolini}}, \bibinfo {author} {\bibfnamefont {A.}~\bibnamefont {Iovino},
  \bibfnamefont {Junior.}}, \ and\ \bibinfo {author} {\bibfnamefont
  {A.}~\bibnamefont {Urbano}},\ }\href {\doibase 10.1088/1475-7516/2023/06/057}
  {\bibfield  {journal} {\bibinfo  {journal} {JCAP}\ }\textbf {\bibinfo
  {volume} {06}},\ \bibinfo {pages} {057} (\bibinfo {year} {2023})},\ \Eprint
  {http://arxiv.org/abs/2305.13382} {arXiv:2305.13382 [astro-ph.CO]}
  \BibitemShut {NoStop}%
\bibitem [{\citenamefont {Cai}\ \emph {et~al.}(2023)\citenamefont {Cai},
  \citenamefont {Zhu},\ and\ \citenamefont {Piao}}]{Cai:2023uhc}%
  \BibitemOpen
  \bibfield  {author} {\bibinfo {author} {\bibfnamefont {Y.}~\bibnamefont
  {Cai}}, \bibinfo {author} {\bibfnamefont {M.}~\bibnamefont {Zhu}}, \ and\
  \bibinfo {author} {\bibfnamefont {Y.-S.}\ \bibnamefont {Piao}},\ }\href@noop
  {} {\  (\bibinfo {year} {2023})},\ \Eprint {http://arxiv.org/abs/2305.10933}
  {arXiv:2305.10933 [gr-qc]} \BibitemShut {NoStop}%
\bibitem [{\citenamefont {Franciolini}\ \emph {et~al.}(2023)\citenamefont
  {Franciolini}, \citenamefont {Iovino}, \citenamefont {Vaskonen},\ and\
  \citenamefont {Veermae}}]{Franciolini:2023pbf}%
  \BibitemOpen
  \bibfield  {author} {\bibinfo {author} {\bibfnamefont {G.}~\bibnamefont
  {Franciolini}}, \bibinfo {author} {\bibfnamefont {A.}~\bibnamefont {Iovino},
  \bibfnamefont {Junior.}}, \bibinfo {author} {\bibfnamefont {V.}~\bibnamefont
  {Vaskonen}}, \ and\ \bibinfo {author} {\bibfnamefont {H.}~\bibnamefont
  {Veermae}},\ }\href {\doibase 10.1103/PhysRevLett.131.201401} {\bibfield
  {journal} {\bibinfo  {journal} {Phys. Rev. Lett.}\ }\textbf {\bibinfo
  {volume} {131}},\ \bibinfo {pages} {201401} (\bibinfo {year} {2023})},\
  \Eprint {http://arxiv.org/abs/2306.17149} {arXiv:2306.17149 [astro-ph.CO]}
  \BibitemShut {NoStop}%
\bibitem [{\citenamefont {Balaji}\ \emph {et~al.}(2023)\citenamefont {Balaji},
  \citenamefont {Dom\`enech},\ and\ \citenamefont
  {Franciolini}}]{Balaji:2023ehk}%
  \BibitemOpen
  \bibfield  {author} {\bibinfo {author} {\bibfnamefont {S.}~\bibnamefont
  {Balaji}}, \bibinfo {author} {\bibfnamefont {G.}~\bibnamefont {Dom\`enech}},
  \ and\ \bibinfo {author} {\bibfnamefont {G.}~\bibnamefont {Franciolini}},\
  }\href@noop {} {\  (\bibinfo {year} {2023})},\ \Eprint
  {http://arxiv.org/abs/2307.08552} {arXiv:2307.08552 [gr-qc]} \BibitemShut
  {NoStop}%
\bibitem [{\citenamefont {Liu}\ \emph {et~al.}(2024)\citenamefont {Liu},
  \citenamefont {Chen},\ and\ \citenamefont {Huang}}]{Liu:2023ymk}%
  \BibitemOpen
  \bibfield  {author} {\bibinfo {author} {\bibfnamefont {L.}~\bibnamefont
  {Liu}}, \bibinfo {author} {\bibfnamefont {Z.-C.}\ \bibnamefont {Chen}}, \
  and\ \bibinfo {author} {\bibfnamefont {Q.-G.}\ \bibnamefont {Huang}},\ }\href
  {\doibase 10.1103/PhysRevD.109.L061301} {\bibfield  {journal} {\bibinfo
  {journal} {Phys. Rev. D}\ }\textbf {\bibinfo {volume} {109}},\ \bibinfo
  {pages} {L061301} (\bibinfo {year} {2024})},\ \Eprint
  {http://arxiv.org/abs/2307.01102} {arXiv:2307.01102 [astro-ph.CO]}
  \BibitemShut {NoStop}%
\bibitem [{\citenamefont {Vagnozzi}(2023)}]{Vagnozzi:2023lwo}%
  \BibitemOpen
  \bibfield  {author} {\bibinfo {author} {\bibfnamefont {S.}~\bibnamefont
  {Vagnozzi}},\ }\href {\doibase 10.1016/j.jheap.2023.07.001} {\bibfield
  {journal} {\bibinfo  {journal} {JHEAp}\ }\textbf {\bibinfo {volume} {39}},\
  \bibinfo {pages} {81} (\bibinfo {year} {2023})},\ \Eprint
  {http://arxiv.org/abs/2306.16912} {arXiv:2306.16912 [astro-ph.CO]}
  \BibitemShut {NoStop}%
\bibitem [{\citenamefont {Franciolini}\ \emph {et~al.}(2024)\citenamefont
  {Franciolini}, \citenamefont {Racco},\ and\ \citenamefont
  {Rompineve}}]{Franciolini:2023wjm}%
  \BibitemOpen
  \bibfield  {author} {\bibinfo {author} {\bibfnamefont {G.}~\bibnamefont
  {Franciolini}}, \bibinfo {author} {\bibfnamefont {D.}~\bibnamefont {Racco}},
  \ and\ \bibinfo {author} {\bibfnamefont {F.}~\bibnamefont {Rompineve}},\
  }\href {\doibase 10.1103/PhysRevLett.132.081001} {\bibfield  {journal}
  {\bibinfo  {journal} {Phys. Rev. Lett.}\ }\textbf {\bibinfo {volume} {132}},\
  \bibinfo {pages} {081001} (\bibinfo {year} {2024})},\ \Eprint
  {http://arxiv.org/abs/2306.17136} {arXiv:2306.17136 [astro-ph.CO]}
  \BibitemShut {NoStop}%
\bibitem [{\citenamefont {Madge}\ \emph {et~al.}(2023)\citenamefont {Madge},
  \citenamefont {Morgante}, \citenamefont {Puchades-Ib\'a\~nez}, \citenamefont
  {Ramberg}, \citenamefont {Ratzinger}, \citenamefont {Schenk},\ and\
  \citenamefont {Schwaller}}]{Madge:2023dxc}%
  \BibitemOpen
  \bibfield  {author} {\bibinfo {author} {\bibfnamefont {E.}~\bibnamefont
  {Madge}}, \bibinfo {author} {\bibfnamefont {E.}~\bibnamefont {Morgante}},
  \bibinfo {author} {\bibfnamefont {C.}~\bibnamefont {Puchades-Ib\'a\~nez}},
  \bibinfo {author} {\bibfnamefont {N.}~\bibnamefont {Ramberg}}, \bibinfo
  {author} {\bibfnamefont {W.}~\bibnamefont {Ratzinger}}, \bibinfo {author}
  {\bibfnamefont {S.}~\bibnamefont {Schenk}}, \ and\ \bibinfo {author}
  {\bibfnamefont {P.}~\bibnamefont {Schwaller}},\ }\href {\doibase
  10.1007/JHEP10(2023)171} {\bibfield  {journal} {\bibinfo  {journal} {JHEP}\
  }\textbf {\bibinfo {volume} {10}},\ \bibinfo {pages} {171} (\bibinfo {year}
  {2023})},\ \Eprint {http://arxiv.org/abs/2306.14856} {arXiv:2306.14856
  [hep-ph]} \BibitemShut {NoStop}%
\bibitem [{\citenamefont {Figueroa}\ \emph {et~al.}(2023)\citenamefont
  {Figueroa}, \citenamefont {Pieroni}, \citenamefont {Ricciardone},\ and\
  \citenamefont {Simakachorn}}]{Figueroa:2023zhu}%
  \BibitemOpen
  \bibfield  {author} {\bibinfo {author} {\bibfnamefont {D.~G.}\ \bibnamefont
  {Figueroa}}, \bibinfo {author} {\bibfnamefont {M.}~\bibnamefont {Pieroni}},
  \bibinfo {author} {\bibfnamefont {A.}~\bibnamefont {Ricciardone}}, \ and\
  \bibinfo {author} {\bibfnamefont {P.}~\bibnamefont {Simakachorn}},\
  }\href@noop {} {\  (\bibinfo {year} {2023})},\ \Eprint
  {http://arxiv.org/abs/2307.02399} {arXiv:2307.02399 [astro-ph.CO]}
  \BibitemShut {NoStop}%
\bibitem [{\citenamefont {Garcia-Bellido}\ \emph {et~al.}(2024)\citenamefont
  {Garcia-Bellido}, \citenamefont {Papageorgiou}, \citenamefont {Peloso},\ and\
  \citenamefont {Sorbo}}]{Garcia-Bellido:2023ser}%
  \BibitemOpen
  \bibfield  {author} {\bibinfo {author} {\bibfnamefont {J.}~\bibnamefont
  {Garcia-Bellido}}, \bibinfo {author} {\bibfnamefont {A.}~\bibnamefont
  {Papageorgiou}}, \bibinfo {author} {\bibfnamefont {M.}~\bibnamefont
  {Peloso}}, \ and\ \bibinfo {author} {\bibfnamefont {L.}~\bibnamefont
  {Sorbo}},\ }\href {\doibase 10.1088/1475-7516/2024/01/034} {\bibfield
  {journal} {\bibinfo  {journal} {JCAP}\ }\textbf {\bibinfo {volume} {01}},\
  \bibinfo {pages} {034} (\bibinfo {year} {2024})},\ \Eprint
  {http://arxiv.org/abs/2303.13425} {arXiv:2303.13425 [astro-ph.CO]}
  \BibitemShut {NoStop}%
\bibitem [{\citenamefont {Murai}\ and\ \citenamefont
  {Yin}(2023)}]{Murai:2023gkv}%
  \BibitemOpen
  \bibfield  {author} {\bibinfo {author} {\bibfnamefont {K.}~\bibnamefont
  {Murai}}\ and\ \bibinfo {author} {\bibfnamefont {W.}~\bibnamefont {Yin}},\
  }\href {\doibase 10.1007/JHEP10(2023)062} {\bibfield  {journal} {\bibinfo
  {journal} {JHEP}\ }\textbf {\bibinfo {volume} {10}},\ \bibinfo {pages} {062}
  (\bibinfo {year} {2023})},\ \Eprint {http://arxiv.org/abs/2307.00628}
  {arXiv:2307.00628 [hep-ph]} \BibitemShut {NoStop}%
\bibitem [{\citenamefont {Konoplya}\ and\ \citenamefont
  {Zhidenko}(2023)}]{Konoplya:2023fmh}%
  \BibitemOpen
  \bibfield  {author} {\bibinfo {author} {\bibfnamefont {R.~A.}\ \bibnamefont
  {Konoplya}}\ and\ \bibinfo {author} {\bibfnamefont {A.}~\bibnamefont
  {Zhidenko}},\ }\href@noop {} {\  (\bibinfo {year} {2023})},\ \Eprint
  {http://arxiv.org/abs/2307.01110} {arXiv:2307.01110 [gr-qc]} \BibitemShut
  {NoStop}%
\bibitem [{\citenamefont {Smarra}\ \emph {et~al.}(2023)\citenamefont {Smarra}
  \emph {et~al.}}]{EPTA:2023xiy}%
  \BibitemOpen
  \bibfield  {author} {\bibinfo {author} {\bibfnamefont {C.}~\bibnamefont
  {Smarra}} \emph {et~al.} (\bibinfo {collaboration} {EPTA}),\ }\href@noop {}
  {\  (\bibinfo {year} {2023})},\ \Eprint {http://arxiv.org/abs/2306.16228}
  {arXiv:2306.16228 [astro-ph.HE]} \BibitemShut {NoStop}%
\bibitem [{\citenamefont {Hazboun}\ \emph {et~al.}(2019)\citenamefont
  {Hazboun}, \citenamefont {Romano},\ and\ \citenamefont
  {Smith}}]{Hazboun:2019vhv}%
  \BibitemOpen
  \bibfield  {author} {\bibinfo {author} {\bibfnamefont {J.~S.}\ \bibnamefont
  {Hazboun}}, \bibinfo {author} {\bibfnamefont {J.~D.}\ \bibnamefont {Romano}},
  \ and\ \bibinfo {author} {\bibfnamefont {T.~L.}\ \bibnamefont {Smith}},\
  }\href {\doibase 10.1103/PhysRevD.100.104028} {\bibfield  {journal} {\bibinfo
   {journal} {Phys. Rev. D}\ }\textbf {\bibinfo {volume} {100}},\ \bibinfo
  {pages} {104028} (\bibinfo {year} {2019})},\ \Eprint
  {http://arxiv.org/abs/1907.04341} {arXiv:1907.04341 [gr-qc]} \BibitemShut
  {NoStop}%
\bibitem [{cod()}]{code_repo}%
  \BibitemOpen
  \href@noop {} {}\bibinfo {howpublished}
  {\url{https://github.com/Mauropieroni/fastPTA/}}\BibitemShut {NoStop}%
\bibitem [{\citenamefont {Anholm}\ \emph {et~al.}(2009)\citenamefont {Anholm},
  \citenamefont {Ballmer}, \citenamefont {Creighton}, \citenamefont {Price},\
  and\ \citenamefont {Siemens}}]{Anholm:2008wy}%
  \BibitemOpen
  \bibfield  {author} {\bibinfo {author} {\bibfnamefont {M.}~\bibnamefont
  {Anholm}}, \bibinfo {author} {\bibfnamefont {S.}~\bibnamefont {Ballmer}},
  \bibinfo {author} {\bibfnamefont {J.~D.~E.}\ \bibnamefont {Creighton}},
  \bibinfo {author} {\bibfnamefont {L.~R.}\ \bibnamefont {Price}}, \ and\
  \bibinfo {author} {\bibfnamefont {X.}~\bibnamefont {Siemens}},\ }\href
  {\doibase 10.1103/PhysRevD.79.084030} {\bibfield  {journal} {\bibinfo
  {journal} {Phys. Rev. D}\ }\textbf {\bibinfo {volume} {79}},\ \bibinfo
  {pages} {084030} (\bibinfo {year} {2009})},\ \Eprint
  {http://arxiv.org/abs/0809.0701} {arXiv:0809.0701 [gr-qc]} \BibitemShut
  {NoStop}%
\bibitem [{\citenamefont {Chamberlin}\ \emph {et~al.}(2015)\citenamefont
  {Chamberlin}, \citenamefont {Creighton}, \citenamefont {Siemens},
  \citenamefont {Demorest}, \citenamefont {Ellis}, \citenamefont {Price},\ and\
  \citenamefont {Romano}}]{Chamberlin:2014ria}%
  \BibitemOpen
  \bibfield  {author} {\bibinfo {author} {\bibfnamefont {S.~J.}\ \bibnamefont
  {Chamberlin}}, \bibinfo {author} {\bibfnamefont {J.~D.~E.}\ \bibnamefont
  {Creighton}}, \bibinfo {author} {\bibfnamefont {X.}~\bibnamefont {Siemens}},
  \bibinfo {author} {\bibfnamefont {P.}~\bibnamefont {Demorest}}, \bibinfo
  {author} {\bibfnamefont {J.}~\bibnamefont {Ellis}}, \bibinfo {author}
  {\bibfnamefont {L.~R.}\ \bibnamefont {Price}}, \ and\ \bibinfo {author}
  {\bibfnamefont {J.~D.}\ \bibnamefont {Romano}},\ }\href {\doibase
  10.1103/PhysRevD.91.044048} {\bibfield  {journal} {\bibinfo  {journal} {Phys.
  Rev. D}\ }\textbf {\bibinfo {volume} {91}},\ \bibinfo {pages} {044048}
  (\bibinfo {year} {2015})},\ \Eprint {http://arxiv.org/abs/1410.8256}
  {arXiv:1410.8256 [astro-ph.IM]} \BibitemShut {NoStop}%
\bibitem [{\citenamefont {Rosado}\ \emph {et~al.}(2015)\citenamefont {Rosado},
  \citenamefont {Sesana},\ and\ \citenamefont {Gair}}]{Rosado:2015epa}%
  \BibitemOpen
  \bibfield  {author} {\bibinfo {author} {\bibfnamefont {P.~A.}\ \bibnamefont
  {Rosado}}, \bibinfo {author} {\bibfnamefont {A.}~\bibnamefont {Sesana}}, \
  and\ \bibinfo {author} {\bibfnamefont {J.}~\bibnamefont {Gair}},\ }\href
  {\doibase 10.1093/mnras/stv1098} {\bibfield  {journal} {\bibinfo  {journal}
  {Mon. Not. Roy. Astron. Soc.}\ }\textbf {\bibinfo {volume} {451}},\ \bibinfo
  {pages} {2417} (\bibinfo {year} {2015})},\ \Eprint
  {http://arxiv.org/abs/1503.04803} {arXiv:1503.04803 [astro-ph.HE]}
  \BibitemShut {NoStop}%
\bibitem [{\citenamefont {DeRocco}\ and\ \citenamefont
  {Dror}(2024)}]{DeRocco:2022irl}%
  \BibitemOpen
  \bibfield  {author} {\bibinfo {author} {\bibfnamefont {W.}~\bibnamefont
  {DeRocco}}\ and\ \bibinfo {author} {\bibfnamefont {J.~A.}\ \bibnamefont
  {Dror}},\ }\href {\doibase 10.1103/PhysRevLett.132.101403} {\bibfield
  {journal} {\bibinfo  {journal} {Phys. Rev. Lett.}\ }\textbf {\bibinfo
  {volume} {132}},\ \bibinfo {pages} {101403} (\bibinfo {year} {2024})},\
  \Eprint {http://arxiv.org/abs/2212.09751} {arXiv:2212.09751 [astro-ph.HE]}
  \BibitemShut {NoStop}%
\bibitem [{\citenamefont {DeRocco}\ and\ \citenamefont
  {Dror}(2023)}]{DeRocco:2023qae}%
  \BibitemOpen
  \bibfield  {author} {\bibinfo {author} {\bibfnamefont {W.}~\bibnamefont
  {DeRocco}}\ and\ \bibinfo {author} {\bibfnamefont {J.~A.}\ \bibnamefont
  {Dror}},\ }\href {\doibase 10.1103/PhysRevD.108.103011} {\bibfield  {journal}
  {\bibinfo  {journal} {Phys. Rev. D}\ }\textbf {\bibinfo {volume} {108}},\
  \bibinfo {pages} {103011} (\bibinfo {year} {2023})},\ \Eprint
  {http://arxiv.org/abs/2304.13042} {arXiv:2304.13042 [astro-ph.HE]}
  \BibitemShut {NoStop}%
\bibitem [{\citenamefont {Chalumeau}\ \emph {et~al.}(2021)\citenamefont
  {Chalumeau} \emph {et~al.}}]{EPTA:2021fqa}%
  \BibitemOpen
  \bibfield  {author} {\bibinfo {author} {\bibfnamefont {A.}~\bibnamefont
  {Chalumeau}} \emph {et~al.} (\bibinfo {collaboration} {EPTA}),\ }\href
  {\doibase 10.1093/mnras/stab3283} {\bibfield  {journal} {\bibinfo  {journal}
  {Mon. Not. Roy. Astron. Soc.}\ }\textbf {\bibinfo {volume} {509}},\ \bibinfo
  {pages} {5538} (\bibinfo {year} {2021})},\ \Eprint
  {http://arxiv.org/abs/2111.05186} {arXiv:2111.05186 [astro-ph.HE]}
  \BibitemShut {NoStop}%
\bibitem [{\citenamefont {Caballero}\ \emph {et~al.}(2016)\citenamefont
  {Caballero} \emph {et~al.}}]{EPTA:2015ike}%
  \BibitemOpen
  \bibfield  {author} {\bibinfo {author} {\bibfnamefont {R.~N.}\ \bibnamefont
  {Caballero}} \emph {et~al.} (\bibinfo {collaboration} {EPTA}),\ }\href
  {\doibase 10.1093/mnras/stw179} {\bibfield  {journal} {\bibinfo  {journal}
  {Mon. Not. Roy. Astron. Soc.}\ }\textbf {\bibinfo {volume} {457}},\ \bibinfo
  {pages} {4421} (\bibinfo {year} {2016})},\ \Eprint
  {http://arxiv.org/abs/1510.09194} {arXiv:1510.09194 [astro-ph.IM]}
  \BibitemShut {NoStop}%
\bibitem [{\citenamefont {Caprini}\ and\ \citenamefont
  {Figueroa}(2018)}]{Caprini:2018mtu}%
  \BibitemOpen
  \bibfield  {author} {\bibinfo {author} {\bibfnamefont {C.}~\bibnamefont
  {Caprini}}\ and\ \bibinfo {author} {\bibfnamefont {D.~G.}\ \bibnamefont
  {Figueroa}},\ }\href {\doibase 10.1088/1361-6382/aac608} {\bibfield
  {journal} {\bibinfo  {journal} {Class. Quant. Grav.}\ }\textbf {\bibinfo
  {volume} {35}},\ \bibinfo {pages} {163001} (\bibinfo {year} {2018})},\
  \Eprint {http://arxiv.org/abs/1801.04268} {arXiv:1801.04268 [astro-ph.CO]}
  \BibitemShut {NoStop}%
\bibitem [{\citenamefont {Kehagias}\ and\ \citenamefont
  {Riotto}(2024)}]{Kehagias:2024plp}%
  \BibitemOpen
  \bibfield  {author} {\bibinfo {author} {\bibfnamefont {A.}~\bibnamefont
  {Kehagias}}\ and\ \bibinfo {author} {\bibfnamefont {A.}~\bibnamefont
  {Riotto}},\ }\href@noop {} {\  (\bibinfo {year} {2024})},\ \Eprint
  {http://arxiv.org/abs/2401.10680} {arXiv:2401.10680 [gr-qc]} \BibitemShut
  {NoStop}%
\bibitem [{\citenamefont {Gair}\ \emph {et~al.}(2014)\citenamefont {Gair},
  \citenamefont {Romano}, \citenamefont {Taylor},\ and\ \citenamefont
  {Mingarelli}}]{Gair:2014rwa}%
  \BibitemOpen
  \bibfield  {author} {\bibinfo {author} {\bibfnamefont {J.}~\bibnamefont
  {Gair}}, \bibinfo {author} {\bibfnamefont {J.~D.}\ \bibnamefont {Romano}},
  \bibinfo {author} {\bibfnamefont {S.}~\bibnamefont {Taylor}}, \ and\ \bibinfo
  {author} {\bibfnamefont {C.~M.~F.}\ \bibnamefont {Mingarelli}},\ }\href
  {\doibase 10.1103/PhysRevD.90.082001} {\bibfield  {journal} {\bibinfo
  {journal} {Phys. Rev. D}\ }\textbf {\bibinfo {volume} {90}},\ \bibinfo
  {pages} {082001} (\bibinfo {year} {2014})},\ \Eprint
  {http://arxiv.org/abs/1406.4664} {arXiv:1406.4664 [gr-qc]} \BibitemShut
  {NoStop}%
\bibitem [{\citenamefont {Romano}\ and\ \citenamefont
  {Allen}(2023)}]{Romano:2023zhb}%
  \BibitemOpen
  \bibfield  {author} {\bibinfo {author} {\bibfnamefont {J.~D.}\ \bibnamefont
  {Romano}}\ and\ \bibinfo {author} {\bibfnamefont {B.}~\bibnamefont {Allen}},\
  }\href@noop {} {\  (\bibinfo {year} {2023})},\ \Eprint
  {http://arxiv.org/abs/2308.05847} {arXiv:2308.05847 [gr-qc]} \BibitemShut
  {NoStop}%
\bibitem [{\citenamefont {Agazie}\ \emph
  {et~al.}(2023{\natexlab{d}})\citenamefont {Agazie} \emph
  {et~al.}}]{NANOGrav:2023ctt}%
  \BibitemOpen
  \bibfield  {author} {\bibinfo {author} {\bibfnamefont {G.}~\bibnamefont
  {Agazie}} \emph {et~al.} (\bibinfo {collaboration} {NANOGrav}),\ }\href
  {\doibase 10.3847/2041-8213/acda88} {\bibfield  {journal} {\bibinfo
  {journal} {Astrophys. J. Lett.}\ }\textbf {\bibinfo {volume} {951}},\
  \bibinfo {pages} {L10} (\bibinfo {year} {2023}{\natexlab{d}})},\ \Eprint
  {http://arxiv.org/abs/2306.16218} {arXiv:2306.16218 [astro-ph.HE]}
  \BibitemShut {NoStop}%
\bibitem [{\citenamefont {Agazie}\ \emph
  {et~al.}(2023{\natexlab{e}})\citenamefont {Agazie} \emph
  {et~al.}}]{InternationalPulsarTimingArray:2023mzf}%
  \BibitemOpen
  \bibfield  {author} {\bibinfo {author} {\bibfnamefont {G.}~\bibnamefont
  {Agazie}} \emph {et~al.} (\bibinfo {collaboration} {International Pulsar
  Timing Array}),\ }\href@noop {} {\  (\bibinfo {year} {2023}{\natexlab{e}})},\
  \Eprint {http://arxiv.org/abs/2309.00693} {arXiv:2309.00693 [astro-ph.HE]}
  \BibitemShut {NoStop}%
\bibitem [{\citenamefont {Contaldi}\ \emph {et~al.}(2020)\citenamefont
  {Contaldi}, \citenamefont {Pieroni}, \citenamefont {Renzini}, \citenamefont
  {Cusin}, \citenamefont {Karnesis}, \citenamefont {Peloso}, \citenamefont
  {Ricciardone},\ and\ \citenamefont {Tasinato}}]{Contaldi:2020rht}%
  \BibitemOpen
  \bibfield  {author} {\bibinfo {author} {\bibfnamefont {C.~R.}\ \bibnamefont
  {Contaldi}}, \bibinfo {author} {\bibfnamefont {M.}~\bibnamefont {Pieroni}},
  \bibinfo {author} {\bibfnamefont {A.~I.}\ \bibnamefont {Renzini}}, \bibinfo
  {author} {\bibfnamefont {G.}~\bibnamefont {Cusin}}, \bibinfo {author}
  {\bibfnamefont {N.}~\bibnamefont {Karnesis}}, \bibinfo {author}
  {\bibfnamefont {M.}~\bibnamefont {Peloso}}, \bibinfo {author} {\bibfnamefont
  {A.}~\bibnamefont {Ricciardone}}, \ and\ \bibinfo {author} {\bibfnamefont
  {G.}~\bibnamefont {Tasinato}},\ }\href {\doibase 10.1103/PhysRevD.102.043502}
  {\bibfield  {journal} {\bibinfo  {journal} {Phys. Rev. D}\ }\textbf {\bibinfo
  {volume} {102}},\ \bibinfo {pages} {043502} (\bibinfo {year} {2020})},\
  \Eprint {http://arxiv.org/abs/2006.03313} {arXiv:2006.03313 [astro-ph.CO]}
  \BibitemShut {NoStop}%
\bibitem [{\citenamefont {Bond}\ \emph {et~al.}(1998)\citenamefont {Bond},
  \citenamefont {Jaffe},\ and\ \citenamefont {Knox}}]{Bond:1998zw}%
  \BibitemOpen
  \bibfield  {author} {\bibinfo {author} {\bibfnamefont {J.~R.}\ \bibnamefont
  {Bond}}, \bibinfo {author} {\bibfnamefont {A.~H.}\ \bibnamefont {Jaffe}}, \
  and\ \bibinfo {author} {\bibfnamefont {L.}~\bibnamefont {Knox}},\ }\href
  {\doibase 10.1103/PhysRevD.57.2117} {\bibfield  {journal} {\bibinfo
  {journal} {Phys. Rev. D}\ }\textbf {\bibinfo {volume} {57}},\ \bibinfo
  {pages} {2117} (\bibinfo {year} {1998})},\ \Eprint
  {http://arxiv.org/abs/astro-ph/9708203} {arXiv:astro-ph/9708203} \BibitemShut
  {NoStop}%
\bibitem [{\citenamefont {Foreman-Mackey}\ \emph {et~al.}(2013)\citenamefont
  {Foreman-Mackey}, \citenamefont {Hogg}, \citenamefont {Lang},\ and\
  \citenamefont {Goodman}}]{Foreman-Mackey:2012any}%
  \BibitemOpen
  \bibfield  {author} {\bibinfo {author} {\bibfnamefont {D.}~\bibnamefont
  {Foreman-Mackey}}, \bibinfo {author} {\bibfnamefont {D.~W.}\ \bibnamefont
  {Hogg}}, \bibinfo {author} {\bibfnamefont {D.}~\bibnamefont {Lang}}, \ and\
  \bibinfo {author} {\bibfnamefont {J.}~\bibnamefont {Goodman}},\ }\href
  {\doibase 10.1086/670067} {\bibfield  {journal} {\bibinfo  {journal} {Publ.
  Astron. Soc. Pac.}\ }\textbf {\bibinfo {volume} {125}},\ \bibinfo {pages}
  {306} (\bibinfo {year} {2013})},\ \Eprint {http://arxiv.org/abs/1202.3665}
  {arXiv:1202.3665 [astro-ph.IM]} \BibitemShut {NoStop}%
\bibitem [{\citenamefont {Janssen}\ \emph {et~al.}(2015)\citenamefont {Janssen}
  \emph {et~al.}}]{Janssen:2014dka}%
  \BibitemOpen
  \bibfield  {author} {\bibinfo {author} {\bibfnamefont {G.}~\bibnamefont
  {Janssen}} \emph {et~al.},\ }\href {\doibase 10.22323/1.215.0037} {\bibfield
  {journal} {\bibinfo  {journal} {PoS}\ }\textbf {\bibinfo {volume}
  {AASKA14}},\ \bibinfo {pages} {037} (\bibinfo {year} {2015})},\ \Eprint
  {http://arxiv.org/abs/1501.00127} {arXiv:1501.00127 [astro-ph.IM]}
  \BibitemShut {NoStop}%
\bibitem [{\citenamefont {Lazio}(2013)}]{Lazio:2013mea}%
  \BibitemOpen
  \bibfield  {author} {\bibinfo {author} {\bibfnamefont {T.~J.~W.}\
  \bibnamefont {Lazio}},\ }\href {\doibase 10.1088/0264-9381/30/22/224011}
  {\bibfield  {journal} {\bibinfo  {journal} {Class. Quant. Grav.}\ }\textbf
  {\bibinfo {volume} {30}},\ \bibinfo {pages} {224011} (\bibinfo {year}
  {2013})}\BibitemShut {NoStop}%
\bibitem [{\citenamefont {van Haasteren}\ and\ \citenamefont
  {Vallisneri}(2014)}]{vanHaasteren:2014qva}%
  \BibitemOpen
  \bibfield  {author} {\bibinfo {author} {\bibfnamefont {R.}~\bibnamefont {van
  Haasteren}}\ and\ \bibinfo {author} {\bibfnamefont {M.}~\bibnamefont
  {Vallisneri}},\ }\href {\doibase 10.1103/PhysRevD.90.104012} {\bibfield
  {journal} {\bibinfo  {journal} {Phys. Rev. D}\ }\textbf {\bibinfo {volume}
  {90}},\ \bibinfo {pages} {104012} (\bibinfo {year} {2014})},\ \Eprint
  {http://arxiv.org/abs/1407.1838} {arXiv:1407.1838 [gr-qc]} \BibitemShut
  {NoStop}%
\bibitem [{\citenamefont {Antoniadis}\ \emph
  {et~al.}(2023{\natexlab{d}})\citenamefont {Antoniadis} \emph
  {et~al.}}]{EPTA:2023akd}%
  \BibitemOpen
  \bibfield  {author} {\bibinfo {author} {\bibfnamefont {J.}~\bibnamefont
  {Antoniadis}} \emph {et~al.} (\bibinfo {collaboration} {EPTA, InPTA}),\
  }\href {\doibase 10.1051/0004-6361/202346842} {\bibfield  {journal} {\bibinfo
   {journal} {Astron. Astrophys.}\ }\textbf {\bibinfo {volume} {678}},\
  \bibinfo {pages} {A49} (\bibinfo {year} {2023}{\natexlab{d}})},\ \Eprint
  {http://arxiv.org/abs/2306.16225} {arXiv:2306.16225 [astro-ph.HE]}
  \BibitemShut {NoStop}%
\bibitem [{\citenamefont {Ellis}\ \emph {et~al.}(2020)\citenamefont {Ellis},
  \citenamefont {Vallisneri}, \citenamefont {Taylor},\ and\ \citenamefont
  {Baker}}]{enterprise}%
  \BibitemOpen
  \bibfield  {author} {\bibinfo {author} {\bibfnamefont {J.~A.}\ \bibnamefont
  {Ellis}}, \bibinfo {author} {\bibfnamefont {M.}~\bibnamefont {Vallisneri}},
  \bibinfo {author} {\bibfnamefont {S.~R.}\ \bibnamefont {Taylor}}, \ and\
  \bibinfo {author} {\bibfnamefont {P.~T.}\ \bibnamefont {Baker}},\ }\href
  {\doibase 10.5281/zenodo.4059815} {\enquote {\bibinfo {title} {Enterprise:
  Enhanced numerical toolbox enabling a robust pulsar inference suite},}\
  }\bibinfo {howpublished} {Zenodo} (\bibinfo {year} {2020})\BibitemShut
  {NoStop}%
\bibitem [{\citenamefont {Taylor}\ \emph {et~al.}(2021)\citenamefont {Taylor},
  \citenamefont {Baker}, \citenamefont {Hazboun}, \citenamefont {Simon},\ and\
  \citenamefont {Vigeland}}]{enterprise_ext}%
  \BibitemOpen
  \bibfield  {author} {\bibinfo {author} {\bibfnamefont {S.~R.}\ \bibnamefont
  {Taylor}}, \bibinfo {author} {\bibfnamefont {P.~T.}\ \bibnamefont {Baker}},
  \bibinfo {author} {\bibfnamefont {J.~S.}\ \bibnamefont {Hazboun}}, \bibinfo
  {author} {\bibfnamefont {J.}~\bibnamefont {Simon}}, \ and\ \bibinfo {author}
  {\bibfnamefont {S.~J.}\ \bibnamefont {Vigeland}},\ }\href
  {https://github.com/nanograv/enterprise_extensions} {\enquote {\bibinfo
  {title} {enterprise\_extensions},}\ } (\bibinfo {year} {2021}),\ \bibinfo
  {note} {v2.3.3}\BibitemShut {NoStop}%
\bibitem [{\citenamefont {Ellis}\ and\ \citenamefont {van
  Haasteren}(2017)}]{justin_ellis_2017_1037579}%
  \BibitemOpen
  \bibfield  {author} {\bibinfo {author} {\bibfnamefont {J.}~\bibnamefont
  {Ellis}}\ and\ \bibinfo {author} {\bibfnamefont {R.}~\bibnamefont {van
  Haasteren}},\ }\href {\doibase 10.5281/zenodo.1037579} {\enquote {\bibinfo
  {title} {jellis18/ptmcmcsampler: Official release},}\ } (\bibinfo {year}
  {2017})\BibitemShut {NoStop}%
\bibitem [{\citenamefont {Siemens}\ \emph {et~al.}(2013)\citenamefont
  {Siemens}, \citenamefont {Ellis}, \citenamefont {Jenet},\ and\ \citenamefont
  {Romano}}]{Siemens:2013zla}%
  \BibitemOpen
  \bibfield  {author} {\bibinfo {author} {\bibfnamefont {X.}~\bibnamefont
  {Siemens}}, \bibinfo {author} {\bibfnamefont {J.}~\bibnamefont {Ellis}},
  \bibinfo {author} {\bibfnamefont {F.}~\bibnamefont {Jenet}}, \ and\ \bibinfo
  {author} {\bibfnamefont {J.~D.}\ \bibnamefont {Romano}},\ }\href {\doibase
  10.1088/0264-9381/30/22/224015} {\bibfield  {journal} {\bibinfo  {journal}
  {Class. Quant. Grav.}\ }\textbf {\bibinfo {volume} {30}},\ \bibinfo {pages}
  {224015} (\bibinfo {year} {2013})},\ \Eprint {http://arxiv.org/abs/1305.3196}
  {arXiv:1305.3196 [astro-ph.IM]} \BibitemShut {NoStop}%
\bibitem [{\citenamefont {Pol}\ \emph {et~al.}(2021)\citenamefont {Pol} \emph
  {et~al.}}]{NANOGrav:2020spf}%
  \BibitemOpen
  \bibfield  {author} {\bibinfo {author} {\bibfnamefont {N.~S.}\ \bibnamefont
  {Pol}} \emph {et~al.} (\bibinfo {collaboration} {NANOGrav}),\ }\href
  {\doibase 10.3847/2041-8213/abf2c9} {\bibfield  {journal} {\bibinfo
  {journal} {Astrophys. J. Lett.}\ }\textbf {\bibinfo {volume} {911}},\
  \bibinfo {pages} {L34} (\bibinfo {year} {2021})},\ \Eprint
  {http://arxiv.org/abs/2010.11950} {arXiv:2010.11950 [astro-ph.HE]}
  \BibitemShut {NoStop}%
\bibitem [{\citenamefont {Liang}\ \emph {et~al.}(2023)\citenamefont {Liang},
  \citenamefont {Lin},\ and\ \citenamefont {Trodden}}]{Liang:2023ary}%
  \BibitemOpen
  \bibfield  {author} {\bibinfo {author} {\bibfnamefont {Q.}~\bibnamefont
  {Liang}}, \bibinfo {author} {\bibfnamefont {M.-X.}\ \bibnamefont {Lin}}, \
  and\ \bibinfo {author} {\bibfnamefont {M.}~\bibnamefont {Trodden}},\ }\href
  {\doibase 10.1088/1475-7516/2023/11/042} {\bibfield  {journal} {\bibinfo
  {journal} {JCAP}\ }\textbf {\bibinfo {volume} {11}},\ \bibinfo {pages} {042}
  (\bibinfo {year} {2023})},\ \Eprint {http://arxiv.org/abs/2304.02640}
  {arXiv:2304.02640 [astro-ph.CO]} \BibitemShut {NoStop}%
\bibitem [{\citenamefont {Roebber}\ and\ \citenamefont
  {Holder}(2017)}]{Roebber:2016jzl}%
  \BibitemOpen
  \bibfield  {author} {\bibinfo {author} {\bibfnamefont {E.}~\bibnamefont
  {Roebber}}\ and\ \bibinfo {author} {\bibfnamefont {G.}~\bibnamefont
  {Holder}},\ }\href {\doibase 10.3847/1538-4357/835/1/21} {\bibfield
  {journal} {\bibinfo  {journal} {Astrophys. J.}\ }\textbf {\bibinfo {volume}
  {835}},\ \bibinfo {pages} {21} (\bibinfo {year} {2017})},\ \Eprint
  {http://arxiv.org/abs/1609.06758} {arXiv:1609.06758 [astro-ph.CO]}
  \BibitemShut {NoStop}%
\bibitem [{\citenamefont {Allen}(2023)}]{Allen:2022dzg}%
  \BibitemOpen
  \bibfield  {author} {\bibinfo {author} {\bibfnamefont {B.}~\bibnamefont
  {Allen}},\ }\href {\doibase 10.1103/PhysRevD.107.043018} {\bibfield
  {journal} {\bibinfo  {journal} {Phys. Rev. D}\ }\textbf {\bibinfo {volume}
  {107}},\ \bibinfo {pages} {043018} (\bibinfo {year} {2023})},\ \Eprint
  {http://arxiv.org/abs/2205.05637} {arXiv:2205.05637 [gr-qc]} \BibitemShut
  {NoStop}%
\bibitem [{\citenamefont {Aghanim}\ \emph {et~al.}(2020)\citenamefont {Aghanim}
  \emph {et~al.}}]{Planck:2018vyg}%
  \BibitemOpen
  \bibfield  {author} {\bibinfo {author} {\bibfnamefont {N.}~\bibnamefont
  {Aghanim}} \emph {et~al.} (\bibinfo {collaboration} {Planck}),\ }\href
  {\doibase 10.1051/0004-6361/201833910} {\bibfield  {journal} {\bibinfo
  {journal} {Astron. Astrophys.}\ }\textbf {\bibinfo {volume} {641}},\ \bibinfo
  {pages} {A6} (\bibinfo {year} {2020})},\ \bibinfo {note} {[Erratum:
  Astron.Astrophys. 652, C4 (2021)]},\ \Eprint
  {http://arxiv.org/abs/1807.06209} {arXiv:1807.06209 [astro-ph.CO]}
  \BibitemShut {NoStop}%
\bibitem [{\citenamefont {Ellis}\ \emph {et~al.}(2024)\citenamefont {Ellis},
  \citenamefont {Fairbairn}, \citenamefont {Franciolini}, \citenamefont
  {H\"utsi}, \citenamefont {Iovino}, \citenamefont {Lewicki}, \citenamefont
  {Raidal}, \citenamefont {Urrutia}, \citenamefont {Vaskonen},\ and\
  \citenamefont {Veerm\"ae}}]{Ellis:2023oxs}%
  \BibitemOpen
  \bibfield  {author} {\bibinfo {author} {\bibfnamefont {J.}~\bibnamefont
  {Ellis}}, \bibinfo {author} {\bibfnamefont {M.}~\bibnamefont {Fairbairn}},
  \bibinfo {author} {\bibfnamefont {G.}~\bibnamefont {Franciolini}}, \bibinfo
  {author} {\bibfnamefont {G.}~\bibnamefont {H\"utsi}}, \bibinfo {author}
  {\bibfnamefont {A.}~\bibnamefont {Iovino}}, \bibinfo {author} {\bibfnamefont
  {M.}~\bibnamefont {Lewicki}}, \bibinfo {author} {\bibfnamefont
  {M.}~\bibnamefont {Raidal}}, \bibinfo {author} {\bibfnamefont
  {J.}~\bibnamefont {Urrutia}}, \bibinfo {author} {\bibfnamefont
  {V.}~\bibnamefont {Vaskonen}}, \ and\ \bibinfo {author} {\bibfnamefont
  {H.}~\bibnamefont {Veerm\"ae}},\ }\href {\doibase
  10.1103/PhysRevD.109.023522} {\bibfield  {journal} {\bibinfo  {journal}
  {Phys. Rev. D}\ }\textbf {\bibinfo {volume} {109}},\ \bibinfo {pages}
  {023522} (\bibinfo {year} {2024})},\ \Eprint
  {http://arxiv.org/abs/2308.08546} {arXiv:2308.08546 [astro-ph.CO]}
  \BibitemShut {NoStop}%
\bibitem [{\citenamefont {et~al.}(2024)}]{inprepCGWB}%
  \BibitemOpen
  \bibfield  {author} {\bibinfo {author} {\bibfnamefont {C.~C.}\ \bibnamefont
  {et~al.}},\ }\href@noop {} {\  (\bibinfo {year} {2024})},\ \bibinfo {note}
  {in preparation}\BibitemShut {NoStop}%
\bibitem [{\citenamefont {Babak}\ and\ \citenamefont
  {Sesana}(2012)}]{Babak:2011mr}%
  \BibitemOpen
  \bibfield  {author} {\bibinfo {author} {\bibfnamefont {S.}~\bibnamefont
  {Babak}}\ and\ \bibinfo {author} {\bibfnamefont {A.}~\bibnamefont {Sesana}},\
  }\href {\doibase 10.1103/PhysRevD.85.044034} {\bibfield  {journal} {\bibinfo
  {journal} {Phys. Rev. D}\ }\textbf {\bibinfo {volume} {85}},\ \bibinfo
  {pages} {044034} (\bibinfo {year} {2012})},\ \Eprint
  {http://arxiv.org/abs/1112.1075} {arXiv:1112.1075 [astro-ph.CO]} \BibitemShut
  {NoStop}%
\bibitem [{\citenamefont {Ellis}\ \emph {et~al.}(2012)\citenamefont {Ellis},
  \citenamefont {Siemens},\ and\ \citenamefont {Creighton}}]{Ellis:2012zv}%
  \BibitemOpen
  \bibfield  {author} {\bibinfo {author} {\bibfnamefont {J.~A.}\ \bibnamefont
  {Ellis}}, \bibinfo {author} {\bibfnamefont {X.}~\bibnamefont {Siemens}}, \
  and\ \bibinfo {author} {\bibfnamefont {J.~D.~E.}\ \bibnamefont {Creighton}},\
  }\href {\doibase 10.1088/0004-637X/756/2/175} {\bibfield  {journal} {\bibinfo
   {journal} {Astrophys. J.}\ }\textbf {\bibinfo {volume} {756}},\ \bibinfo
  {pages} {175} (\bibinfo {year} {2012})},\ \Eprint
  {http://arxiv.org/abs/1204.4218} {arXiv:1204.4218 [astro-ph.IM]} \BibitemShut
  {NoStop}%
\bibitem [{\citenamefont {Ali-Ha\"\i{}moud}\ \emph {et~al.}(2020)\citenamefont
  {Ali-Ha\"\i{}moud}, \citenamefont {Smith},\ and\ \citenamefont
  {Mingarelli}}]{Ali-Haimoud:2020ozu}%
  \BibitemOpen
  \bibfield  {author} {\bibinfo {author} {\bibfnamefont {Y.}~\bibnamefont
  {Ali-Ha\"\i{}moud}}, \bibinfo {author} {\bibfnamefont {T.~L.}\ \bibnamefont
  {Smith}}, \ and\ \bibinfo {author} {\bibfnamefont {C.~M.~F.}\ \bibnamefont
  {Mingarelli}},\ }\href {\doibase 10.1103/PhysRevD.102.122005} {\bibfield
  {journal} {\bibinfo  {journal} {Phys. Rev. D}\ }\textbf {\bibinfo {volume}
  {102}},\ \bibinfo {pages} {122005} (\bibinfo {year} {2020})},\ \Eprint
  {http://arxiv.org/abs/2006.14570} {arXiv:2006.14570 [gr-qc]} \BibitemShut
  {NoStop}%
\bibitem [{\citenamefont {Ali-Ha\"\i{}moud}\ \emph {et~al.}(2021)\citenamefont
  {Ali-Ha\"\i{}moud}, \citenamefont {Smith},\ and\ \citenamefont
  {Mingarelli}}]{Ali-Haimoud:2020iyz}%
  \BibitemOpen
  \bibfield  {author} {\bibinfo {author} {\bibfnamefont {Y.}~\bibnamefont
  {Ali-Ha\"\i{}moud}}, \bibinfo {author} {\bibfnamefont {T.~L.}\ \bibnamefont
  {Smith}}, \ and\ \bibinfo {author} {\bibfnamefont {C.~M.~F.}\ \bibnamefont
  {Mingarelli}},\ }\href {\doibase 10.1103/PhysRevD.103.042009} {\bibfield
  {journal} {\bibinfo  {journal} {Phys. Rev. D}\ }\textbf {\bibinfo {volume}
  {103}},\ \bibinfo {pages} {042009} (\bibinfo {year} {2021})},\ \Eprint
  {http://arxiv.org/abs/2010.13958} {arXiv:2010.13958 [gr-qc]} \BibitemShut
  {NoStop}%
\bibitem [{\citenamefont {Cruz}\ \emph {et~al.}(2024)\citenamefont {Cruz},
  \citenamefont {Malhotra}, \citenamefont {Tasinato},\ and\ \citenamefont
  {Zavala}}]{Cruz:2024svc}%
  \BibitemOpen
  \bibfield  {author} {\bibinfo {author} {\bibfnamefont {N.~M.~J.}\
  \bibnamefont {Cruz}}, \bibinfo {author} {\bibfnamefont {A.}~\bibnamefont
  {Malhotra}}, \bibinfo {author} {\bibfnamefont {G.}~\bibnamefont {Tasinato}},
  \ and\ \bibinfo {author} {\bibfnamefont {I.}~\bibnamefont {Zavala}},\
  }\href@noop {} {\  (\bibinfo {year} {2024})},\ \Eprint
  {http://arxiv.org/abs/2402.17312} {arXiv:2402.17312 [gr-qc]} \BibitemShut
  {NoStop}%
\end{thebibliography}%
\end{document}